\documentclass[a4paper,11pt]{article}
\usepackage{jheppub} % for details on the use of the package, please see the JINST-author-manual
\usepackage{lineno}
\usepackage{braket}
\usepackage{amsmath}
\usepackage{csquotes}
\usepackage{tikz}
\usepackage{hyperref}
\DeclareMathOperator{\sn}{sn}

\DeclareMathOperator{\arcsinh}{arcsinh}
\DeclareMathOperator{\Tr}{Tr}

\title{\boldmath  
Blowing-up the edge: connection formulae and stability chart of the Lam\'e equation}
\author{Giulio Bonelli, Pavlo Gavrylenko, Tommaso Pedroni and Alessandro Tanzini }
\affiliation{SISSA, Via Bonomea 265, 34136 Trieste, Italy}
\affiliation{INFN, Sezione di Trieste, Trieste, Italy}
\affiliation{Institute for Geometry and Physics, IGAP, via Beirut 2, 34136 Trieste, Italy}

\emailAdd{bonelli@sissa.it, pasha.145@gmail.com, tpedroni@sissa.it, tanzini@sissa.it}

\abstract{
We study periodic spectral problems through their connection with supersymmetric gauge theories and two-dimensional conformal field theory. To characterize the associated stability chart, we develop a novel and systematic approach for analyzing semi-classical Virasoro blocks near their poles. Via the AGT correspondence, these blocks correspond to $SU(2)$ Nekrasov partition functions in the Nekrasov–Shatashvili limit, which we propose to resum using an appropriate limit of {\it blow-up equations}. We show that the analytic structure of the resulting resummed partition functions features branch cuts located precisely {\it at the edges between bands and gaps in the spectrum of the associated quantum integrable system with periodic potential}. We examine the Nekrasov partition functions of $\mathcal{N}=2$ SQCD with $N_f\le 4$ flavors and of the $\mathcal{N}=2^*$ theory, which are related to the Heun equation, its confluent forms, and the Lamé equation. In the latter case, we analyze the spectrum in detail and solve the associated connection problem. Finally, we compare our results with those obtained via isomonodromic deformation techniques and the computation of orbifold surface defect partition functions in the $\mathcal{N}=2^*$ gauge theory, finding perfect agreement.}

\begin{document}
\maketitle
\flushbottom

\section{Introduction}
\label{sec:intro}
Periodic spectral problems naturally arise in the study of oscillatory systems throughout Mathematics and Physics. A key aspect of their analysis is understanding how the stability of solutions depends on the parameters of the problem, as captured by the so-called \textit{stability chart}. Recent advances in the study of differential equations, driven by new special functions emerging from supersymmetric localization, offer a fresh perspective on this class of problems. The simplest, yet highly non-trivial, case of a periodic spectral problem is Hill’s equation, a second-order ordinary differential equation with a periodic potential. A particularly notable instance is the {\it Lam\'e equation}:
\begin{equation}
\label{eq: first lame}
\left(\partial_{z}^2+E-\mu(\mu-1) \wp(z)\right)\psi(z)=0,
\end{equation}
presented here in its Weierstrass form. This equation, originally derived in \cite{lame1837} from the separation of variables of the Euclidean Laplacian in three dimensions using ellipsoidal coordinates, continues to find impactful applications in Physics, including in condensed matter, quantum field theory, and cosmology. More specifically, the Lamé equation arises, for instance, in the study of one-dimensional crystals \cite{PhysRevLett.50.873}; in the derivation of one-loop corrections to the quark–antiquark potential in $\mathcal{N}=4$ SYM \cite{Forini:2010ek}; in the computation of correlation functions in the one-dimensional defect CFT defined by a 1/2-BPS Wilson loop in $\mathcal{N}=4$ SYM, in the large-charge limit \cite{Giombi:2021zfb, Giombi:2022anm}; in the study of resonant particle production during post-inflationary preheating; and in the analysis of classical stability of periodic solitons \cite{Liang:1992mw}.

In the study of linear differential equations, solving the connection problem, i.e. relating local solutions expanded around different singularities, is a fundamental step in understanding the global behavior of solutions. The Lamé equation arises in the study of the semi-classical limit of Virasoro conformal blocks on the one-punctured torus with a single degenerate insertion. In this limit, the BPZ equation \cite{Belavin:1984vu} for the Virasoro conformal block reduces precisely to the Lam\'e equation. One can thus exploit the crossing symmetry of the underlying two-dimensional conformal field theory (2d CFT) to solve the connection problem. The very same procedure was used in \cite{Bonelli:2022ten} to study the Heun equation, which also admits a doubly periodic formulation -- \cite{NIST:DLMF} Section \href{https://dlmf.nist.gov/31.2.iv}{31.2}, and also the related works \cite{Litvinov:2013sxa,Jeong:2018qpc,Consoli:2022eey,Lisovyy:2022flm}.
An explicit combinatorial solution to the Lamé equation can be obtained via the AGT correspondence \cite{Alday:2009aq,LeFloch:2020uop, Alday:2010vg}, which relates Virasoro conformal blocks to supersymmetric partition functions computed through equivariant localization \cite{Nekrasov:2002qd}. In this framework, the semi-classical limit of the conformal block corresponds to the Nekrasov–Shatashvili limit \cite{Nekrasov:2009rc} of the associated supersymmetric partition function. Several studies have explored this correspondence in related contexts; see, for example, \cite{Bonelli:2010gk,Bonelli:2011na,Piatek:2013ifa,Piatek:2015jva,Basar:2015xna,Desiraju:2024fmo}.

A distinctive feature of the Lamé equation, as a specific instance of Hill equation, is the characteristic band-gap structure of its energy spectrum, where the edges between bands and gaps are determined by the eigenvalues associated with periodic and anti-periodic solutions. On the conformal field theory side, it is well known that conformal blocks exhibit a lattice of poles in the internal momentum. In the semi-classical limit, the internal momentum corresponds to the Floquet exponent of solutions to the Lamé equation, and the poles lie precisely at the edges between bands and gaps. As a consequence, a direct application of the standard methods breaks down in the vicinity of these points. In the Nekrasov–Shatashvili (NS) limit, the lattice of poles collapses along one direction, giving rise to a branch cut structure. To accurately capture this analytic behavior, an appropriate resummation of the conformal block—or equivalently, of the Nekrasov partition function—is required. A proposal for such a resummation was put forward in \cite{Beccaria:2016nnb,Beccaria:2016vxq,Beccaria:2016wop,Gorsky:2017ndg}, based on extrapolations from a direct analysis of the NS free energy at special values of the masses, where the algebraic integrability of the system allows for high-order computations. In the Mathieu case, although no algebraically integrable points exist, the eigenvalue expansion can nevertheless be computed to high orders, e.g. using the continued fraction method, making it possible to infer the resummed structure from the series data.

\begin{figure}[t!]
    \centering
\includegraphics[width=1\linewidth]{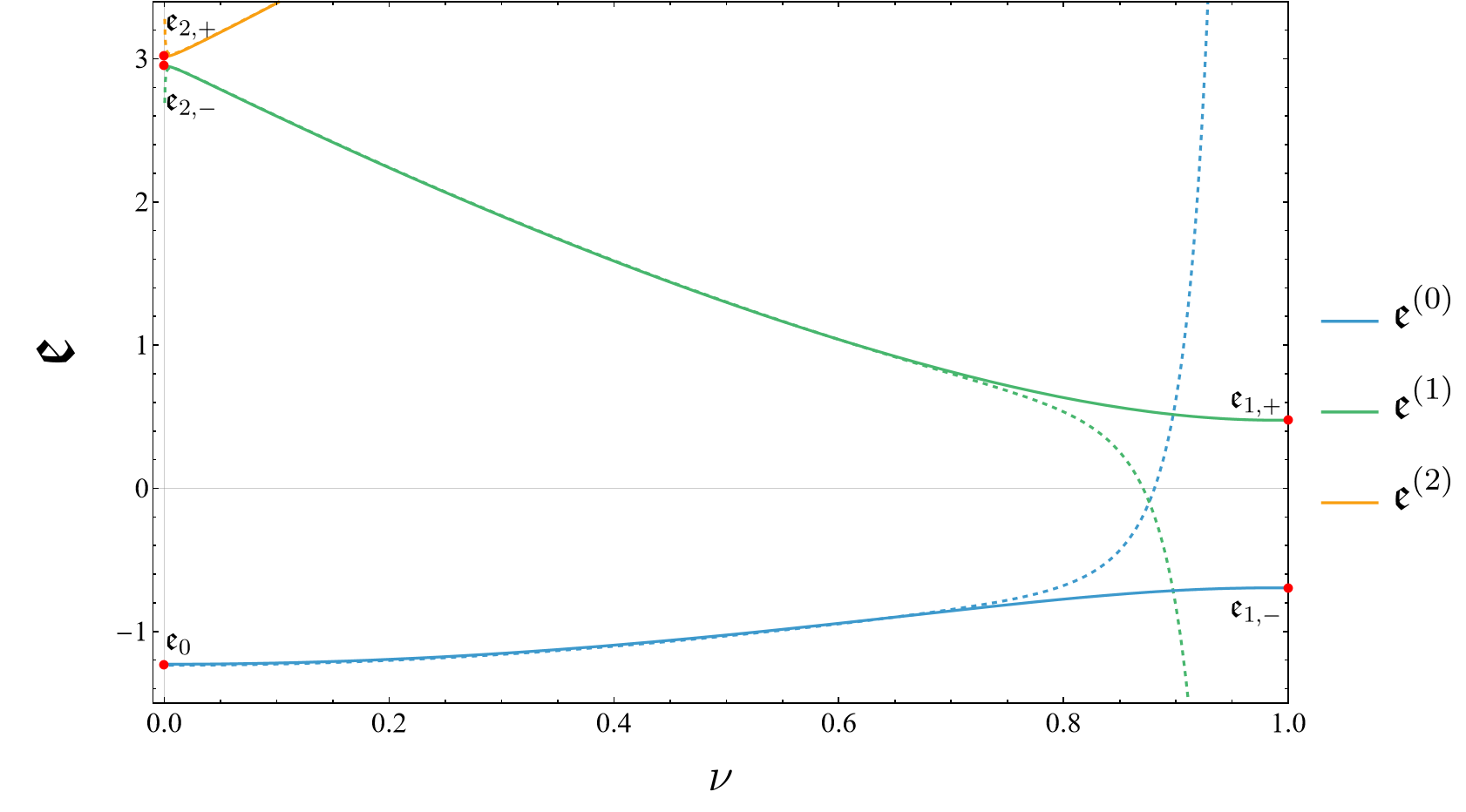}
    \caption{The first three bands obtained from equation (\ref{eq: bands k}) using the standard NS prepotential (dashed lines) are compared with those obtained using the resummed prepotential (solid lines), for the coupling $\mu=2.4$ and $\tau=i$. The two results show good agreement away from the band-gap edges, which are indicated by red dots.}
    \label{fig: bands k}
\end{figure}

In this paper, we propose a new systematic method to compute the resummed NS free energy using the Nakajima–Yoshioka blow-up equations \cite{nakajima2003lectures} and their $\mathbb{C}^2/\mathbb{Z}^2$ variant \cite{Bonelli:2011jx,Bonelli:2011kv}, which was first derived in \cite{Bershtein:2021uts} for the ${\mathcal N}=2^*$ theory. More precisely, we show that in the NS limit, the blow-up equations involve only the leading-order NS free energy $\mathcal{W}_0$ and its first $\epsilon_2$-correction $\mathcal{W}_1$. Indeed, the NS limit of the $\mathbb{C}^2$ blow-up equations for $SU(2)$ theories with $N_f$ fundamental hypermultiplets take the form:
\begin{equation}
    \sum_{n \in \mathbb{Z}+\frac{1}{2}}e^{ 2\mathcal{W}_{1}^{[N_{f}]}(a+n\hbar,\{\tilde{\mu}_i\},\hbar; t)-n\partial_{a}\mathcal{W}_{0}^{[N_{f}]}(a+n\hbar,\{\tilde{\mu}_i\},\hbar; t)-\partial_{\hbar}\mathcal{W}_{0}^{[N_{f}]}(a+n\hbar,\{\tilde{\mu}_i\},\hbar; t)}=0,
\end{equation}
while, for the $\mathcal{N}=2^*$ theory we obtain:
\begin{equation}
\begin{aligned}
& \theta_{2}(0|2\tau)\sum_{n \in \mathbb{Z}}e^{ 2\mathcal{W}_{1}(a+n\hbar,m,\hbar; \mathfrak{q})-n\partial_{a}\mathcal{W}_{0}(a+n\hbar,m,\hbar; \mathfrak{q})-\partial_{\hbar}\mathcal{W}_{0}(a+n\hbar,m,\hbar; \mathfrak{q})}\\
&-\theta_{3}(0|2\tau)\sum_{n \in \mathbb{Z}+\frac{1}{2}}e^{ 2\mathcal{W}_{1}(a+n\hbar,m,\hbar; \mathfrak{q})-n\partial_{a}\mathcal{W}_{0}(a+n\hbar,m,\hbar; \mathfrak{q})-\partial_{\hbar}\mathcal{W}_{0}(a+n\hbar,m,\hbar; \mathfrak{q})}=0.
\end{aligned}
\end{equation}
For the $\mathbb{C}^2/\mathbb{Z}^2$ case, see Eqs.~\eqref{eq: blow-up 1 pure ns} and \eqref{eq: blow-up 1 NS lame 1}.
We construct an ansatz for the solutions of the above equations in terms of suitable profile functions. For instance, in the case of the $\mathcal{N}=2^*$ theory, this takes the form:
\begin{subequations}
\begin{align}
\mathcal{W}_{0,\, inst}(a,\mu,\hbar; \mathfrak{q}) 
&= \hbar\sum_{j,k=1}^{\infty}\left[\sum_{\pm}
g_{k,j}\left(\frac{\mathfrak{q}^{j/2}}{j\pm2a/\hbar},\frac{\mu}{\hbar}\right)\right]\mathfrak{q}^{k+j/2-1}, \\
\mathcal{W}_{1,\, inst}(a, \mu,\hbar; \mathfrak{q}) 
&= \log\varphi(\mathfrak{q})^{-1}
+ \sum_{j,k=1}^{\infty}\left[\sum_{\pm}
\tilde{g}_{k,j}\left(\frac{\mathfrak{q}^{j/2}}{j\pm 2a/\hbar},\frac{\mu}{\hbar}\right)\right]\mathfrak{q}^{k+j/2-1} \nonumber\\
&\quad + \sum_{j,k=1}^{\infty}\left[\sum_{\pm}
\tilde{f}_{k,j}\left(\frac{\mathfrak{q}^{j/2}}{j\pm2a/\hbar},\frac{\mu}{\hbar}\right)\right]\mathfrak{q}^{k-1},
\end{align}
\end{subequations}
and similarly for the case with $N_f$ hypermultiplets, see Eqs.~\eqref{eq: resumm ansatz F0} and \eqref{eq: resumm ansatz F1}. The blow-up equations in the NS limit, when supplemented with appropriate boundary conditions, fully determine the profile functions, which turn out to coincide with those inferred in \cite{Gorsky:2017ndg,Beccaria:2016vxq}. These profile functions accurately capture the analytic properties of the resummed semi-classical blocks, exhibiting a branch cut structure that arises directly from solving the NS blow-up equations; see in particular the discussion around Eqs.~\eqref{eq: C2 log branch}, \eqref{eq: C2 root branch} and \eqref{eq: C2Z2 log branch}, \eqref{eq: C2Z2 root branch}. Indeed, the approach described above is quite general and can be extended to gauge theories with fundamental hypermultiplets, starting with the case $N_f = 4$, which corresponds to the four-point Virasoro conformal block on the sphere. In the special case $N_f = 0$, it provides a powerful tool for analyzing the Mathieu equation. Remarkably, a single blow-up equation suffices to fully determine both the resummed $\mathcal{W}_0$ and $\mathcal{W}_1$. Through the resummed NS free energy, we can approach the edges between bands and gaps—namely, the eigenvalues corresponding to (anti-)periodic solutions—in a controlled way. We provide explicit evidence for this in both the Lamé and Mathieu equations. Moreover, the resummed semi-classical block enables the construction of an infinite set of periodic functions of the Floquet exponent, corresponding to the bands in the spectrum; see Figure~\ref{fig: bands k}. 

As a further check, we compute the eigenvalues corresponding to (anti-)periodic wave functions using more traditional techniques, such as isomonodromic deformations \cite{Bershtein:2021uts} and orbifold surface defect partition functions, following the approach of \cite{Jeong:2017pai}. In all cases, we find perfect agreement. It is important to note, however, that these methods do not provide a full resummation of the NS free energy. As such, they do not yield a single analytic object—namely, the resummed semi-classical block—that encapsulates the complete spectral information of the equation.

\paragraph {Content of the paper.} 

In Section~\ref{sec:connection}, we begin by reviewing the connection problem for the degenerate 4-point conformal blocks on the sphere and the degenerate 2-point conformal blocks on the torus. We then analyze the semi-classical limit of the 2-point blocks and revisit their relation to the Lamé equation, solving the connection problem in terms of conformal field theory data. In Section~\ref{sec: blow-up C2}, we begin by reviewing the AGT correspondence and its dictionary relating conformal blocks to Nekrasov partition functions. We then introduce the $\mathbb{C}^{2}$ blow-up equations and take their NS limit, obtaining a system that involves only the NS prepotential $\mathcal{W}_0$ and its first $\epsilon_2$-correction $\mathcal{W}_1$. Finally, by formulating a specific ansatz for both functions, we derive a set of algebraic equations for the functions parameterizing the prepotentials, demonstrate how these can be solved, and analyze the analytic structure of their solutions. In Section~\ref{sec: blow-up C2Z2}, we repeat the analysis of the previous section for the case of the $\mathbb{C}^{2}/\mathbb{Z}_{2}$ blow-up equations. In this setting, the profile functions satisfy simple ordinary differential equations, which can be integrated explicitly in terms of elementary functions. As expected, the resulting functions coincide with those obtained in the previous section, since the two variants of the blow-up equations encode the same amount of information. In Section~\ref{sec: spectral problem}, we analyze the real periodic spectral problems associated with the Lamé equation, beginning with the one along the A-cycle. We describe how the resummed NS function $\mathcal{W}_0$ parametrizes the full spectrum, including the band and gap structure. We then explore alternative approaches to studying the periodic spectral problem. First, we examine its relation to isomonodromic deformation problems on the torus, which allow for the computation of periodic and anti-periodic eigenvalues. Next, we turn to orbifold defect partition functions, which provide a method to compute both eigenvalues and eigenfunctions, including the periodic and anti-periodic cases. Finally, we show how the connection formulas for Lamé functions can be used to investigate the spectral problem along the B-cycle, viewed as a different limit of the A-cycle problem. In Appendix \ref{app: special functions}, we summarize our conventions for special functions and Nekrasov partition functions. 
Appendix \ref{app: checks} contains two consistency checks on the results of the paper, carried out using standard textbook methods. 
Finally, in Appendix~\ref{app: blow-up}, we explicitly present and solve the first few equations derived from the blow-up equations for the $\mathcal{N}=2^{*}$ theory in the $\mathbb{C}^2$ case, as well as for the $N_f = 0$ and $\mathcal{N}=2^{*}$ theories in the $\mathbb{C}^2/\mathbb{Z}_2$ case.

\paragraph{Open questions:}
\begin{itemize}

    \item The gauge theory partition function in the presence of a surface defect is known to yield a constructive solution to the wave equations of the corresponding quantum integrable system. However, it exhibits the same formal divergence issues at the (anti-)periodic points of the spectrum. The resummation procedure developed in this paper should, in principle, also apply to the wave function itself, which displays the same lattice of poles.

    \item The resummation procedure we discuss has a representation-theoretic counterpart in the study of Painlev\'e tau functions with resonant initial conditions. This requires fine-tuning the latter through a special kind of double-scaling limit, leading to logarithmic terms in the time expansion of the Painlev\'e tau function. The periodic spectral problems we discuss in this paper can also be approached from this viewpoint.

    \item The AGT dual of $SU(2)$ gauge theories on $\mathbb{C}^2/\mathbb{Z}_2$ is known to be $\mathcal{N}=1$ super-Liouville CFT \cite{Belavin:2011pp,Bonelli:2011jx,Bonelli:2011kv,Belavin:2012eg}. Consequently, the results presented in this paper can be applied to analyze the analytic structure of super-Virasoro conformal blocks in the semi-classical limit.

    \item The present analysis can be extended to more general classes of supersymmetric gauge theories. In particular, one can consider linear, circular, or more general quiver gauge theories, as well as higher-rank unitary gauge groups. Furthermore, it would be natural to study theories with arbitrary simple Lie groups and their associated isomonodromic deformation problems \cite{Bonelli:2021rrg, Bonelli:2022iob}.

    \item It would also be interesting to investigate semi-classical conformal blocks in strong coupling regimes, including at Argyres–Douglas points \cite{Nagoya:2015cja, Bonelli:2016qwg,Nagoya:2016mlj, Nagoya:2018pgp,Gavrylenko:2020gjb,Fucito:2023plp,Poghosyan:2023zvy,Bonelli:2024wha,Bonelli:2025owb,Poghossian:2025nef,Iorgov:2025hxt}, as well as in $\mathcal{N}=1$ and $\mathcal{N}=1^*$ vacua \cite{Fucito:2005wc,Bonelli:2013pva}.
    
    \item 
    Five-dimensional gauge theory partition functions also exhibit a similar structure of poles, which we expect can be analyzed through an analogous resummation procedure. In this context, blow-up equations are also known \cite{Nakajima:2005fg}, along with their connection to Painlevé $\tau$-functions \cite{Bershtein:2016aef,Bonelli:2017gdk,Bershtein:2017swf,Bonelli:2020dcp,Bershtein:2018srt}. Moreover, the connection formulae for the associated $q$-difference equations can be studied in terms of (degenerate) $q$-Virasoro conformal blocks; see \cite{Gavrylenko:2025nuo} for the case of $q$-Painlevé ${\rm III}_3$. The spectrum in these systems is extracted from Wilson loop expectation values \cite{Sciarappa:2016ctj,Grassi:2017qee,Wang:2023zcb}.

    \item It would be interesting to investigate the ODE/IM correspondence in relation to the Lamé equation and its spectrum; see \cite{Lukyanov:2011wd, Litvinov:2013sxa, Fioravanti:2019vxi} for studies connecting the ODE/IM framework to Painlevé equations. We note that our computational scheme, based on the blow-up equations in the NS limit, provides a direct algorithm for computing the connection coefficients of the associated differential equation, thereby conceptually simplifying a key step in the analysis. Nonetheless, deriving this equation purely from the perspective of differential equation theory would likely require techniques similar to those underlying the ODE/IM correspondence.

    \item Besides the applications of the Lamé equation to holography already mentioned in the introduction, semi-classical Virasoro conformal blocks themselves have intriguing applications in the holographic description of three-dimensional gravitational systems. The analytic structure uncovered in this paper is expected to have a counterpart in the gravitational dual picture \cite{Alkalaev:2017bzx,Alkalaev:2016ptm,Alekseev:2019gkl}.

    \item The results presented in this paper have applications to concrete physical systems. For instance, the detailed band-gap structure of the Lamé equation—or more generally, of Hill equations—is phenomenologically relevant in inflationary cosmology, particularly in the context of the preheating problem \cite{Greene:1997fu} and in models describing the early inflationary phase of the universe \cite{Amin:2011hj}.
    
\end{itemize}

\acknowledgments
We would like to thank 
P.~Arnaudo,
M.~Bershtein,
F.~Del Monte,
H.~Desiraju,
G.~Dunne,
M.~François,
A.~Grassi,
C.~Iossa,
S.~Jeong,
O.~Lisovyy,
I.~Majtara,
V. Roubtsov and
A.~Shchechkin
for useful discussions and comments. T.P. gratefully acknowledges the hospitality of CERN and the University of Geneva during the completion of this work, and in particular thanks A. Grassi for making the visit possible. The research of G.B.  is partly supported by the INFN Iniziativa Specifica ST\&FI and by the PRIN project “Non-perturbative Aspects Of Gauge Theories And Strings”. The research of  P.G., T.P. and A.T. is partly supported by the INFN Iniziativa Specifica GAST and Indam GNFM. 
The research is partly supported by the MIUR PRIN Grant 2020KR4KN2 ``String Theory as a bridge between Gauge Theories and Quantum Gravity''.  
All the authors acknowledge funding from the EU project Caligola (HORIZON-MSCA-2021-SE-01), Project ID: 101086123, and CA21109 - COST Action CaLISTA.

\newpage
\section{Connecting solutions of the Lamé equation}
\label{sec:connection}
In this section, we show how the Lamé equation emerges in the semi-classical limit of the BPZ equation \cite{Belavin:1984vu}, satisfied by Virasoro conformal blocks on the torus, and we solve the associated connection problem following the approach of \cite{Bonelli:2022ten}. To fix notation, we begin by reviewing the connection problem for the degenerate 4-point conformal blocks on the sphere, deriving their braiding and fusion matrices. We then turn to the degenerate 2-point blocks on the torus, whose connection formulas we compute using the data obtained in the spherical case. Finally, we take the semi-classical limit, in which the BPZ equation reduces to the Lamé equation, and show that the resulting connection formulas for the semi-classical conformal blocks coincide with those of the Lamé equation's solutions.

\subsection{Degenerate 4-point blocks on the sphere}
Consider the 4-point conformal block on the Riemann sphere:
\begin{equation}
    \mathcal{F}(a_4,a_3,-b/2,a_1;x)=\bra{\Delta_{4}}V_{a_{3}}(1)\phi_{(2,1)}(x)\ket{\Delta_{1}},
\end{equation}
with $V_{a_{i}}$ chiral fields of conformal dimension $\Delta_{i}=a_{i}(Q-a_{i})$, $Q=b+b^{-1}$, and $\phi_{(2,1)}$ the degenerate field of Liouville momentum $a_{(2,1)}=-b/2$\footnote{All parameters are considered to be complex unless otherwise stated.}. The block $\mathcal{F}(\{a_i\};x)$ satisfies the following hypergeometric BPZ equation:
\begin{equation}
    \left(b^{-2}\frac{d^{2}}{dx^{2}}+\frac{2x-1}{x(1-x)}\frac{d}{dx}+\frac{\Delta_1}{x^{2}} + \frac{\Delta_3}{(1-x)^{2}}+\frac{\Delta_1+\Delta_2+\Delta_3-\Delta_4}{x(1-x)}\right)\mathcal{F}(\{a_i\};x)=0,
\end{equation}
which can be put in standard form by the ansatz $\mathcal{F}\left(\{a_i\};x\right)=x^{b a_1}(1-x)^{b a_3}F(\{a_i\};x)$. By considering power series solutions near the point $x=0$, one finds a basis of $s$-$channel$ conformal blocks:
\begin{subequations}
\begin{align}
&\mathcal{F}^{(s)}_{+}(\{a_i\};x)
    =x^{b a_{1}}(1-x)^{b a_3}\ _2F_1(\alpha,\beta,\gamma;x),\\
&\mathcal{F}^{(s)}_{-}(\{a_i\};x)=x^{b(Q-a_{1})}(1-x)^{ba_3}\ _2F_1(\alpha-\gamma+1,\beta-\gamma+1,2-\gamma;x),
\end{align}
\end{subequations}
where $\alpha=b(a_1+a_3+a_4-3b/2)-1$, $\beta = b(a_{1}+a_{3}-a_{4}-b/2)$ and $\gamma = b(2a_{1}-b)$. Similarly, by looking for solutions near $x=1$ we find a basis of $t$-$channel$ conformal blocks:
\begin{subequations}
\begin{align}
&\mathcal{F}^{(t)}_{+}(\{a_i\};x)=x^{ba_{1}}(1-x)^{ba_3}\ _2F_1(\alpha,\beta,\alpha+\beta+1-\gamma;1-x),\\
&\mathcal{F}^{(t)}_{-}(\{a_i\};x)=x^{ba_{1}}(1-x)^{b(Q-a_3)}\ _2F_1(\gamma-\alpha,\gamma-\beta,1+\gamma-\alpha-\beta;1-x),
\end{align}
\end{subequations}
and finally, near the point at infinity, we find a basis of $u$-$channel$ conformal blocks:
\begin{subequations}
\begin{align}
&\mathcal{F}^{(u)}_{+}(\{a_i\};x)=e^{-i\pi ba_{3}}x^{ba_{1}-\beta}(1-x)^{ba_{3}}\ _2F_1(\beta,\beta-\gamma+1,\beta-\alpha+1;x^{-1}),\\
&\mathcal{F}^{(u)}_{-}(\{a_i\};x)=e^{-i\pi ba_{3}}x^{ba_{1}-\alpha}(1-x)^{ba_{3}}\ _2F_1(\alpha,\alpha-\gamma+1,\alpha-\beta+1;x^{-1}),
\end{align}
\end{subequations}
where the normalization has been chosen appropriately for the study of conformal blocks on the torus. Indeed, we want to identify solutions near the points $x = 0$ and $x=\infty$ by matching their local monodromies. With the previous choice we can see that, when $a_{4}=a_{1}=a/b$ and we consider the $semi$-$classical$ limit $b\to 0$, the behavior of these local solutions around their respective expansion points matches:
\begin{subequations}
\begin{align*}
&\mathcal{F}^{(s)}_{+} \sim x^{a},\quad \mathcal{F}^{(s)}_{-}\sim x^{1-a},\\[4pt]
&\mathcal{F}^{(u)}_{+} \sim x^{a},\quad \mathcal{F}^{(u)}_{-}\sim x^{1-a}.
\end{align*}
\end{subequations}
Due to their explicit expressions in terms of hypergeometric functions, we can obtain connection formulas between different local basis of solutions, see e.g. \cite{Iorgov:2014vla} and \cite{NIST:DLMF} (Section \href{https://dlmf.nist.gov/15.10.ii}{15.10}). Namely, we find the fusion matrix:
\begin{equation}
\mathcal{F}^{(s)}_{\epsilon}(\{a_i\};x) =\sum_{\epsilon'=\pm}\mathbb{F}_{\epsilon \epsilon'}\begin{bmatrix}
a_{3} & -b/2\\
a_{4} & a_{1}
\end{bmatrix} \mathcal{F}^{(t)}_{\epsilon'}(\{a_i\};x),
\end{equation}
expressing the following change of local basis:
\vspace{1em}
\begin{center}
\begin{tikzpicture}[baseline={(current bounding box.center)}]

  \begin{scope}[shift={(-2.5,0)}, scale=0.8]
    \draw[thick] (-2,0) -- (2,0);
    \draw[thick] (-1,0) -- (-1,2);
    \node at (-1,2.2) {\( a_3 \)};
    \node at (-1.6, 0.2) {\( a_4 \)};
    \draw[dashed] (1,0) -- (1,2);
    \node at (1,2.2) {\( -b/2 \)};
    \node at (1.6, 0.2) {\( a_{1} \)};
    \node at (0, 0.2) {\( a_{1\epsilon} \)};
  \end{scope}

  \node at (0.5,0.5) {\( =\sum_{\epsilon'=\pm}\mathbb{F}_{\epsilon \epsilon'} \)};

  \begin{scope}[shift={(3.2,0)}, scale=0.8]
    % base orizzontale
    \draw[thick] (-2,0) -- (2,0);
    % asta centrale
    \draw[thick] (0,0) -- (0,1);
    % biforcazione in due rami
    \draw[thick] (0,1) -- (-1,2);
    \draw[dashed] (0,1) -- (1,2);
    % etichette sui rami
    \node at (-1.2,2.2) {\( a_3 \)};
    \node at (1.2,2.2) {\( -b/2 \)};
    % etichette alla base
    \node at (-1.6, 0.2) {\( a_4 \)};
    \node at (1.6, 0.2) {\( a_1 \)};
    % etichetta sul punto di innesto
    \node at (0.5, 0.5) {\( a_{3\epsilon'} \)};
  \end{scope}

\end{tikzpicture}
\end{center}
\vspace{1em}
and the braiding matrix:
\begin{equation}
\mathcal{F}^{(s)}_{\epsilon}(\{a_i\};x)=\sum_{\epsilon'=\pm}\mathbb{B}_{\epsilon \epsilon'}\begin{bmatrix}
a_{3} & -b/2\\
a_{4} & a_{1}
\end{bmatrix}\mathcal{F}^{(u)}_{\epsilon'}(\{a_i\};x),
\end{equation}
expressing the remaining possible change of local basis:
\vspace{1em}
\begin{center}
\begin{tikzpicture}[baseline={(current bounding box.center)}]

  \begin{scope}[shift={(-2.5,0)}, scale=0.8]
    \draw[thick] (-2,0) -- (2,0);
    \draw[thick] (-1,0) -- (-1,2);
    \node at (-1,2.2) {\( a_3 \)};
    \node at (-1.6, 0.2) {\( a_4 \)};
    \draw[dashed] (1,0) -- (1,2);
    \node at (1,2.2) {\( -b/2 \)};
    \node at (1.6, 0.2) {\( a_{1} \)};
    \node at (0, 0.2) {\( a_{1\epsilon} \)};
  \end{scope}

  \node at (0.5,0.5) {\( =\sum_{\epsilon'=\pm}\mathbb{B}_{\epsilon \epsilon'} \)};

  \begin{scope}[shift={(3.2,0)}, scale=0.8]
    \draw[thick] (-2,0) -- (2,0);
    \draw[dashed] (-1,0) -- (-1,2);
    \node at (-1,2.2) {\( -b/2 \)};
    \node at (-1.6, 0.2) {\( a_4 \)};
    \draw[thick] (1,0) -- (1,2);
    \node at (1,2.2) {\( a_3 \)};
    \node at (1.6, 0.2) {\( a_1 \)};
    \node at (0, 0.2) {\( a_{4\epsilon'} \)};
  \end{scope}

\end{tikzpicture}
\end{center}
\vspace{1em}
These two connection matrices read respectively:
\begin{subequations}
\begin{align}
&\mathbb{F}\begin{bmatrix}
a_{3} & -b/2\\
a_{4} & a_{1}
\end{bmatrix}=\begin{pmatrix}
\frac{\Gamma(\gamma)\Gamma(\gamma-\alpha-\beta)}{\Gamma(\gamma-\alpha)\Gamma(\gamma-\beta)}
& \frac{\Gamma(\gamma)\Gamma(\alpha+\beta-\gamma)}{\Gamma(\alpha)\Gamma(\beta)}\\
\frac{\Gamma(2-\gamma)\Gamma(\gamma-\alpha-\beta)}{\Gamma(1-\alpha)\Gamma(1-\beta)} & \frac{\Gamma(2-\gamma)\Gamma(\alpha+\beta-\gamma)}{\Gamma(1+\alpha-\gamma)\Gamma(1+\beta-\gamma)}
\end{pmatrix},\\
&\mathbb{B}\begin{bmatrix}
a_{3} & -b/2\\
a_{4} & a_{1}
\end{bmatrix}=\begin{pmatrix}
\frac{e^{i\pi(ba_{3}-\beta)}\Gamma(\gamma)\Gamma(\alpha-\beta)}{\Gamma(\gamma-\beta)\Gamma(\alpha)}&\frac{e^{i\pi(ba_{3}-\alpha)}\Gamma(\gamma)\Gamma(\beta-\alpha)}{\Gamma(\gamma-\alpha)\Gamma(\beta)}\\
\frac{e^{i\pi(ba_{3}-\beta+\gamma-1)}\Gamma(2-\gamma)\Gamma(\alpha-\beta)}{\Gamma(1-\beta)\Gamma(\alpha-\gamma+1)} & \frac{e^{i\pi(ba_{3}-\alpha+\gamma-1)}\Gamma(2-
\gamma)\Gamma(\beta-\alpha)}{\Gamma(1-\alpha)\Gamma(\beta-\gamma+1)}
\end{pmatrix}.\label{eq: braiding}
\end{align}
\end{subequations}
\subsection{Degenerate 2-point blocks on the torus}
Consider the 2-point conformal block on the torus $\mathbb{C}/(\mathbb{Z}+\tau\mathbb{Z})$ \cite{Itzykson:1986pj, Bonelli:2019boe, Hadasz:2009db, Cho:2017oxl}, with again an insertion of the $\phi_{(2,1)}$ degenerate field: 
\begin{equation}
\label{eq: torusblock}
    \Phi(z,a,m; \mathfrak{q})= \frac{1}{Z(\tau)}Tr_{\mathcal{V}_{a}}\left(\mathfrak{q}^{L_{0}-c/24}V_{m}(0)\phi_{(2,1)}(z)\right),
\end{equation}
Here $Z(\tau)$ is the torus partition function, $\mathfrak{q}=e^{2\pi i \tau}$ is the modular parameter, $L_{0}$ the diagonal Virasoro generator, the central charge is $c= 1+6Q^{2}$ and finally the trace is taken over the Verma module $\mathcal{V}_{a}$ generated by the primary $V_{a}$ of momentum $a$. Precisely as before, due to the fusion rules of the degenerate primary:
\begin{equation}
\big[\phi_{(2,1)}\big]\times\big[V_{p}\big]=\sum_{\epsilon=\pm}\big[V_{p_{\epsilon}}\big],
\end{equation}
with $p_{\epsilon}=p-\epsilon b/2$, the space of such conformal blocks is two dimensional and we can denote a basis by: 
\begin{equation}
\label{eq: torusbasis}
    \Phi_{\epsilon}(z,a,m; \mathfrak{q})= \frac{1}{Z(\tau)}Tr_{\mathcal{V}_{a}}\left(\mathfrak{q}^{L_{0}-c/24}V_{m}(0)\phi_{\epsilon}(z)\right),
\end{equation}
where $\epsilon=\pm$ specifies the choice of intermediate momentum. Making the change of variable\footnote{This change of variable maps the torus to the annulus, where Virasoro generators are defined, and introduces a factor of $x^{\Delta_{2,1}}$ which is relevant for the local monodromy of the block.} $x=e^{2\pi i z}$ and taking the OPE between the degenerate field and $V_{a}$, we find for the corresponding $necklace$ channel basis of blocks:
\begin{equation}
\label{eq: necklace blocks}
    \Phi_{\epsilon}^{(n)}(x,a,m; \mathfrak{q}) \sim x^{\Delta_{a_{\epsilon}}-\Delta_{a}}\Big(1+\mathcal{O}(x)\Big),
\end{equation}
which are quasi-periodic functions along the A-cycle of the torus:
\begin{equation}
        \Phi_{\epsilon}^{(n)}(z+1,a,m; \mathfrak{q})=e^{2\pi i(\Delta_{a_{\epsilon}}-\Delta_{a})}\Phi_{\epsilon}^{(n)}(z,a,m; \mathfrak{q}).
\end{equation}
Meanwhile, along the B-cycle the monodromy matrix is non-diagonal and can be computed via the Moore-Seiberg formalism \cite{Moore:1988qv, Bonelli:2019boe} by knowing the braiding matrix on the sphere (\ref{eq: braiding}). Indeed, performing the following sequence of transformations:
\vspace{1em}
\begin{center}
\begin{tikzpicture}[baseline={(current bounding box.center)}]

  % First figure
  \begin{scope}[shift={(-4,0)}]
    \draw[thick] (0,0) ellipse (1.2 and 0.5);
    \draw[dashed] (0.5,0.5) -- (0.5,1.8);
    \draw[thick] (-0.5,0.5) -- (-0.5,1.8);
    \node at (-0.5,2) {\( m \)};
    \node at (0.5,2) {\( -b/2 \)};
    \node at (0,0.7) {\( a_{\epsilon} \)};
    \node at (0,-0.7) {\( a \)};
  \end{scope}

  % First "="
  \node at (-2.2,1) {\( = \)};

  % Second figure (closer)
  \begin{scope}[shift={(-0.5,0)}]
    \draw[thick] (0,0) ellipse (1.2 and 0.5);
    \draw[thick] (0.5,0.5) -- (0.5,1.8);
    \draw[dashed] (-0.5,0.5) -- (-0.5,1.8);
    \node at (-0.5,2) {\( -b/2 \)};
    \node at (0.5,2) {\( m \)};
    \node at (0,0.7) {\( a \)};
    \node at (0,-0.7) {\( a_{\epsilon} \)};
  \end{scope}

  % Second "=" with expression
  \node at (2.2,1) {\( =\quad \sum_{\epsilon'=\pm } \mathbb{B}^{-1}_{\epsilon  \epsilon'}  \)};

  % Third figure
  \begin{scope}[shift={(4.5,0)}]
    \draw[thick] (0,0) ellipse (1.2 and 0.5);
    \draw[dashed] (0.5,0.5) -- (0.5,1.8);
    \draw[thick] (-0.5,0.5) -- (-0.5,1.8);
    \node at (-0.5,2) {\( m \)};
    \node at (0.5,2) {\( -b/2 \)};
    \node at (0,0.7) {\( a_{\epsilon+\epsilon'} \)};
    \node at (0,-0.7) {\( a_{\epsilon} \)};
  \end{scope}

\end{tikzpicture}
\end{center}
\vspace{1em}
we find:
\begin{equation}
\Phi_{\epsilon}^{(n)}(z+\tau,a,m; \mathfrak{q})=\sum_{\epsilon'=\pm}M_{B,\,\epsilon \epsilon'}^{(n)}\Phi_{\epsilon'}^{(n)}(z,a,m; \mathfrak{q}),
\end{equation}
where the B-cycle monodromy matrix is: 
\begin{equation}
\label{eq: Mb necklace}
M_{B,\,\epsilon \epsilon'}^{(n)}=\mathbb{B}^{-1}_{\epsilon\epsilon'}\begin{bmatrix}
m & -b/2\\
a_{\epsilon} & a_{\epsilon} 
\end{bmatrix}e^{-\epsilon\frac{b}{2}\partial_{a}},
\end{equation}
and $e^{k\partial_{a}}f(a)=f(a+k)$.
Taking instead the OPE between $V_{m}$ and $\phi_{(2,1)}$, and going to the annulus $x=e^{2\pi i z}$, we see that for the OPE channel basis:
\begin{equation}
\label{eq: ope blocks}
    \Phi_{\epsilon}^{(o)}(x,a,m; \mathfrak{q})\sim (1-x)^{\Delta_{m_{\epsilon}}-\Delta_{m}-\Delta_{(2,1)}}\Big(1+\mathcal{O}(1-x)\Big),
\end{equation}
thus the blocks are diagonal with respect to the analytic continuation around the $z=0$ point: 
\begin{equation}
\label{eq: 0mono}
    \Phi_{\epsilon}^{(o)}(e^{2\pi i}z,a,m; \mathfrak{q})=e^{2\pi i(\Delta_{m_{\epsilon}}-\Delta_{m}-\Delta_{(2,1)})}\Phi_{\epsilon}^{(o)}(z,a,m; \mathfrak{q}).
\end{equation}
We can again compute their A-cycle and B-cycle monodromy matrices just by knowing the connection matrices on the sphere. Indeed, we can simply write the OPE channel blocks in terms of the necklace ones, of which we know the monodromy properties:
\vspace{1em}
\begin{center}
\begin{tikzpicture}[baseline={(current bounding box.center)}]

  \begin{scope}[shift={(-2.5,0)}, scale=0.8]
    % base orizzontale
    \draw[thick] (0,-0.6) ellipse (1.5 and 0.6);
    % asta centrale
    \draw[thick] (0,0) -- (0,0.8);
    % biforcazione in due rami
    \draw[thick] (0,0.8) -- (-1,1.8);
    \draw[dashed] (0,0.8) -- (1,1.8);
    % etichette sui rami
    \node at (-1.3,2.1) {\( m \)};
    \node at (1.3,2.1) {\( -b/2 \)};
    % etichette alla base
    \node at (0, -1.5) {\( a \)};
    % etichetta sul punto di innesto
    \node at (0.4, 0.3) {\( a_{\epsilon} \)};
  \end{scope}

  \node at (0.5,0.5) {\( =\sum_{\epsilon'=\pm}\mathbb{F}^{-1}_{\epsilon \epsilon'} \)};

  \begin{scope}[shift={(3.2,0)}, scale=0.8]
    \draw[thick] (0,-0.6) ellipse (1.5 and 0.6);
    \draw[thick] (-0.8,0) -- (-0.8,1.8);
    \node at (-0.8,2.1) {\( m \)};
    \draw[dashed] (0.8,0) -- (0.8,1.8);
    \node at (0.8,2.1) {\( -b/2 \)};
    \node at (0, 0.3) {\( a_{\epsilon'} \)};
    \node at (0, -1.5) {\( a \)};
  \end{scope}

\end{tikzpicture}
\end{center}
\vspace{1em}
Thus, we find:
\begin{subequations}
\begin{align}
&\Phi_{\epsilon}^{(o)}(z+1,a,m; \mathfrak{q})=\sum_{\epsilon''=\pm}M_{A,\,\epsilon \epsilon''}^{(o)}\Phi_{\epsilon''}^{(o)}(z,a,m; \mathfrak{q}),\\
&\Phi_{\epsilon}^{(o)}(z+\tau,a,m; \mathfrak{q})=\sum_{\epsilon'''=\pm}M_{B,\,\epsilon \epsilon'''}^{(o)}\Phi_{\epsilon'''}^{(o)}(z,a,m; \mathfrak{q}),
\end{align}
\end{subequations}
where 
\begin{subequations}
\label{eq: Ma, Mb ope}
\begin{align}
&M_{A,\,\epsilon \epsilon''}^{(o)}=\sum_{\epsilon'=\pm}\mathbb{F}^{-1}_{\epsilon\epsilon'}\begin{bmatrix}
m & -b/2\\
a & a
\end{bmatrix}e^{2\pi i(\Delta_{a_{\epsilon'}}-\Delta_{a})}\mathbb{F}_{\epsilon'\epsilon''}\begin{bmatrix}
m & -b/2\\
a & a
\end{bmatrix},\\
&M_{B,\,\epsilon \epsilon'''}^{(o)}=\sum_{\epsilon',\,\epsilon''=\pm}\mathbb{F}^{-1}_{\epsilon\epsilon'}\begin{bmatrix}
m & -b/2\\
a & a
\end{bmatrix}M_{B,\,\epsilon' \epsilon''}^{(n)}\mathbb{F}_{\epsilon''\epsilon'''}\begin{bmatrix}
m & -b/2\\
a & a
\end{bmatrix},
\end{align}
\end{subequations}
and the shift operator $e^{k\partial_{a}}$ in $M_{B}^{(n)}$ acts on everything on its right. By going to the OPE channel basis (\ref{eq: 0mono}), we can also compute the monodromy matrix for the analytic continuation of necklace channel blocks around the point $z=0$, which is given by:
\begin{equation}
\label{eq: M0 necklace}
M^{(n)}_{0,\,\epsilon\epsilon''}=\sum_{\epsilon'=\pm}\mathbb{F}_{\epsilon\epsilon'}\begin{bmatrix}
m & -b/2\\
a & a
\end{bmatrix}e^{2\pi i(\Delta_{m_{\epsilon'}}-\Delta_{m}-\Delta_{(2,1)})}\mathbb{F}^{-1}_{\epsilon'\epsilon''}\begin{bmatrix}
m & -b/2\\
a & a
\end{bmatrix}.
\end{equation}
Similarly, by decomposing the computation in terms of the elementary fusion and braiding matrices, one can obtain the monodromy matrix for the analytic continuation of any n-point block, with the insertion of at least one degenerate $\phi_{(2,1)}$, around any point of insertion of other primaries and around the two non-trivial cycles of the torus.
\subsection{The Lamé and the Mathieu equations}
The torus conformal blocks (\ref{eq: torusblock}) satisfy the BPZ equation:
\begin{equation}
\label{eqn:BPZt1}
    \Big(-b^{-2}\partial_{z}^{2}+\zeta_{*}(z)\partial_{z}-\Delta_{m}\wp_{*}(z)-2\Delta_{(2,1)}\eta_{1}-2\pi i\partial_{\tau}\log Z(\tau)-2\pi i\partial_{\tau}\Big)\Phi(z,a,m; \mathfrak{q})=0,
\end{equation}
where $\zeta_{*}(z)=\zeta(z; 1,\tau)-2\eta_{1}z$ and $\wp_{*}(z)=\wp(z;1, \tau)+2\eta_{1}$ \cite{Piatek:2013ifa}, $\wp$ and $\zeta$ being the Weierstrass elliptic functions, see Appendix \ref{app: special functions} for the conventions adopted. In the $semi$-$classical$ limit $b\rightarrow 0$ \cite{Zamolodchikov:1995aa, Desiraju:2024fmo}, considering $a=\alpha/b$ and $m=\mu/b$, the conformal blocks have the asymptotic behavior:
\begin{equation}
    \Phi\left(z,a,m; \mathfrak{q}\right) \simeq e^{\frac{1}{b^{2}}f(\alpha,\mu;\mathfrak{q})}\psi(z,\alpha, \mu;\mathfrak{q}),
\end{equation}
where $f(\alpha,\mu; \mathfrak{q})$ is the semi-classical 1-point torus conformal block:
\begin{equation}
f(\alpha,\mu; \mathfrak{q})=\lim_{b\to 0}b^{2}\log \mathcal{F}\left(\frac{\alpha}{b},\frac{\mu}{b}; \mathfrak{q}\right),
\end{equation}
with $\mathcal{F}(a,m; \mathfrak{q})$ being the 1-point torus conformal block, while the $semi$-$classical$ 2-point block\footnote{Due to the AGT correspondence \cite{Alday:2009aq}, the 2-point block on the torus, and thus its semi-classical limit, can be computed via simple combinatorial formulas, see equations \eqref{eq: block = inst} and (\ref{eq: classical defect}).} is given by:
\begin{equation}
\label{eq: 2pt classical}
\psi(z,\alpha,\mu; \mathfrak{q})=\lim_{b\to 0}\frac{\Phi(z,\alpha/b,\mu/b; \mathfrak{q})}{\mathcal{F}(\alpha/b,\mu/b; \mathfrak{q})}.
\end{equation}
In this limit, equation (\ref{eqn:BPZt1}) reduces to the Lamé equation in Weierstrass form: 
\begin{equation}
\label{eq: lame1}
    \left(\partial_{z}^{2}+E-\mu(\mu-1) \wp(z)\right)\psi(z,\alpha,\mu; \mathfrak{q})=0,
\end{equation}
and the eigenvalue $E$ is given in terms of the semi-classical 1-point block as:
\begin{equation}
\label{eq: Ens}
E=-2\mu(\mu-1)\eta_{1}-4\pi^{2}\mathfrak{q}\partial_{\mathfrak{q}}f(\alpha,\mu; \mathfrak{q}).
\end{equation}
To obtain connection formulas for the analytic continuation of solutions to the Lamé equation, we can simply take the semi-classical limit of the monodromy matrices \eqref{eq: Mb necklace}, \eqref{eq: Ma, Mb ope} and \eqref{eq: M0 necklace} derived in the previous section. More precisely, the necklace channel blocks (\ref{eq: necklace blocks}) descend to a basis of Floquet solutions to the Hill equation\footnote{Indeed, this equation defines a $real$ periodic spectral problem for $\tau \in i\mathbb{R}$.}
\begin{equation}
\label{eq: hill}
    \left(\partial_{x}^2+E-\mu(\mu-1)\wp(x+\tau/2)\right)\psi(x+\tau/2,\alpha,\mu; \mathfrak{q})=0,
\end{equation}
with Floquet exponent $\nu =2\alpha-1$, as can be seen by their quasi-periodicity along the A-cycle:
\begin{equation}
\label{eqn:A-n}
    \psi_{\epsilon}^{(n)}(z+1,\alpha,\mu;\mathfrak{q})=e^{2\pi i \epsilon(\alpha-1/2)}\psi_{\epsilon}^{(n)}(z,\alpha,\mu;\mathfrak{q}).
\end{equation}
In terms of $\nu$, the $\mathfrak{q}$-expansion of the eigenvalue (\ref{eq: Ens}) takes a particularly simple form:
\begin{equation}
\label{eq: expansion ns lame}
E/\pi^{2} = \nu^2 -\frac{1}{3}\mu(\mu-1) + \frac{8\mu(\mu-1)(\nu^2-1+\mu(\mu-1))}{\nu^2-1}\mathfrak{q}+\mathcal{O}(\mathfrak{q}^{2}),
\end{equation}
which can easily be verified by standard methods, see appendix \ref{app: checks 1}. Concerning the B-cycle monodromy \eqref{eq: Mb necklace}, in the semi-classical limit we have:
\begin{equation*}
\Phi_{\epsilon'}(z,a-\epsilon b/2,m; \mathfrak{q})\simeq e^{\frac{1}{b^{2}}f( \alpha-\epsilon b^{2}/2,\mu; \mathfrak{q})}\psi_{\epsilon'}(z, \alpha-\epsilon b^{2}/2,\mu; \mathfrak{q}),
\end{equation*}
so that, by expanding the semi-classical 1-point block at leading order in $b^{2}$, we find:
\begin{equation}
\label{eqn:B-n}
    \psi_{\epsilon}^{(n)}(z+\tau,\alpha,\mu;\mathfrak{q})=\sum_{\epsilon'=\pm}M^{(n)}_{B,\,\epsilon \epsilon'}\psi_{\epsilon'}^{(n)}(z,\alpha,\mu;\mathfrak{q}),
\end{equation}
with the B-cycle monodromy matrix $M_{B}^{(n)}$ given by:
\begin{equation}
\label{eq: B-cycle mm}
M_{B}^{(n)}=
\begin{pmatrix}
-e^{-\frac{1}{2}\frac{\partial f}{\partial \alpha}}(i+\cot2\pi\alpha)\sin\pi\mu & e^{-\frac{1}{2}\frac{\partial f}{\partial \alpha}}\frac{\cos4\pi\alpha-\cos2\pi\mu}{\cos4\pi\alpha-1}\frac{\Gamma(2-2\alpha-\mu)\Gamma(1-2\alpha+\mu)}{\Gamma(2-2\alpha)\Gamma(1-2\alpha)}\\
e^{\frac{1}{2}\frac{\partial f}{\partial \alpha}}\frac{\cos4\pi\alpha-\cos2\pi\mu}{\cos4\pi\alpha-1}\frac{\Gamma(2-2\alpha-\mu)\Gamma(2\alpha+\mu-1)}{\Gamma(2\alpha)\Gamma(2\alpha-1)}\,\, &\,\, e^{\frac{1}{2}\frac{\partial f}{\partial \alpha}}(-i+\cot2\pi\alpha)\sin\pi\mu 
\end{pmatrix}.
\end{equation}
Meanwhile, the OPE channel blocks (\ref{eq: ope blocks}) descend to a basis of Frobenius solutions near the point $z=0$. This too can be seen by their monodromy around these points:
\begin{equation}
    \psi_{\epsilon}^{(o)}(e^{2\pi i}z,\alpha,\mu;\mathfrak{q})=e^{2\pi i \epsilon\mu}\,\psi_{\epsilon}^{(o)}(z,\alpha,\mu;\mathfrak{q}).
\end{equation}
All other connection matrices can be obtained in a similar way\footnote{The connection formulas for the Lamé equation can be obtained also via the probabilistic construction of conformal blocks \cite{connections}.}. As a quick check of the consistency of our results one can compute the combination $M_{A}M_{B}M_{A}^{-1}M_{B}^{-1}$, which results in the same conjugacy class of $M_{0}$, as expected. Moreover, the commutator $[M_{A},M_{B}]$ is zero if and only if $\mu \in \mathbb{Z}$, meaning that this is the only case in which the A-cycle and B-cycle monodromy matrices can be simultaneously diagonalized\footnote{Monodromy matrices are defined up to conjugation, while the commutator is invariant.}. In other words, we rediscover that the Lamé equation admits elliptic solutions only for $\mu \in \mathbb{Z}$, which go under the name of Lamé polynomials \cite{Wang:1989}. By a further limit, it is known that the Lamé equation degenerates into the Mathieu one. Specifically, using the representation (\ref{eq: wp q-exp}) of the Weierstrass $\wp$ function in equation (\ref{eq: hill}) and considering the limit $\mathfrak{q}\to 0$, $\mu\to \infty$ with $t= \mathfrak{q}\mu^{4}$ fixed, one finds:
\begin{equation}
\label{eq: mathieu}
    \left(\partial_x^2+\lambda+8\pi^{2}\sqrt{t} \cos2\pi x\right)\psi(x,\alpha;t)=0,
\end{equation}
so that the eigenvalue is obtained from equation (\ref{eq: Ens}) as the limit:
\begin{equation}
    \lambda =  \Big[ E + 2\mu(\mu - 1)\eta_1 \Big]_{q \to 0,\, \mu \to \infty} = -4\pi^2 t \, \partial_t f(\alpha; t),
\end{equation}
and similarly for the eigenfunctions. The function $f(\alpha ;t)$ obtained as the limit of the semi-classical torus 1-point block corresponds to the semi-classical irregular block on the sphere \cite{Gaiotto:2009ma}. In terms of the Floquet exponent $\nu=2\alpha-1$, the first few terms of the small $t$ expansion of the eigenvalue read\footnote{In order to compare with the expansion presented in \cite{NIST:DLMF}, Section \href{https://dlmf.nist.gov/28.15.i}{28.15}, one must identify $t=\frac{1}{16}q^2$.}:
\begin{equation}
    \lambda/\pi^{2}=\nu^{2}+\frac{8t}{(\nu^{2}-1)}+\frac{8(7+5\nu^{2})t^{2}}{(\nu^{2}-1)^{3}(\nu^{2}-4)}+\mathcal{O}(t^{3}),
\end{equation}
which again can be verified via standard methods.
\section{\texorpdfstring{Resummation via $\mathbb{C}^{2}$ blow-up equations}{C2 blow-up equations and resummation}}
\label{sec: blow-up C2}
In this section we investigate the analytic structure of semi-classical conformal blocks as functions of the semi-classical intermediate momentum $\alpha$ or better, the Floquet exponent $\nu=2\alpha-1$. Since in the following we will discuss the $\mathbb{C}^{2}$ and the $\mathbb{C}^{2}/\mathbb{Z}_{2}$ blow-up equations, following the conventions of \cite{Bershtein:2021uts,Bonelli:2025owb}, we briefly review the dictionary of the AGT correspondence \cite{Alday:2009aq}, leaving definitions to Appendix \ref{app: special functions}. The Nekrasov partition function \cite{Nekrasov:2002qd, Nekrasov:2003rj} of the $SU(2)$ gauge theory with $N_f$ fundamentals of masses $\{\mu_{i}\}_{i=1}^{N_f}$ is the product of a semi-classical, a 1-loop and an instanton part:
\begin{equation}
\label{eq: Z fundamentals}
    \mathcal{Z}^{[N_{f}]}(a,\{\mu_{i}\};\epsilon_{1},\epsilon_{2}; t)=\mathcal{Z}^{[N_{f}]}_{cl}(a;\epsilon_{1},\epsilon_{2}; t)\mathcal{Z}^{[N_{f}]}_{1\text{-}loop}(a,\{\mu_{i}\};\epsilon_{1},\epsilon_{2})\mathcal{Z}^{[N_{f}]}_{inst}(a,\{\mu_{i}\};\epsilon_{1},\epsilon_{2}; t),
\end{equation}
which depend on the vev of the adjoint scalar $a$, the two equivariant parameters $\epsilon_{1}$, $\epsilon_{2}$ and the coupling constant $t$. The AGT correspondence relates the $N_f =4$ Nekrasov partition function to the 4-point conformal block on the sphere by:
\begin{equation}
\begin{aligned}
\mathcal{F}(\tilde{a};\alpha_{0},m_0,m_1,\alpha_{1};t)=&\,(1-t)^{-2\frac{m_0(Q-m_1)}{\epsilon_1 \epsilon_2}}t^{-\frac{m_1(Q-m_1)+\alpha_1(Q-\alpha_1)}{\epsilon_1 \epsilon_2}}\\
&\cdot\mathcal{Z}^{[4]}_{cl}(a;\epsilon_{1},\epsilon_{2}; t)\mathcal{Z}^{[4]}_{inst}(a,\{\mu_{i}\};\epsilon_{1},\epsilon_{2}; t),
\end{aligned}
\end{equation}
where $\tilde{a}$, $Q$ are related to the gauge theory parameters by:
\begin{equation}
\label{eq: atilde and Q}
\Tilde{a}=a+Q/2,\quad Q=b+b^{-1}=\epsilon_1 + \epsilon_2,
\end{equation}
while for the remaining ones the dictionary reads:
\begin{equation}
\begin{aligned}
\alpha_{0}&=(Q+\mu_1-\mu_2)/2 ,\quad m_0 =(Q+\mu_1+\mu_2 )/2,\\
\alpha_{1}&=(Q-\mu_3 +\mu_4)/2 ,\quad m_1 =(Q-\mu_3 -\mu_4)/2.
\end{aligned}
\end{equation}
The partition function can be equivalently written as:
\begin{equation}
\label{eq: Z with F}
    \mathcal{Z}^{[N_{f}]}(a,\{\mu_{i}\};\epsilon_{1},\epsilon_{2}; t)=\exp\bigg( F^{[N_{f}]}(a,\{\mu_{i}\};\epsilon_{1},\epsilon_{2}; t)\bigg),
\end{equation}
where the prepotential:
\begin{equation}
    F^{[N_{f}]}(a,\{\mu_{i}\};\epsilon_{1},\epsilon_{2}; t)=F^{[N_{f}]}_{cl}(a;\epsilon_{1},\epsilon_{2}; t)+F^{[N_{f}]}_{1\text{-}loop}(a,\{\mu_{i}\};\epsilon_{1},\epsilon_{2})+F^{[N_{f}]}_{inst}(a,\{\mu_{i}\};\epsilon_{1},\epsilon_{2}; t),
\end{equation}
admits an expansion in powers of $\epsilon_{1}$ and $\epsilon_{2}$. In particular, writing:
\begin{equation}
    F^{[N_{f}]}(a,\{\mu_{i}\};\epsilon_{1},\epsilon_{2}; t)\simeq\sum_{n=0}^{\infty}\mathcal{W}^{[N_{f}]}_{n}(a,\{\mu_{i}\},\epsilon_{1}; t)\epsilon_{2}^{n-1},
\end{equation}
and denoting the semi-classical 4-point block on the sphere by $f^{[4]}( \nu, \{\mu_{i}\};t)$, the latter can be expressed in terms of the NS prepotential \cite{Nekrasov:2009rc} $\mathcal{W}^{[4]}_{0}(a,\{\mu_{i}\};\epsilon_{1}; t)$ as:
\begin{equation}
    \epsilon_1 f^{[4]}\left(\frac{\nu}{\epsilon_1}, \left\{\frac{\mu_{i}}{\epsilon_1}\right\}; t\right)=-\frac{\beta_1}{2\epsilon_1}\log (1-t)-\frac{\beta_2}{2\epsilon_1}\log t+\mathcal{W}^{[4]}_{0,\,cl}\left(\frac{\nu}{2},\epsilon_{1}; t\right)+\mathcal{W}^{[4]}_{0,\,inst}\left(\frac{\nu}{2},\{\mu_{i}\},\epsilon_{1}; t\right),
\end{equation}
where $\beta_1=(\epsilon_1+\mu_1+\mu_2)(\epsilon_1+\mu_3+\mu_4)$, $\beta_2=\epsilon_1^{2}-\mu_3^{2}-\mu_4^{2}$, and $\nu=2a$. The remaining equivariant parameter $\epsilon_{1}$ is then identified with the reduced Planck constant $\epsilon_{1}=\hbar$ of the relevant quantum mechanical problem. Then, from the $N_f=4$ partition function we can obtain the $N_f<4$ ones by decoupling the appropriate number of hypermultiplets. This is achieved by sending their masses to infinity and the coupling constant $t$ to zero, while keeping their product fixed:
\begin{equation}
    \mathcal{Z}^{[N_f < 4]}(a,\{\mu_{i}\};\epsilon_{1},\epsilon_{2}; t)=\lim_{\mu_{N_f+1}\to \infty}\cdots\lim_{\mu_{4}\to \infty}\mathcal{Z}^{[4]}\left(a,\{\mu_{i}\};\epsilon_{1},\epsilon_{2}; \frac{t}{\prod_{i=N_f+1}^{4}\mu_{i}}\right).
\end{equation}
At the level of the semi-classical 4-point conformal block $f^{[4]}( \nu, \{\mu_{i}\};t)$, these decoupling limits result in the irregular semi-classical blocks $f^{[N_f<4]}( \nu, \{\mu_{i}\};t)$ \cite{Gaiotto:2009ma}. Moving to the torus, the Nekrasov partition function $\mathcal{Z}(a,m;\epsilon_{1},\epsilon_{2};\mathfrak{q})$ of the $\mathcal{N}=2^{*}$ $SU(2)$ gauge theory depends instead on the mass of the adjoint hypermultiplet $m$ and the coupling constant $\mathfrak{q}=e^{2\pi i \tau}$. Via the AGT correspondence, the 1-point torus conformal block is identified with the Nekrasov instanton partition function as:
\begin{equation}
\label{eq: block = inst}
    \mathcal{F}(m, \Tilde{a}; \mathfrak{q})=\mathcal{Z}_{cl}(a;\epsilon_{1},\epsilon_{2};\mathfrak{q})\mathcal{Z}_{inst}(a,m;\epsilon_1,\epsilon_2;\mathfrak{q}),
\end{equation}
where $\tilde{a}$ and $Q$ are as in (\ref{eq: atilde and Q}). Thus, the NS prepotential $\mathcal{W}_{0}(a,m;\epsilon_1;\mathfrak{q})$ is related to the semi-classical conformal block $f(\nu,\mu;\mathfrak{q})$ by:
\begin{equation}
\label{eq: F0 = f}
    \epsilon
    _{1}f\left(\frac{\nu}{\epsilon_{1}},\frac{\mu}{\epsilon_{1}};\mathfrak{q}\right)=\mathcal{W}_{0,\,cl}\left(\frac{\nu}{2},\epsilon_{1};\mathfrak{q}\right)+\mathcal{W}_{0,\, inst}\left(\frac{\nu}{2},\mu,\epsilon_{1};\mathfrak{q}\right).
\end{equation}
In the limit $m \to \infty$, $\mathfrak{q}\to 0$ with $t=m^{4}\mathfrak{q}$ fixed, the $\mathcal{N}=2^{*}$ partition function $\mathcal{Z}(a,m;\epsilon_{1},\epsilon_{2};\mathfrak{q})$ reduces to the pure gauge theory one, $\mathcal{Z}^{[0]}(a;\epsilon_{1},\epsilon_{2};t)$, and similarly for the prepotential and the semi-classical block. Leveraging the identifications due to the AGT correspondence, in the following we will denote by $f_{inst}^{[N_f]}$ and $f_{inst}$ the $instanton$ part of the semi-classical conformal blocks.

\subsection{\texorpdfstring{NS limit: $N_f\leq4$ and $\mathcal{N}=2^{*}$ theories}{C2 blow-up equations}}
The blow-up equations are a set of extremely rigid constraints satisfied by Nekrasov partition functions \cite{nakajima2005instanton, Jeong:2020uxz}, and thus conformal blocks \cite{bershtein2013coupling}. We consider the vanishing $\mathbb{C}^{2}$ blow-up equations, which, for the partition functions of $SU(2)$ gauge theories with $N_f \leq 4$ fundamental hypermultiplets, take the form:
\begin{equation}
\label{eq: C2 blow-up 1 Nf}
\sum_{n\in \mathbb{Z}+\frac{1}{2}}\mathcal{Z}^{[N_{f}]}(a+n\epsilon_{1},\{\tilde{\mu}_{i}\};\epsilon_{1},\epsilon_{2}-\epsilon_{1}; t)\mathcal{Z}^{[N_{f}]}(a+n\epsilon_{2},\{\tilde{\mu}_{i}\};\epsilon_{2},\epsilon_{1}-\epsilon_{2}; t)=0,
\end{equation}
where $\tilde{\mu}_{1,2}=\mu_{1,2}-(\epsilon_1+\epsilon_2)/2$ and $\tilde{\mu}_{3,4}=-\mu_{3,4}+(\epsilon_1+\epsilon_2)/2$, while for the $\mathcal{N}=2^{*}$ $SU(2)$ gauge theory, we consider \cite{Bershtein:2021uts}:
\begin{equation}
\label{eq: C2 blow-up 1 *}
\begin{aligned}
& \theta_{2}(0|2\tau)\sum_{n\in \mathbb{Z}}\mathcal{Z}(a+n\epsilon_{1},m;\epsilon_1,\epsilon_2-\epsilon_1;\mathfrak{q})\mathcal{Z}(a+2n\epsilon_{2},m;\epsilon_2,\epsilon_1-\epsilon_2;\mathfrak{q})\\
&-\theta_{3}(0|2\tau)\sum_{n\in\mathbb{Z}+\frac{1}{2}}\mathcal{Z}(a+n\epsilon_{1},m;\epsilon_1,\epsilon_2-\epsilon_1;\mathfrak{q})\mathcal{Z}(a+n\epsilon_{2},m;\epsilon_2,\epsilon_1-\epsilon_2;\mathfrak{q})=0.
\end{aligned}
\end{equation}
In particular, in the pure gauge theory limit $\mathfrak{q} \to 0$, $m \to \infty$, with $t = \mathfrak{q} m^4$ held fixed, equation~(\ref{eq: C2 blow-up 1 *}) reduces to equation~\eqref{eq: C2 blow-up 1 Nf} for $N_f = 0$. To obtain relations involving the semi-classical conformal blocks, we consider the $\epsilon_1 - \epsilon_2 \to 0$ limit of the previous blow-up equations, starting with \eqref{eq: C2 blow-up 1 Nf}. Expressing the partition function in terms of the prepotential, as in equation~\eqref{eq: Z with F}, we obtain:
\begin{equation}
    \sum_{n\in \mathbb{Z}+\frac{1}{2}}\exp\left(F^{[N_{f}]}(a+n\epsilon_{1},\{\tilde{\mu}_{i}\};\epsilon_{1},\epsilon_{2}-\epsilon_{1}; t)+F^{[N_{f}]}(a+n\epsilon_{2},\{\tilde{\mu}_{i}\};\epsilon_{2},\epsilon_{1}-\epsilon_{2}; t)\right)=0.
\end{equation}
Then, expanding the prepotential as:
\begin{equation}
F^{[N_{f}]}(a,\{\mu_i\};\epsilon_{1},\epsilon_{2}; t)=\frac{1}{\epsilon_{2}}\mathcal{W}^{[N_{f}]}_{0}(a,\{\mu_i\},\epsilon_{1}; t)+\mathcal{W}^{[N_{f}]}_{1}(a,\{\mu_i\},\epsilon_{1}; t)+ \mathcal{O}(\epsilon_{2}),
\end{equation}
in the limit $\epsilon_1 - \epsilon_2 \to 0$, we obtain:
\begin{equation*}
\begin{aligned}
& F^{[N_{f}]}(a+n\epsilon_{1},\{\tilde{\mu}_{i}\};\epsilon_{1},\epsilon_{2}-\epsilon_{1}; t)+F^{[N_{f}]}(a+n\epsilon_{2},\{\tilde{\mu}_{i}\};\epsilon_{2},\epsilon_{1}-\epsilon_{2}; t)\\[4pt]
\simeq& \frac{1}{\epsilon_{2}-\epsilon_{1}}\mathcal{W}^{[N_{f}]}_{0}(a+n\epsilon_{1},\{\tilde{\mu}_i\},\epsilon_{1}; t)+\mathcal{W}^{[N_{f}]}_{1}(a+n\epsilon_{1},\{\tilde{\mu}_i\},\epsilon_{1}; t)+\epsilon_{1}\leftrightarrow\epsilon_{2}\\[4pt]
=&2\mathcal{W}^{[N_{f}]}_{1}(a+n\hbar,\{\tilde{\mu}_i\},\hbar; t)-n\partial_{a}\mathcal{W}^{[N_{f}]}_{0}(a+n\hbar,\{\tilde{\mu}_i\},\hbar; t)-\partial_{\hbar}\mathcal{W}^{[N_{f}]}_{0}(a+n\hbar,\{\tilde{\mu}_i\},\hbar; t),
\end{aligned}
\end{equation*}
where we set $\epsilon_{1}=\epsilon_{2}=\hbar$. Thus, in this limit, equation \eqref{eq: C2 blow-up 1 Nf} becomes:
\begin{equation}
\label{eq: C2 blow-up 1 pure ns}
    \sum_{n \in \mathbb{Z}+\frac{1}{2}}e^{ 2\mathcal{W}_{1}^{[N_{f}]}(a+n\hbar,\{\tilde{\mu}_i\},\hbar; t)-n\partial_{a}\mathcal{W}_{0}^{[N_{f}]}(a+n\hbar,\{\tilde{\mu}_i\},\hbar; t)-\partial_{\hbar}\mathcal{W}_{0}^{[N_{f}]}(a+n\hbar,\{\tilde{\mu}_i\},\hbar; t)}=0,
\end{equation}
where the Coulomb branch parameter $a$ and the masses $\mu_i$ are shifted after taking the $\hbar$-derivative. In particular, after taking the limit, the shifted masses read $\tilde{\mu}_{1,2} = \mu_{1,2} - \hbar$ and $\tilde{\mu}_{3,4} = -\mu_{3,4} + \hbar$. Similarly, for the $\mathcal{N}=2^*$ gauge theory, the NS limit of the blow-up equation~(\ref{eq: C2 blow-up 1 *}) takes the form:
\begin{equation}
\label{eq: C2 blow-up 1 NS lame 1}
\begin{aligned}
& \theta_{2}(0|2\tau)\sum_{n \in \mathbb{Z}}e^{ 2\mathcal{W}_{1}(a+n\hbar,m,\hbar; \mathfrak{q})-n\partial_{a}\mathcal{W}_{0}(a+n\hbar,m,\hbar; \mathfrak{q})-\partial_{\hbar}\mathcal{W}_{0}(a+n\hbar,m,\hbar; \mathfrak{q})}\\
&-\theta_{3}(0|2\tau)\sum_{n \in \mathbb{Z}+\frac{1}{2}}e^{ 2\mathcal{W}_{1}(a+n\hbar,m,\hbar; \mathfrak{q})-n\partial_{a}\mathcal{W}_{0}(a+n\hbar,m,\hbar; \mathfrak{q})-\partial_{\hbar}\mathcal{W}_{0}(a+n\hbar,m,\hbar; \mathfrak{q})}=0.
\end{aligned}
\end{equation}

\subsection{\texorpdfstring{Solving the blow-up equations: $N_f\leq 4$ theories}{Solving the blow-up equations 1}}
Let us begin by considering equation (\ref{eq: C2 blow-up 1 pure ns}). To make it more explicit, we split the two prepotentials into classical, 1-loop and instanton parts, using the definitions \eqref{eq: classical part Nf} and \eqref{eq: 1-loop part Nf} . Then, dividing by a common factor of $t^{\frac{a^2}{\hbar ^2}-\frac{3}{4}}$, we find:
\begin{equation}
\label{eq: C2 blow-up 1 explicit}
\sum_{n\in \mathbb{Z}+\frac{1}{2}}t^{n^{2}} e^{ 2\mathcal{W}^{[N_f]}_{1,\,1\text{-}loop} -\left( \partial_{\hbar}+n \partial_{a}\right) \mathcal{W}^{[N_f]}_{0,\,1\text{-}loop} +2\mathcal{W}_{1,\, inst} - \left(\partial_{\hbar}+n \partial_{a}\right) \mathcal{W}^{[N_f]}_{0,\, inst}}=0,
\end{equation}
where $\mathcal{W}^{[N_f]}_{j,\,1\text{-}loop}=\mathcal{W}^{[N_f]}_{j,\,1\text{-}loop}\left(a + n\hbar,\{\tilde{\mu}_{i}\}, \hbar\right)$ and $\mathcal{W}^{[N_f]}_{j,\, inst}=\mathcal{W}^{[N_f]}_{j,\, inst}\left(a + n\hbar,\{\tilde{\mu}_{i}\}, \hbar ; t\right)$. Normalizing by the 1-loop term for $n=0$, what remains are just simple rational functions of $a$, $\{\mu_{i}\}_{i=1}^{N_f}$ and $\hbar$, with poles for $a \in \frac{\hbar}{2} \mathbb{Z}$. Indeed, with this normalization and setting $\hbar=1$, equation (\ref{eq: C2 blow-up 1 explicit}) reads:
\begin{equation}
\label{eq: C2 blow-up 1 explicit 2}
    \sum_{n\in \mathbb{Z}+\frac{1}{2}}t^{n^{2}}\mathcal{G}^{[N_f]}_{1\text{-}loop}\big(a,\{\mu_{i}\},n\big)\,e^{ 2\mathcal{W}^{[N_f]}_{1,\,inst}- \left( \partial_{\hbar}+n \partial_{a}\right) \mathcal{W}^{[N_f]}_{0,\, inst}  }=0,
\end{equation}
where $\mathcal{G}_{1\text{-}loop}^{[N_f]}(a,\{\mu_{i}\},n)$ can be computed as the $\epsilon_{1}=\epsilon_{2}=1$ limit of the $l$-factors $L^{(n,-n),1}$ in Appendix B of \cite{Jeong:2020uxz}. Namely, defining the functions:
\begin{subequations}
\begin{align}
g_{1\text{-}loop}^{hyp}(a, \mu, n) &= \prod _{j=0}^{n-\frac{1}{2}} (-a-j+\mu +1)^j (a+j+\mu +1)^j, \quad n\geq \frac{1}{2},\\
g_{1\text{-}loop}^{vec}(a, n) &=
\begin{cases}
\prod _{j=1}^{2 n} (-2 a-j+1)^j (2 a+j)^{j-1}, & \text{if } n \geq \frac{1}{2},\\[4pt]
 1, & \text{if } n =0,
\end{cases}\label{eq: gvec}
\end{align}
\end{subequations}
the 1-loop factor can be written as:
\begin{equation}
\label{eq: C2 G1L general}
\mathcal{G}^{[N_f]}_{1\text{-}loop}(a,\{\mu_{i}\},n)=
\begin{cases}
\frac{\prod_{i=1}^{N_{f}}g_{1\text{-}loop}^{hyp}(a, \mu_{i}, n)}{g_{1\text{-}loop}^{vec}(a, n)}, & \text{if } n \geq \frac{1}{2},\\[4pt]
 \frac{\prod_{i=1}^{N_{f}}g_{1\text{-}loop}^{hyp}(-a, \mu_{i}, -n)}{g_{1\text{-}loop}^{vec}(-a, -n)}, & \text{if } n \leq -\frac{1}{2}.
\end{cases}
\end{equation}
Generalizing the considerations of \cite{Beccaria:2016wop,Gorsky:2017ndg}, to solve equation (\ref{eq: C2 blow-up 1 explicit 2}) we make the following ansatz:
\begin{subequations}
\begin{align}
\mathcal{W}_{0,\,inst}^{[N_{f}]}(a,\{\mu_{i}\},\hbar;t) 
&= R_{0}^{[N_{f}]}(a,\{\mu_{i}\},\hbar;t) \nonumber\\
&\quad + \hbar\sum_{k,j=1}^{\infty}\left[\sum_{\pm}
    g^{[N_f]}_{k,j}\left(\frac{\left(t\hbar^{N_f -4}\right)^{j/2}}{j\pm2a/\hbar},\left\{\frac{\mu_{i}}{\hbar}\right\}\right)\right]\left(t\hbar^{N_f -4}\right)^{k+j/2-1}, \label{eq: resumm ansatz F0} \\
\mathcal{W}_{1,\,inst}^{\,[N_{f}]}(a,\{\mu_{i}\},\hbar;t) 
&= R_{1}^{[N_{f}]}(\{\mu_{i}\},\hbar;t) \nonumber\\
&\quad + \sum_{k,j=1}^{\infty}\left[\sum_{\pm}
\tilde{g}^{[N_f]}_{k,j}\left(\frac{\left(t\hbar^{N_f -4}\right)^{j/2}}{j\pm2a/\hbar},\left\{\frac{\mu_{i}}{\hbar}\right\}\right)\right]\left(t\hbar^{N_f -4}\right)^{k+j/2-1} \nonumber\\
&\quad + \sum_{k,j=1}^{\infty}\left[\sum_{\pm}
\tilde{f}^{[N_f]}_{k,j}\left(\frac{\left(t\hbar^{N_f -4}\right)^{j/2}}{j\pm2a/\hbar},\left\{\frac{\mu_{i}}{\hbar}\right\}\right)\right]\left(t\hbar^{N_f -4}\right)^{k-1}, \label{eq: resumm ansatz F1}
\end{align}
\end{subequations}
where the $resummation$ $functions$ $g^{[N_f]}_{k,j}(x,\{\mu_i\})$, $\tilde{g}^{[N_f]}_{k,j}(x,\{\mu_i\})$ and $\tilde{f}^{[N_f]}_{k,j}(x,\{\mu_i\})$ must be analytic in a neighborhood of $x=0$, where they vanish. Moreover, in order for the ansatz to admit a power series expansion in integer powers of $t$, the functions must have definite parity in their first argument, namely, the $g^{[N_f]}_{k,j}$ and $\tilde{g}^{[N_f]}_{k,j}$ are assumed to be odd, while the functions $\tilde{f}_{k,j}$ are assumed to be even. The functions $R_i^{[N_f]}$ are also vanishing at $t=0$ and take into account the regular part of the instanton expansion. They can be computed explicitly from $R_{0}^{[4]}$ and $R_{1}^{[4]}$ in equations (\ref{eq: reg0}) and (\ref{eq: reg1}) by taking appropriate decoupling limits. To obtain equations for the resummation functions we proceed in the following way. The blow-up equation (\ref{eq: C2 blow-up 1 explicit 2}) holds for any value of the parameter $a$, allowing us to set:
\begin{equation}
\label{eq: a pole}
    a = \frac{\hbar}{2}\left(m_{1}+\frac{1}{x}t^{\frac{m_{2}}{2}} \right),\quad m_{1}\in \mathbb{Z},\quad m_{2}\in \mathbb{N},
\end{equation}
with $x$ a complex variable, and expand the resulting expression in $t$. Due to the symmetry of the equation for $a \to -a$ (accompanied by $n \to -n$), we may restrict to $m_{1}\geq 0$. This choice allows us to focus on the functions with $j=m_{2}$, as can be seen by noting that, when $j=m_{2}$ and $2n=\mp j-m_{1}$, these functions and their derivatives enter the equations as: 
\begin{equation*}
    g_{k,j}^{[N_f]}\left(\frac{\left(t\hbar^{N_f -4}\right)^{j/2}}{j\pm (2n+m_{1})\pm t^{m_{2}/2}/x},\left\{\frac{\mu_{i}}{\hbar}\right\}\right)=g^{[N_f]}_{k,m_{2}}\left(\pm x\hbar^{m_{2}(N_f-4)/2},\left\{\frac{\mu_{i}}{\hbar}\right\}\right).
\end{equation*}
Meanwhile, when $2n=\mp j-m_{1}$ and $j\neq m_{2}$ we have two possibilities: if $m_{2}<j$ the functions will enter through their expansion around zero, while if $m_{2}>j$ they will enter through their expansion around infinity. Finally, if $j\pm (2n+m_{1})\neq 0$, the functions will enter simply through their expansion around zero. We now begin by deriving the equations satisfied by the resummation functions and show that they can be easily solved. Finally, we discuss the structure of the blow-up equation's expansion, taking into account the behavior of the resummation functions expanded around infinity. 
\subsubsection{Algebraic equations for the resummation functions}
First of all, we consider the simple rational functions $\mathcal{G}^{[N_f]}_{1\text{-}loop}(a,\{\mu_i\},n)$ in equation \eqref{eq: C2 G1L general}. When both $a>0$ and $n>0$, its expansion in $a$ around some fixed positive value will start from a constant term. Then, for any value of $n\in \mathbb{Z}+\frac{1}{2}$, $\mathcal{G}_{1\text{-}loop}(a,n)$ has always a simple pole for $a=0$ and finally, for $n<0$, the function has poles at $a=j/2$, $j\in \{1,...,2|n|-1\}$, of order $2j$ except for $j=2|n|$, for which the order is $2|n|-1$. Setting $2a = m_{1}+t^{m_{2}/2}/x$, expanding in $t$ and combining with the overall power of $t^{n^{2}}$, results in a predictable leading order behavior. In particular, for $2n=\mp m_{2}-m_{1}$ we find:
\begin{equation*}
\begin{aligned}
&t^{(m_{1}+m_{2})^{2}/4}\mathcal{G}^{[N_f]}_{1\text{-}loop}\left(\frac{m_{1}x+t^{m_{2}/2}}{2x},\{\mu_{i}\},-\frac{m_{2}+m_{1}}{2}\right)\sim \mathcal{O}\big(t^{(m_{1}-m_{2})^{2}/4}\big),\\
& t^{(m_{1}-m_{2})^{2}/4}\mathcal{G}^{[N_f]}_{1\text{-}loop}\left(\frac{m_{1}x+t^{m_{2}/2}}{2x},\{\mu_{i}\},\frac{m_{2}-m_{1}}{2}\right)\sim \mathcal{O}\big(t^{(m_{1}-m_{2})^{2}/4}\big),
\end{aligned}
\end{equation*}
while for values of $n$ between $-(m_{2}+m_{1})/2$ and $(m_{2}-m_{1})/2$, the power of $t$ at leading order can be negative. Let us now consider the instanton contributions. To understand the structure of the expansion and the form of the resulting equations, we can examine the $n$-th term of the blow-up equation (\ref{eq: C2 blow-up 1 explicit 2}), where $n$ is such that $j\pm (2n+m_{1})=0$, for some $j \leq m_{2}$. These terms give the most interesting contributions, since for $j<m_{2}$ we will find the non-trivial expansion of the resummation functions around infinity and for $j=m_{2}$ we will find the functions valued at $\pm x$, while for $j>m_{2}$ the functions are expanded around zero. For example, we have:
\begin{equation*}
\begin{aligned}
\mathcal{W}_{1,\, inst}^{\mp,\,[N_f]}&=\mathcal{W}^{[N_f]}_{1,\, inst}\left(\frac{\mp jx+t^{m_{2}/2}}{2x},\{\tilde{\mu}_{i}\}, 1; t\right)\\
&= \sum_{k=1}^{\infty}t^{k-1}\left[\Tilde{f}^{[N_f]}_{k,j}\left(y,\{\tilde{\mu}_{i}\}\right)\pm t^{j/2}
\Tilde{g}^{[N_f]}_{k,j}\left(y,\{\tilde{\mu}_{i}\}\right)\right]+h^{\mp,\,[N_f]}(x,\{\mu_{i}\},t),
\end{aligned}
\end{equation*}
and similarly for $\mathcal{W}^{[N_f]}_{0,\, inst}$, where $y=x t^{(j-m_{2})/2}$, while the functions $h^{\mp,\,[N_f]}$ take into account the remaining part of the expression, which will have an expansion starting from some positive power of $t$. The parity of the resummation functions has been used to simplify the previous formula. Thus, isolating only the relevant terms at the exponent in equation \eqref{eq: C2 blow-up 1 explicit 2}, we find the following expression:
\begin{equation}
\label{eq: C2 Gexp gen}
\begin{aligned}
&2\mathcal{W}^{\mp,\,[N_f]}_{1,\,inst}- \left( \partial_{\hbar}+n \partial_{a}\right) \mathcal{W}^{\mp,\, [N_f]}_{0,\, inst}=\sum_{k\geq 1}t^{k-1}\bigg[2\tilde{f}^{[N_f]}_{k,j}(y,\{\tilde{\mu}_{i}\})\mp m_{1}y^{2} \partial_{y}g^{[N_f]}_{k,j}(y,\{\tilde{\mu}_{i}\})\\
&\pm t^{j/2}\bigg(2\tilde{g}^{[N_f]}_{k,j}(y,\{\tilde{\mu}_{i}\})+\gamma_{k,j}^{[N_f]}g^{[N_f]}_{k,j}(y,\{\tilde{\mu}_{i}\})+  \delta_j^{[N_f]}y\, \partial_{y}g^{[N_f]}_{k,j}(y,\{\tilde{\mu}_{i}\})\\
&+\sum_{i=1}^{N_f}\tilde{\mu}_{i}\partial_{\tilde{\mu}_i}g^{[N_f]}_{k,j}(y,\{\tilde{\mu}_{i}\})\bigg)\bigg] + s^{\mp,\, [N_f]}(x,\{\mu_{i}\},t),
\end{aligned}
\end{equation}
where:
\begin{subequations}
\begin{align}
\gamma_{k,j}^{[N_f]}&=(4-N_f)(k+j/2)+N_f-5,\\
\delta_j^{[N_f]}&=(4-N_f)j/2-1,
\end{align}
\end{subequations}
and the functions $s^{\mp,\,[N_f]}(x,\{\mu_{i}\}, t)$ vanish at $t=0$. By collecting terms of the same order in $t$ from the expansion of the blow-up equation, these expressions, together with the 1-loop term, allow us to derive the form of the equations satisfied by the resummation functions and show that the $t$-expansion around the poles (\ref{eq: a pole}) is well-behaved. Namely, by setting $j = m_2$, assuming $m_1 > 0$ for simplicity\footnote{The equations for $m_{1}=0$ are automatically satisfied once a solution for the $m_1>0$ case is found.}, omitting the $[N_f]$ label from the resummation functions, and performing the change of variables:
\begin{equation}
\label{eq: C2 log branch}
\tilde{f}_{1,m_2}=\frac{1}{2}\log u_{m_2}, \quad\quad\quad
\partial_x g_{1,m_2}=\frac{1}{x^{2}}\log(x\, v_{m_2}) ,
\end{equation}
we obtain at leading order:
\begin{equation}
\label{eq: 1st linear eq}
u_{m_2}\big(v_{m_2}^{m_1}+c_{m_1,m_2} v_{m_2}^{-m_1}\big)+r_{m_1,m_2,1}=0
\end{equation}
where the constant $c_{m_1,m_2}$ can be computed from the 1-loop factor \eqref{eq: C2 G1L general} and reads:
\begin{equation}
    c_{m_1,m_2}(\{\mu_i\})=(-1)^{m_1+1}\frac{\prod_{i=1}^{N_f}\tilde{c}_{m_1,m_2}(\mu_i+1)}{m_2^{2m1}[(m_2-1)!]^{4m_1}},
\end{equation}
with $\tilde{c}_{m_1,m_2}(\mu)$ defined in equation \eqref{eq: c tilde}. We observe here, via the change of variables \eqref{eq: C2 log branch}, the appearance of logarithmic branching, necessary to obtain a linear equation \eqref{eq: 1st linear eq} in the function $u_{m_2}$. Meanwhile, at orders $t^{k-1}$ and $t^{k+j/2-1}$, respectively, compared to the leading order equations, we find:
\begin{equation}
\label{eq: linear equations fk}
2(v_{m_2}^{m_1}+c_{m_1,m_2}v_{m_2}^{-m_1})\tilde{f}_{k,m_2}+m_1x^2(v_{m_2}^{m_1}-c_{m_1,m_2}v_{m_2}^{-m_1})\partial_x g_{k,m_2}+r_{m_1,m_2,k}=0
\end{equation}
\begin{equation}
\label{eq: linear equations gtk}
2\tilde{g}_{k,j}+\gamma_{k,j}^{[N_f]} g_{k,j}+ \delta_{j}^{[N_f]} x\partial_x g_{k,j} +\sum_{i=1}^{N_f}\tilde{\mu}_{i}\partial_{\tilde{\mu}_{i}}g_{k,j}+\tilde{r}_{m_1,m_2,k}=0
\end{equation}
The functions $r_{m_{1},m_{2},k}(x,\{\mu_i\})$ and $\tilde{r}_{m_{1},m_{2},k}(x,\{\mu_i\})$ account for the remaining contributions of the $t$-expansion at each fixed order and can be explicitly computed for any choice of $m_1$ and $m_2$. In particular, they also include contributions from resummation functions that are assumed to have been determined from equations at lower orders in the expansion. For example, in equation \eqref{eq: linear equations gtk}, the terms involving the function $g_{k,j}$ could, in principle, be absorbed into $\tilde{r}_{m_{1},m_{2},k}$. However, we write them explicitly to highlight the simple relations that connect the functions at each step. The three families of equations \eqref{eq: 1st linear eq}, \eqref{eq: linear equations fk} and \eqref{eq: linear equations gtk}, determine the resummation functions up to a branching ambiguity, which can be fixed by imposing the appropriate behavior at zero and at infinity, see Appendix \ref{app: blow-up example 1} for an explicit example. Indeed, equation \eqref{eq: 1st linear eq} determines $u_{m_2}$ as a function of $v_{m_2}$, and by equating two such expressions for different values of $m_1$, one finds:
\begin{equation}
\label{eq: C2 root branch}
    v_{m_2}^{\tilde{m}_1}\,r_{\tilde{m}_1,m_2,1}-v_{m_2}^{m_1}\,r_{m_1,m_2,1}=0\implies \begin{cases}
v_{m2}^{2}=\frac{r_{2,m_2,1}}{r_{4,m_2,1}}, & \text{for } m_2 \text{ odd},\\[4pt]
 v_{m2}^{2}=\frac{r_{1,m_2,1}}{r_{3,m_2,1}}, & \text{for } m_2 \text{ even},
\end{cases}
\end{equation}
where in the last step we set $m_1=2$, $\tilde{m}_1=4$ for odd $m_2$, and $m_1=1$, $\tilde{m}_1=3$ for even $m_2$, respectively. Thus, $v_{m_2}$ is determined by a simple quadratic equation, which gives rise to a square root branching structure in the resummation functions. Then, two copies of equation \eqref{eq: linear equations fk}, evaluated at different values of $m_1$, determine the functions $\tilde{f}_{k,m_2}$ and $g_{k,m_2}$, with the integration constant of the latter fixed by requiring it to vanish at zero. Finally, equation \eqref{eq: linear equations gtk} determines $\tilde{g}_{k,m_2}$ in terms of the previously determined resummation functions. Thus, we have shown that expanding the blow-up equation \eqref{eq: C2 blow-up 1 explicit 2} for two distinct values of $m_1$ is sufficient to determine all the resummation functions. For example, the first few functions in the $N_f =0$ case are given by:
\begin{center}
\begin{tabular}{c c}
\vspace{1mm}
$g_{1,1}^{[0]}(x) = -\dfrac{1 - \sqrt{1 + 4x^2} + \log\left( \sqrt{\tfrac{1}{4} + x^2} + \tfrac{1}{2} \right)}{x},$
\vspace{1mm}
&
\vspace{1mm}
$g^{[0]}_{2,1}(x) = \dfrac{(2 - x^2)\sqrt{1 + 4x^2} - 3x^2 - 2}{12x^3},$
\vspace{1mm}
\\
\vspace{1mm}
$g^{[0]}_{1,2}(x) = \dfrac{\sqrt{1 + x^2} - 1 - \log\left( \tfrac{1}{2} \sqrt{1 + x^2} + \tfrac{1}{2} \right)}{x},$
\vspace{1mm}
&
\vspace{1mm}
$g^{[0]}_{2,2}(x) = -\dfrac{8(\sqrt{1 + x^2} - 1)}{9x},$
\vspace{1mm}
\\
\vspace{1mm}
$\tilde{g}^{[0]}_{1,1}(x) = \dfrac{1-\sqrt{1 + 4x^2} }{2x},$
\vspace{1mm}
&
\vspace{1mm}
$\tilde{g}^{[0]}_{2,1}(x) = \dfrac{5x^4 - 5x^2 - 1 + \tfrac{12x^4 + 7x^2 + 1}{\sqrt{1 + 4x^2}}}{6x^3},$
\vspace{1mm}
\\
\vspace{1mm}
$\tilde{g}^{[0]}_{1,2}(x) = \dfrac{1-\sqrt{1 + x^2} }{4x},$
\vspace{1mm}
&
\vspace{1mm}
$\tilde{g}^{[0]}_{2,2}(x) = \dfrac{32 - \tfrac{2(13x^2 + 16)}{\sqrt{1 + x^2}}}{27x},$
\vspace{1mm}
\\
\vspace{1mm}
$\tilde{f}^{[0]}_{1,1}(x) = -\dfrac{1}{4} \log(1 + 4x^2),$
\vspace{1mm}
&
\vspace{1mm}
$\tilde{f}^{[0]}_{2,1}(x) = \dfrac{3x^2}{2(1 + 4x^2)},$
\vspace{1mm}
\\
\vspace{1mm}
$\tilde{f}^{[0]}_{1,2}(x) = -\dfrac{1}{4} \log(1 + x^2),$
\vspace{1mm}
&
\vspace{1mm}
$\tilde{f}^{[0]}_{2,2}(x) = \dfrac{4x^2}{9(1 + x^2)}.$
\vspace{1mm}
\\
\end{tabular}
\end{center}
while their general structure will be discussed in Section \ref{sec: results}. However, two points may still appear unclear at this stage. First, the presence of logarithmic terms in the resummation functions could invalidate the expansion of the blow-up equation for $m_2>1$, as these functions will be expanded around infinity. We will show shortly that this is not the case. Second, it is not clear whether knowledge of the Taylor coefficients of the resummation functions with $j>m_2$ is required to determine the functions at the step $j=m_2$. While such terms do appear in the functions $r_{m_1,m_2,k}$ and $\tilde{r}_{m_1,m_2,k}$ and introduce some subtleties, discussed in detail through an example in Appendix \ref{app: blow-up example 1}, these do not affect the overall procedure. In conclusion, we found that a single blow-up equation is enough to determine all the resummation functions $g^{[N_f]}_{k,j}$, $\tilde{g}^{[N_f]}_{k,j}$ and $\tilde{f}^{[N_f]}_{k,j}$. Moreover, since $\mathcal{W}^{[N_f]}_{0,\,inst}$ and $\mathcal{W}^{[N_f]}_{1,\, inst}$, as in equations (\ref{eq: resumm ansatz F0}) and (\ref{eq: resumm ansatz F1}), coincide with the known instanton expansion when re-expanded around $t=0$, they will automatically satisfy any other blow-up equation.

\subsubsection{Structure of the expansion}
\label{sec: structure expansion}
Let us now examine more carefully the structure of the blow-up equation's expansion, taking into account the asymptotic behavior of the resummation functions at infinity. As already mentioned, there is a distinction between the cases $ m_2 = 1 $ and $ m_2 > 1 $, in that in the former no resummation function is expanded around infinity. Moreover, for $ m_2 > 1 $, certain summands in the blow-up equation (\ref{eq: C2 blow-up 1 explicit 2}) begin with negative powers of $ t $; specifically, this occurs for some values of $ n $ within the range $ \{-\frac{m_2 + m_1}{2}, \ldots, \frac{m_2 - m_1}{2}\} $. However, these terms combine in such a way that the first non-zero coefficient in the final expression appears at order $ t^{(m_1 - m_2)^2 / 4} $. To illustrate the difference between the two cases, let us first consider a single function. Take, for instance, a value of $ n $ such that $ 2n + j + m_1 = 0 $ for some $ j < m_2 $. Then, for example:
\begin{equation*}
\mathcal{W}_{1,\, inst}^{\mp,\, [N_f]}= \sum_{k=1}^{\infty}t^{k-1}\left[\Tilde{f}^{[N_f]}_{k,j}\left(\frac{x}{t^{(m_{2}-j)/2}},\{\tilde{\mu}_{i}\}\right)\pm t^{j/2}
\Tilde{g}^{[N_f]}_{k,j}\left(\frac{x}{t^{(m_{2}-j)/2}},\{\tilde{\mu}_{i}\}\right)\right]+h^{\mp,\, [N_f]},
\end{equation*}
By explicitly solving for the resummation functions, as shown in the previous section, we find that the asymptotic behavior of $g^{[N_f]}_{1,j}$ and $\tilde{f}^{[N_f]}_{1,j}$ is given by:
\begin{equation*}
    g^{[N_f]}_{1,j}(y,\{\tilde{\mu}_{i}\})=\text{const}-\frac{1}{2y}\log y^{2}+\mathcal{O}(y^{-1}),\quad  \tilde{f}^{[N_f]}_{1,j}(y,\{\tilde{\mu}_{i}\})=-\frac{1}{4}\log y^{2}+\text{const}+\mathcal{O}(y^{-2}).
\end{equation*}
In contrast, the asymptotic behavior of all other functions is purely power-like, starting from a constant term. Therefore, by absorbing the subleading terms into $h^{\mp,\, [N_f]}$, we obtain:
\begin{equation*}
    \mathcal{W}_{1,\, inst}^{\mp,\, [N_f]}= \frac{1}{4}\log\!\left(\frac{t^{m_{2} - j}}{x^{2}}\right)+\tilde{h}^{\mp,\, [N_f]}(x,\{\mu_{i}\},t).
\end{equation*}
Such logarithmic terms remain at the exponent in equation \eqref{eq: C2 blow-up 1 explicit 2}, effectively decreasing or increasing the overall power of $t$ from which the expansion of a particular term in the blow-up equation begins. Indeed, considering the expression \eqref{eq: C2 Gexp gen}, we can isolate the terms containing logarithms, i.e. the $k=1$ ones, and obtain:
\begin{equation*}
2\mathcal{W}^{\mp,\, [N_f]}_{1,\,inst}- \left( \partial_{\hbar}+n \partial_{a}\right) \mathcal{W}^{\mp,\, [N_f]}_{0,\, inst}=(\mp m_{1}-1/2)\log \frac{x^{2}}{t^{m_{2}-j}}\, + \tilde{s}^{\mp,\, [N_f]}(x,\{\mu_i\},t),
\end{equation*}
where $\tilde{s}^{\mp,\, [N_f]}$ is regular at $t=0$. In particular, we notice that all the $t\log t$ terms cancel in the final expression. Thus, we have shown that the expansion of the blow-up equation (\ref{eq: C2 blow-up 1 explicit 2}) considered above indeed leads to a well-defined power series in the parameter $ t $.
\subsection{\texorpdfstring{Solving the blow-up equations: $\mathcal{N}=2^{*}$ theory}{Solving the blow-up equations 2}}
It is straightforward to repeat the previous analysis in the $\mathcal{N}=2^{*}$ case. Denoting by $\mu$ the mass of the adjoint hypermultiplet, to avoid confusion with the integers $m_{1}$ and $m_{2}$, and setting $\hbar=1$, we can write the blow-up equation (\ref{eq: C2 blow-up 1 NS lame 1}) as:
\begin{equation}
\label{eq: C2 blow-up lame expl}
\begin{aligned}
& \theta_{2}(0|2\tau)\sum_{n \in \mathbb{Z}}\mathfrak{q}^{n^{2}}\mathcal{G}_{1\text{-}loop}(a,\mu,n)e^{ 2\mathcal{W}_{1,\,inst}-n\partial_{a}\mathcal{W}_{0,\,inst}-\partial_{\hbar}\mathcal{W}_{0,\,inst}}\\
&-\theta_{3}(0|2\tau)\sum_{n \in \mathbb{Z}+\frac{1}{2}}\mathfrak{q}^{n^{2}}\mathcal{G}_{1\text{-}loop}(a,\mu,n)e^{ 2\mathcal{W}_{1,\,inst}-n\partial_{a}\mathcal{W}_{0,\,inst}-\partial_{\hbar}\mathcal{W}_{0,\,inst}}=0,
\end{aligned}
\end{equation}
where $\mathcal{W}_{i,\, inst}=\mathcal{W}_{i,\,inst}\left(a + n,\mu,1; \mathfrak{q}\right)$. Explicitly, the 1-loop term reads:
\begin{equation}
\label{eq: C2 G1L *}
\mathcal{G}_{1\text{-}loop}(a,\mu,n)=
\begin{cases}
\frac{g_{1\text{-}loop}^{adj}(a, \mu, n)}{g_{1\text{-}loop}^{vec}(a, n)}, & \text{for } n \geq 0,\\[4pt]
 \frac{g_{1\text{-}loop}^{adj}(-a, \mu_{i}, -n)}{g_{1\text{-}loop}^{vec}(-a, -n)}, & \text{for } n < -0,
\end{cases}
\end{equation}
where $g_{1\text{-}loop}^{vec}$ is defined in equation \eqref{eq: gvec}, while:
\begin{equation}
    g_{1\text{-}loop}^{adj}(a, \mu, n) =
\begin{cases}
\prod _{j=1}^{2n} (-2a-j-\mu +1)^j (2a+j-\mu )^{j-1}, & \text{for } n > 0,\\[4pt]
 1, & \text{for } n = 0.
\end{cases}
\end{equation}
To solve equation (\ref{eq: C2 blow-up lame expl}), we consider the ansatz:
\begin{subequations}
\begin{align}
\mathcal{W}_{0,\, inst}(a,\mu,\hbar; \mathfrak{q}) 
&= \hbar\sum_{j,k=1}^{\infty}\left[\sum_{\pm}
g_{k,j}\left(\frac{\mathfrak{q}^{j/2}}{j\pm2a/\hbar},\frac{\mu}{\hbar}\right)\right]\mathfrak{q}^{k+j/2-1},\label{ansatzeN2s0}\\
\mathcal{W}_{1,\, inst}(a, \mu,\hbar; \mathfrak{q}) 
&= \log\varphi(\mathfrak{q})^{-1}
+ \sum_{j,k=1}^{\infty}\left[\sum_{\pm}
\tilde{g}_{k,j}\left(\frac{\mathfrak{q}^{j/2}}{j\pm 2a/\hbar},\frac{\mu}{\hbar}\right)\right]\mathfrak{q}^{k+j/2-1} \nonumber\\
&\quad + \sum_{j,k=1}^{\infty}\left[\sum_{\pm}
\tilde{f}_{k,j}\left(\frac{\mathfrak{q}^{j/2}}{j\pm2a/\hbar},\frac{\mu}{\hbar}\right)\right]\mathfrak{q}^{k-1}\label{ansatzeN2s1},
\end{align}
\end{subequations}
and expand in $\mathfrak{q}$ the blow-up equation, setting:
\begin{equation}
\label{eq: a pole q}
    a = \frac{\hbar}{2}\left(m_{1}+\frac{1}{x}\mathfrak{q}^{\frac{m_{2}}{2}} \right),\quad m_{1}\in \mathbb{Z},\quad m_{2}\in \mathbb{N}.
\end{equation}
In this case, the regular parts of the resummation ansatz are determined by considering the $ \mathcal{N}=4 $ limit of the partition function \eqref{eq: SU(2) Z *}, i.e. the $ \mu \to 0 $ limit, which is known exactly. Then, we can compute the analogue of \eqref{eq: C2 Gexp gen}, which reads:
\begin{equation}
\label{eq: C2 Gexp lame}
\begin{aligned}
&2\mathcal{W}^{\mp}_{1,\,inst}- \left( \partial_{\hbar}+n \partial_{a}\right) \mathcal{W}^{\mp}_{0,\, inst}=\sum_{k\geq 1}\mathfrak{q}^{k-1}\bigg[2\tilde{f}_{k,j}(y,\mu)\mp m_{1}y^{2}\partial_{y}g_{k,j}(y,\mu)\\
&\pm \mathfrak{q}^{j/2}\bigg(2\tilde{g}_{k,j}(y,\mu)-g_{k,j}(y,\mu)-y\partial_{y} g_{k,j}(y,\mu)+\mu\partial_{\mu}g_{k,j}(y,\mu)\bigg) \bigg]+s^{\mp}(x,\mu,t)
\end{aligned}
\end{equation}
where $y=x \mathfrak{q}^{(j-m_{2})/2}$. This expression allows us to write the explicit form of the equations, which read:
\begin{subequations}
\label{eq: C2 linear *}
\begin{align}
&u_{m_2}\big(v_{m_2}^{m_1}+c_{m_1,m_2} v_{m_2}^{-m_1}\big)+r_{m_1,m_2,1}=0,\\[4pt]
&2(v_{m_2}^{m_1}+c_{m_1,m_2}v_{m_2}^{-m_1})\tilde{f}_{k,m_2}+m_1x^2(v_{m_2}^{m_1}-c_{m_1,m_2}v_{m_2}^{-m_1})\partial_x g_{k,m_2}+r_{m_1,m_2,k}=0,\\[4pt]
&2\tilde{g}_{k,j}- g_{k,j}- x\partial_x g_{k,j} +\mu\partial_{\mu}g_{k,j}+\tilde{r}_{m_1,m_2,k}=0,
\end{align}
\end{subequations}
where each function depends on the variable $x$ and on the adjoint mass $\mu$ only. In particular, the constants $c_{m_1,m_2}(\mu)$ can be computed from \eqref{eq: C2 G1L *} to be:
\begin{equation}
    c_{m_1,m_2}(\mu)=(-1)^{m_1+1}\frac{\prod_{j=0}^{m_2-1}(j+\mu)^{2m_1}(j-\mu+1)^{2m_1}}{m_2^{2m1}[(m_2-1)!]^{4m_1}}.
\end{equation}
Since the equations \eqref{eq: C2 linear *} are essentially identical to those presented in the previous section, we do not repeat the discussion of their solution here. Instead, we note that the same logic applies and refer the reader to Appendix \ref{app: blow-up example 1} for an explicit example. The general structure of the solutions is likewise essentially unchanged, as can be seen, for instance, from the very first functions:
\begin{subequations}
\begin{align*}
g_{1,1}(x,\mu) &= \frac{
\sqrt{1 + 4 \mu^2 (\mu - 1)^2 x^2} 
- \log\left( \tfrac{1}{2} + \tfrac{1}{2}\sqrt{1 + 4 \mu^2 (\mu - 1)^2 x^2} \right) - 1
}{x}, \\[1ex]
\tilde{g}_{1,1}(x,\mu) &= \frac{(-1 + \mu + \mu^2)\left(1- \sqrt{1 + 4 \mu^2 (\mu - 1)^2 x^2}  \right)}{2 \mu (\mu - 1) x},\\[1ex]
\tilde{f}_{1,1}(x,\mu) &= -\frac{1}{4} \log\left( 1+4 \mu^2(\mu -1)^2  x^2\right).
\end{align*}
\end{subequations}
One can also verify that, given the resulting asymptotic behavior of the resummation functions, the expansion of the blow-up equation \eqref{eq: C2 blow-up lame expl} is well-defined. Finally, we have shown that also for the $\mathcal{N}=2^{*}$ theory, a single blow-up equation completely determines the functions appearing in the resummation ansatz, which, if re-expanded, reproduces the usual Nekrasov expansions. Thus, any other blow-up equation will be satisfied automatically by the ansatz, as in the previous case. 
\subsection{General structure of the resummation functions}
\label{sec: results}
We conclude this section by outlining the general structure exhibited by the resummation functions\footnote{The emerging pattern has been tested for higher or lower values of $k$ and $n$, depending on the computational complexity of the equations involved.}. In particular, our results are compatible with and extend those previously conjectured in the literature \cite{Beccaria:2016wop,Gorsky:2017ndg}. Starting from the NS prepotentials, or semi-classical blocks, we find:
\begin{subequations}
\begin{align}
&g_{1,n}(z,\{\mu_{i}\})=-\frac{\log\left(\frac{1}{2}+\sqrt{\frac{1}{4}+w_{n}z^{2}/\zeta_{n}^{2}}\right)+1-\sqrt{1+4w_{n}z^{2}/\zeta_{n}^{2}}}{z},\\
&g_{k,n}(z,\{\mu_{i}\})=\frac{1}{z^{2k-1}}\left[\left(1+\frac{4w_{n}z^{2}}{\zeta_{n}^{2}}\right)^{5/2-k}Q_{k,n}\left(z^2,\{\mu_{i}\}\right)+P_{k,n}\left(z^2,\{\mu_{i}\}\right)\right],
\end{align}
\end{subequations}
where 
\begin{equation}
w_{n}=
\begin{cases}
w_{n}^{[N_{f}]}=\prod_{k=\frac{1-n}{2}}^{\frac{n-1}{2}}\prod_{i=1}^{N_{f}}(\mu_{i}+k), & \text{for } 0\leq N_f\leq 4,\\[4pt]
c_n^2=\prod_{k=-n}^{n-1}(\mu+k)^2,     & \text{for } \mathcal{N}=2^{*},\\[4pt]
\end{cases}
\end{equation}
and $\zeta_{n}=n!(n-1)!$. The functions $ Q_{k,n} $ and $ P_{k,n} $ are polynomials of degree $ 2k - 3 $ and $ k - 1 $, respectively, with coefficients that are themselves polynomials in the masses. Meanwhile, for the first $ \epsilon_2 $ corrections to the NS prepotentials, or equivalently, the first $ b^2 $ corrections to the semi-classical blocks, we find:
\begin{subequations}
\begin{align}
&\tilde{g}_{k,n}(z,\{\mu_{i}\})=\frac{1}{z^{2k-1}}\left[\left(1+\frac{4w_{n}z^{2}}{\zeta_{n}^{2}}\right)^{\frac{3}{2}-k}U_{k,n}\left(z^{2},\{\mu_{i}\}\right)+V_{k,n}\left(z^{2},\{\mu_{i}\}\right)\right],\\
&\tilde{f}_{1,n}(z,\{\mu_{i}\})=-\frac{1}{4}\log\left(1+4w_{n}z^{2}/\zeta_{n}^{2}\right),\quad \tilde{f}_{k,n}(z,\{\mu_{i}\})=\frac{z^{2}Z_{k,n}(z^2,\{\mu_{i}\})}{\left(1+4w_{n}z^{2}/\zeta_{n}^{2}\right)^{k-1}},
\end{align}
\end{subequations}
where the functions $U_{k,n}$, $V_{k,n}$, and $Z_{k,n}$ are polynomials\footnote{The polynomials $P_{k,n}$ and $V_{k,n}$ are not independent of their partners $Q_{k,n}$ and $U_{k,n}$, since they are determined from them by requiring the functions $g_{k,n}$ and $\tilde{g}_{k,n}$ to be of $\mathcal{O}(z)$ as $z \to 0$.} of degree $2k - 2$, $k - 1$, and $k - 2$, respectively, with coefficients that are again polynomials in the masses. In particular, for the $N_f\leq 4$ cases, we find that also the functions $\tilde{g}_{1,n}$ seem to have a simple universal structure, namely: 
\begin{equation}
    \tilde{g}_{1,n}(z,\{\mu_i\})=\frac{1-\sqrt{1+\frac{4 w_n z^2}{\zeta_n^2}}}{2 n z},
\end{equation}
while the same does not appear to be true in the $\mathcal{N}=2^{*}$ case. As an example, the first few independent polynomials for $N_f=1$ read:
\begin{subequations}
\begin{align*}
&Q_{2,1}(x,\mu_1)=\frac{1}{\mu_1^2}\left(\frac{1}{48} \left(-1+8\mu_1^2\right)+\frac{1}{24} \mu_1 \left(1-2 \mu_1^2\right) x^2\right),\quad Q_{2,2}(x,\mu_1)=-\frac{8}{9} \mu_1 x^2,\\
&U_{2,1}(x,\mu_1)=-\frac{1}{6}-\frac{5}{6}\mu_1 x^2-\left(\frac{1}{4}-\frac{5}{6}\mu_1^2\right) x^4, \quad U_{2,2}(x,\mu_1)= -\frac{13}{54}\mu_1 \left(4 \mu_1^2-1\right) x^4-\frac{32}{27}\mu_1 x^2,\\
&Z_{2,1}(x,\mu_1)=-\frac{1}{4} \left(1-6 \mu_1^2\right),\quad Z_{2,2}(x,\mu_1)=-\frac{1}{9} \mu_1 \left(1-4 \mu_1^2\right).
\end{align*}
\end{subequations}
Moreover, for the $\mathcal{N}=2^{*}$ case, we find that by parametrizing the resummation functions in the following way:
\begin{subequations}
\label{eq: results star}
\begin{align}
&g^{*}_{1,n}(z,\mu)=-\frac{\log\left(\frac{1}{2}+\sqrt{\frac{1}{4}+c_{n}^{2}z^{2}/\zeta_{n}^{2}}\right)+1-\sqrt{1+4c_{n}^{2}z^{2}/\zeta_{n}^{2}}}{z},\\
&g^{*}_{k,n}(z,\mu)=\frac{1}{z\left(c_{n}z\right)^{2k-2}}\left[\left(1+\frac{4c_{n}^{2}z^{2}}{\zeta_{n}^{2}}\right)^{5/2-k}Q^{*}_{k,n}\left(c_{n}^{2}z^{2},\mu\right)+P^{*}_{k,n}\left(c_{n}^{2}z^{2},\mu\right)\right],\\
&\tilde{g}^{*}_{k,n}(z,\mu)=\frac{1}{(c_{n}z)^{2k-1}}\left[\left(1+\frac{4c_{n}^{2}z^{2}}{\zeta_{n}^{2}}\right)^{\frac{3}{2}-k}U^{*}_{k,n}\left(c_{n}^{2}z^{2},\mu\right)+V^{*}_{k,n}\left(c_{n}^{2}z^{2},\mu\right)\right],\\
&\tilde{f}^{*}_{1,n}(z,\mu)=-\frac{1}{4}\log\left(1+4c_{n}^{2}z^{2}/\zeta_{n}^{2}\right), \quad \tilde{f}^{*}_{k\geq 2,n}(z,\mu)=\frac{c_{n}^{2}z^{2}Z^{*}_{k,n}( c_{n}^{2}z^{2},\mu)}{\left(1+4c_{n}^{2}z^{2}/\zeta_{n}^{2}\right)^{k-1}},
\end{align}
\end{subequations}
the coefficients of the functions $Q^{*}_{k,n}$, $P^{*}_{k,n}$, $U^{*}_{k,n}$, $V^{*}_{k,n}$, and $Z^{*}_{k,n}$, which are polynomials of the same degree as their unstarred counterparts, are themselves polynomials in the variable $\mu$, of degree $4(k - 1)$, $4(k - 1)$, $4(k - 1) + 2n$, $4(k - 1) + 2n$, and $4(k - 1)$, respectively. The first few independent polynomials in this case read:
\begin{subequations}
\begin{align*}
&Q^{*}_{2,1}(x,\mu) = \frac{1}{6}\left(1 - 2\mu + \mu^2 + 2\mu^3 - \mu^4\right) - \frac{1}{12}\left(4 - 8\mu + 7\mu^2 + 2\mu^3 - \mu^4\right)x^2,\\
&Q^{*}_{2,2}(x,\mu)=-\frac{8}{9} \mu ^2(\mu -1)^2 x^2, \quad\quad\quad Q^{*}_{2,3}(x,\mu)=-\frac{3}{16} \mu ^2(\mu -1)^2  x^2,\\
&U^{*}_{1,1}(x,\mu)=\frac{1}{2} \left(1-\mu-\mu ^2\right), \quad\quad\quad U^{*}_{1,2}(x,\mu)=-\frac{1}{4} \left(4 - 2\mu - 13\mu^2 + 6\mu^3 + \mu^4\right),\\
&\begin{aligned}
U^{*}_{2,1}(x,\mu) &=\; \frac{1}{6} \left(1 - 3\mu + 3\mu^2 - \mu^3 - 3\mu^4 + 3\mu^5 - \mu^6\right) \\
&+ \frac{1}{12} \left(4 - 12\mu + 15\mu^2 - 10\mu^3 - 15\mu^4 + 24\mu^5 - 10\mu^6\right) x^2 \\
&+ \frac{1}{6} \left(4 - 12\mu + 9\mu^2 + 23\mu^3 - 24\mu^4 - 9\mu^5 + 5\mu^6\right) x^4,
\end{aligned}\\
&Z^{*}_{2,1}(x,\mu) = -\frac{1}{2} \left(4 - 8\mu + 5\mu^2 + 6\mu^3 - 3\mu^4\right),\quad \quad\quad Z^{*}_{2,2}(x,\mu) = \frac{4}{9}\mu ^2 (\mu -1)^2.
\end{align*}
\end{subequations}

\section{\texorpdfstring{Resummation via $\mathbb{C}^{2}/\mathbb{Z}_2$ blow-up equations}{C2/Z2 blow-up equations and resummation}}
\label{sec: blow-up C2Z2}

In this section, we carry out the same computational steps as in the previous section, now in the context of the first-order $\mathbb{C}^2/\mathbb{Z}_2$ blow-up equations \cite{Bonelli:2011jx,Bonelli:2011kv}. This analysis serves both as a consistency check of the previous results and as a means to better understand the interplay between the blow-up equations and the resummation ansatz, with a view toward possible generalizations. On the one hand, the $\mathbb{C}^2$ blow-up equations yield purely algebraic relations, but require more summands due to the sum being restricted to half-integer shifts. On the other hand, the $\mathbb{C}^2/\mathbb{Z}_2$ case leads to differential equations while benefiting from the inclusion of both integer and half-integer shifts. Moreover, the solvability of the ODEs arising from the $\mathbb{C}^2/\mathbb{Z}_2$ blow-up equations can be traced back \cite{Shchechkin:2020ryb} to the simple algebraic relations among the resummation functions, which themselves follow from the $\mathbb{C}^2$ blow-up equations. We continue to follow the conventions of \cite{Bershtein:2021uts, Bonelli:2025owb}, to which we refer the reader for further details, including the discussion of higher-order blow-up equations.
\subsection{\texorpdfstring{NS limit: $N_f\leq4$ and $\mathcal{N}=2^{*}$ theories}{C2/Z2 blow-up equations}}
The first order $\mathbb{C}^{2}/\mathbb{Z}_{2}$ blow-up equations for the partition functions of $SU(2)$ gauge theories with $N_f\leq 4$ fundamental hypermultiplets read: 
\begin{equation}
\label{eq: blow-up 1 Nf}
\sum_{2n\in \mathbb{Z}}D^{1}_{2\epsilon_{1},\,2\epsilon_{2}}\left(\mathcal{Z}^{[N_{f}]}(a+2n\epsilon_{1},\{\mu_{i}\};2\epsilon_{1},\epsilon_{2}-\epsilon_{1}; t),\mathcal{Z}^{[N_{f}]}(a+2n\epsilon_{2},\{\mu_{i}\};2\epsilon_{2},\epsilon_{1}-\epsilon_{2}; t)\right)=0,
\end{equation}
while for the $\mathcal{N}=2^{*}$ $SU(2)$ gauge theory, it takes the form:
\begin{equation}
\label{eq: blow-up1}
\begin{aligned}
& \sum_{2n\in \mathbb{Z}}D^{1}_{2\epsilon_1,\,2\epsilon_2}\big(\mathcal{Z}(a+2n\epsilon_{1},m;2\epsilon_1,\epsilon_2-\epsilon_1;\mathfrak{q}),\mathcal{Z}(a+2n\epsilon_{2},m;\epsilon_1-\epsilon_2,2\epsilon_2;\mathfrak{q})\big)\\
&=(\epsilon_{1}+\epsilon_{2})\gamma_{0}(\mathfrak{q})\sum_{2n\in\mathbb{Z}}\mathcal{Z}(a+2n\epsilon_{1},m;2\epsilon_1,\epsilon_2-\epsilon_1;\mathfrak{q})\mathcal{Z}(a+2n\epsilon_{2},m;\epsilon_1-\epsilon_2,2\epsilon_2;\mathfrak{q}).
\end{aligned}
\end{equation}
The $k$-th asymmetric logarithmic Hirota derivative $D_{\epsilon_{1},\epsilon_{2}}^{k}$ is defined by:
\begin{equation}
   f\left(\mathfrak{q} e^{\hbar\epsilon_{1}}\right)g\left(\mathfrak{q} e^{\hbar\epsilon_{2}}\right)=\sum \frac{\hbar^{k}}{k!}D^{k}_{\epsilon_{1},\epsilon_{2}}\left(f,g\right),
\end{equation}
while $\gamma_{0}(\mathfrak{q}) = 2\mathfrak{q}\partial_{\mathfrak{q}}\log\left[\theta_{3}(0|\tau)/\varphi(\mathfrak{q})\right]$. As expected, in the pure gauge theory limit $\mathfrak{q} \to 0$, $m \to \infty$, with $t = \mathfrak{q} m^4$ held fixed, equation~(\ref{eq: blow-up1}) reduces to equation~\eqref{eq: blow-up 1 Nf} for $N_f = 0$. Taking the $\epsilon_1 - \epsilon_2 \to 0$ limit of the previous equations and performing computations analogous to those in the preceding section, we obtain:
\begin{equation}
\label{eq: blow-up 1 pure ns}
    \sum_{2n \in \mathbb{Z}}e^{2\mathcal{W}_{1}^{[N_f]}-2(n\partial_{a}+\partial_{\hbar})\mathcal{W}_{0}^{[N_f]}}\partial_{\ln t}\left[\mathcal{W}_{1}^{[N_f]}-\left(n\partial_{a}+\partial_{\hbar}+\frac{1}{\hbar}\right)\mathcal{W}_{0}^{[N_f]}\right]=0,
\end{equation}
where $2\epsilon_1 =2\epsilon_2=\hbar$ and the first argument of each function is shifted by $n\hbar$, after taking the $\hbar$-derivative: $\mathcal{W}_{j}^{[N_f]}=\mathcal{W}_{j}^{[N_f]}(a+n\hbar,\{\mu_i\},\hbar;t)$, $j=0,1$. Similarly, for the $\mathcal{N}=2^{*}$ gauge theory, the NS limit of the first order blow-up equation (\ref{eq: blow-up1}) reads:
\begin{equation}
\label{eq: blow-up 1 NS lame 1}
    \sum_{2n \in \mathbb{Z}}e^{2\mathcal{W}_{1}-2(n\partial_{a}+\partial_{\hbar})\mathcal{W}_{0}}\left[\partial_{\ln \mathfrak{q}}\mathcal{W}_{1}-\left(n\partial_{a}+\partial_{\hbar}+\frac{1}{\hbar}\right)\partial_{\ln \mathfrak{q}}\mathcal{W}_{0}-\frac{1}{2}\gamma_{0}(\mathfrak{q})\right]=0,
\end{equation}
where $\mathcal{W}_{j}=\mathcal{W}_{j}(a+n\hbar,m,\hbar;\mathfrak{q})$.

\subsection{\texorpdfstring{Solving the blow-up equations: $N_f\leq 4$ theories}{Solving the blow-up equations 1}}

We begin by considering equation (\ref{eq: blow-up 1 pure ns}) and proceed analogously to the case of the $\mathbb{C}^2$ blow-up equations. Specifically, we explicitly write out the classical and 1-loop parts of the prepotentials $\mathcal{W}_i$, normalize by the 1-loop contribution at $n=0$, divide by the common factor $t^{-2\frac{a^2}{\hbar^2}+\frac{1}{2}}$ and perform the substitution $n \to n/2$. With this normalization and setting $\hbar=1$, equation (\ref{eq: blow-up 1 pure ns}) becomes:
\begin{equation}
\label{eq: blow-up 1 explicit 2}
    \sum_{n\in \mathbb{Z}}t^{n^{2}/2}\widetilde{\mathcal{G}}^{\,[N_f]}_{1\text{-}loop}(a,\{\mu_{i}\},n)\widetilde{\mathcal{G}}^{\,[N_f]}_{exp}(a,\{\mu_{i}\},n;t)\widetilde{\mathcal{G}}^{\,[N_f]}_{inst}(a,\{\mu_{i}\},n;t)=0,
\end{equation}
where 
\begin{subequations}
\begin{align}
&\widetilde{\mathcal{G}}^{\,[N_f]}_{exp}(a,\{\mu_{i}\},n;t)=e^{ 2\mathcal{W}^{[N_f]}_{1,\,inst}- \left(2 \partial_{\hbar}+n \partial_{a}\right) \mathcal{W}^{[N_f]}_{0,\, inst}  }\label{eq: Gexp pure},\\
&\widetilde{\mathcal{G}}^{\,[N_f]}_{inst}(a,\{\mu_{i}\},n;t)=n(a + \frac{n}{2} ) + t\partial_{t} \left[\mathcal{W}^{[N_f]}_{1,\, inst} - \left( \partial_{\hbar}+\frac{n}{2} \partial_{a}+1\right) \mathcal{W}^{[N_f]}_{0,\, inst} \right],\label{eq: Ginst pure}
\end{align}
\end{subequations}
and $\mathcal{W}^{[N_f]}_{j,\,inst}=\mathcal{W}^{[N_f]}_{j,\, inst}(a+n\hbar,\{\mu_{i}\},\hbar;t)$. The 1-loop term $\widetilde{\mathcal{G}}^{\,[N_f]}_{1\text{-}loop}(a,\{\mu_{i}\},n)$ can be computed as the $\epsilon_{1}=\epsilon_{2}=1/2$ limit of the $l$-factors $l_{n}^{[N_{f}]}$ in Eq. (2.15) of \cite{Bonelli:2025owb}. Namely, defining the functions:
\begin{subequations}
\begin{align}
\tilde{g}_{1\text{-}loop}^{\,hyp}(a, \mu, n) =& 
\begin{cases}
        \text{$\prod_{j=\frac{1}{2}}^{\frac{n-1}{2}}(\mu-j-a)^{2j}(\mu+j+a)^{2j}$}, & \text{if } n = 2p, \\
        \text{$\prod_{j=1}^{\frac{n-1}{2}}(\mu-j-a)^{2j}(\mu+j+a)^{2j}$}, & \text{if } n = 2p+1, \\
        \text{1}, & \text{if } n =0,1,\\
\end{cases}\\
\tilde{g}_{1\text{-}loop}^{\,vec}(a, n)=& 
\begin{cases}
        (-1)^{n}2a(n+2a)^{2n-1}\prod_{j=1}^{n-1}(j+2a)^{4j}, & \text{if } n \geq 1, \\
        1, & \text{if } n = 0,
\end{cases}\\
\end{align}
\end{subequations}
for $p \in \mathbb{N}_{\geq 1}$, the 1-loop factor can be written as:
\begin{equation}
\label{eq: G1L general}
\widetilde{\mathcal{G}}^{\,[N_f]}_{1\text{-}loop}(a,\{\mu_{i}\},n)=
\begin{cases}
\frac{\prod_{i=1}^{N_{f}}\tilde{g}_{1\text{-loop}}^{hyp}(a, \mu_{i}, n)}{\tilde{g}_{1\text{-}loop}^{\,vec}(a, n)}, & \text{if } n \geq 0,\\[4pt]
 \frac{\prod_{i=1}^{N_{f}}\tilde{g}_{1\text{-loop}}^{hyp}(-a, \mu_{i}, -n)}{\tilde{g}_{1\text{-}loop}^{\,vec}(-a, -n)}, & \text{if } n < 0.
\end{cases}
\end{equation}
Once again, we consider the ansatz \eqref{eq: resumm ansatz F0}, \eqref{eq: resumm ansatz F1}, and expand the blow-up equations in the parameter $t$, evaluated at the points \eqref{eq: a pole}. We restrict to $m_1 \geq 0$ without loss of generality and focus on the simpler $N_f = 0$ case. We begin by deriving the differential equations satisfied by the resummation functions, and then show that these equations can be solved in closed form due to the specific form of their coefficients. Finally, we analyze the structure of the blow-up equation’s expansion, taking into account the asymptotic behavior of the resummation functions near infinity. On the way, we will comment on the generalization to $N_f>0$, pointing to the relevant formulas collected in Appendix \ref{app: blow-up formulas}. In particular, the form of the differential equations we will study remains exactly the same for every choice of $N_f\leq 4$, due to the fact that the dependence on the masses $\{\mu_i\}_{i=1}^{N_f}$ of the resummation functions is completely fixed by the boundary conditions.
\subsubsection{ODEs from blow-ups}
First of all, let us consider the rational functions $\widetilde{\mathcal{G}}_{1\text{-}loop}(a,n)=\widetilde{\mathcal{G}}^{\,[0]}_{1\text{-}loop}(a,n)$, which we rewrite for convienience:
\begin{equation}
\label{eq: G1L pure}
\widetilde{\mathcal{G}}_{1\text{-}loop}(a,n)=
\begin{cases}
\frac{(-1)^{n}}{2a(n+2a)^{2n-1}}\frac{1}{\prod_{j=1}^{n-1}(j+2a)^{4j}}, & \text{if } n \geq 0,\\[4pt]
\frac{(-1)^{n}}{2a(n+2a)^{-2n-1}}\frac{1}{\prod_{j=1}^{-n-1}(j-2a)^{4j}},     & \text{if } n < 0.
\end{cases}
\end{equation}
Similarly to the $\mathbb{C}^2$ case, setting $2a = m_{1} + t^{m_{2}/2}/x$, expanding in $t$, and combining with the overall factor of $t^{n^{2}/2}$ results in a predictable leading-order behavior. This clearly applies as well to the $N_f>0$ case (\ref{eq: G1L general}). Let us consider, then, the instanton contributions (\ref{eq: Gexp pure}) and (\ref{eq: Ginst pure}). Isolating the relevant terms for the expansion of the equation, similarly to equation \eqref{eq: C2 Gexp gen}, we find:
\begin{equation}
\label{eq: gexp general}
\begin{aligned}
&\log \widetilde{\mathcal{G}}_{exp}^{\,\mp}=\log \widetilde{\mathcal{G}}_{exp}\left(\frac{m_1 x+t^{m_{2}/2}}{2x},\mp j-m_{1};t\right)=2\sum_{k\geq 1}t^{k-1}\bigg[\tilde{f}_{k,j}(y)\mp m_{1}y^{2} g_{k,j}'(y)\\
&\pm t^{j/2}\bigg(\tilde{g}_{k,j}(y)+(2j+4k-5)g_{k,j}(y) +  (2j-1)y\, g_{k,j}'(y)\bigg)\bigg] + \sigma_{exp}^{\mp}(x,t),
\end{aligned}
\end{equation}
\begin{equation}
\label{eq: ginst general}
\begin{aligned}
& \widetilde{\mathcal{G}}_{inst}^{\,\mp}=\sigma_{inst}^{\mp}(x,t)+\frac{j}{2}(j\pm m_{1})+\sum_{k\geq 1}t^{k-1}\Bigg\{ \Bigg[(k-1)\tilde{f}_{k,j}(y)+\frac{j}{2}y  \tilde{f}_{k,j}'(y)\\
&\mp m_{1}(k+j-1)y^{2} g_{k,j}'(y)\mp j\frac{m_{1}}{2}y^{3}g_{k,j}''\Bigg]\pm t^{j/2}\Bigg[(k+j/2-1)\tilde{g}_{k,j}+\frac{j}{2}y \tilde{g}_{k,j}'\\
&+2(k+j/2-1)(2k+j-3)g_{k,j}+(1-k+j (3j+4k-6))y\, g_{k,j}'+\frac{j}{2}(2j-1)y^{2}g_{k,j}''\Bigg] \Bigg\}
\end{aligned}
\end{equation}
where $y=x t^{(j-m_{2})/2}$, a prime denotes a derivative with respect to $y$, and the functions $\sigma_{exp/inst}^{\mp}(x, t)$ vanish at $t=0$. By collecting terms of the same order in $t$ from the expansion of the blow-up equation, these expressions allow us to derive the form of the differential equations and demonstrate that the $t$-expansion around the poles (\ref{eq: a pole}) is well-behaved. Namely, by setting $j=m_2$, assuming $m_{1}>0$ for simplicity\footnote{The equations for $m_{1}=0$ involve all three types of resummation functions simultaneously and are automatically satisfied once a solution for the $m_1>0$ case is found.}, and performing the change of variables\footnote{In particular, the functions $\tilde{v}_{m_2}$ are related to the functions $v_{m_2}$ introduced in the previous section simply by $v_{m_2}^2 = \tilde{v}_{m_2}$, up to a branching ambiguity.}: 
\begin{equation}
\label{eq: C2Z2 log branch}
\tilde{f}_{1,m_2}=\frac{1}{2}\log u_{m_2}, \quad\quad\quad
g_{1,m_2}'=\frac{1}{2x^{2}}\log(x^{2}\tilde{v}_{m_2}) ,
\end{equation}
we obtain at leading order:
\begin{equation}
\label{eq: 1st ode gen}
u_{m_2}'+a_{m_{1},m_{2},1}u_{m_2}+b_{m_{1},m_{2},1}=0,
\end{equation}
with coefficients given by:
\begin{equation*}
\begin{aligned}
a_{m_{1},m_{2},1}&=-\partial_{x}\log \left[x^{2m_2}\left( \tilde{v}_{m_2}^{m_{1}}+d_{m_{1},m_{2}}\tilde{v}_{m_2}^{-m_{1}}\right)\right],\\
b_{m_{1},m_{2},1}&=\frac{4}{m_{2} x^{2m_1 +1}}\frac{\rho_{m_{1},m_{2},1}}{\tilde{v}_{m_2}^{m_{1}}+d_{m_{1},m_{2}}\tilde{v}_{m_2}^{-m_{1}}}.    
\end{aligned}
\end{equation*}
The constants $d_{m_{1},m_{2}}$ can be computed from the 1-loop term (\ref{eq: G1L pure}) and read:
\begin{equation}
\label{eq: dm1m2 pure}
    d_{m_{1},m_{2}}=-m_{2}^{-4m_{1}}[(m_{2}-1)!]^{-8m_{1}}.
\end{equation}
We note here that, with this method, the change of variables \eqref{eq: C2Z2 log branch} is necessary to obtain a linear differential equation \eqref{eq: 1st ode gen} for the function $u_{m_2}$, and it introduces a logarithmic branching. Meanwhile, at orders $t^{k-1}$ and $t^{k+j/2-1}$, respectively, compared to the previous equations, we find:
\begin{equation}
\label{eq: odes fk}
\tilde{f}_{k,m_2}'+a_{m_{1},m_{2},k}\tilde{f}_{k,m_2}+b_{m_{1},m_{2},k}=0,
\end{equation}
\begin{equation}
\label{eq: odes gktilde}
\tilde{g}_{k,m_2}'+\tilde{a}_{m_{1},m_{2},k}\tilde{g}_{k,m_2}+\tilde{b}_{m_{1},m_{2},k}=0,
\end{equation}
with coefficients given by:
\begin{equation*}
\begin{aligned}
&a_{m_{1},m_{2},k}= \partial_{x}\left\{\log \left[u_{m_2}\left(\tilde{v}_{m_2}^{m_{1}}+d_{m_{1},m_{2}}\tilde{v}_{m_2}^{-m_{1}}\right)\right]+2\left(m_{2}+\frac{k-1}{m_2}\right)\log x\right\},\\
&b_{m_{1},m_{2},k}=\bigg\{\partial_{x}\left[m_{1}u_{m_2}(d_{m_{1},m_{2}}\tilde{v}_{m_2}^{-m_{1}}-\tilde{v}_{m_2}^{m_{1}}) \,x^{2\left(m_{2}+1+\frac{k-1}{m_2}\right)} \,g_{k,m_2}'\right]\\
&\quad\quad\quad\,\,\,+2\frac{u_{m_2}}{m_{2}}x^{2\left(m_{2}-m_{1}+\frac{k-1}{m_2}\right)-1}\,\rho_{m_{1},m_{2},k}   \bigg\}e^{-\int a_{m_{1},m_{2},k}\,dx},\\
&\tilde{a}_{m_{1},m_{2},k}= \partial_{x}\left\{\log \left[u_{m_2}\left(\tilde{v}_{m_2}^{m_{1}}+d_{m_{1},m_{2}}\tilde{v}_{m_2}^{-m_{1}}\right)\right]+2\left(m_{2}+\frac{1}{2}+\frac{k-1}{m_2}\right)\log x\right\},\\
&\tilde{b}_{m_{1},m_{2},k}=\frac{2}{m_{2}x^{2m_1 +1}}\frac{\tilde{\rho}_{m_{1},m_{2},k}}{u_{m_2}\left(\tilde{v}_{m_2}^{m_{1}}+d_{m_{1},m_{2}}\tilde{v}_{m_2}^{-m_{1}}\right)}.
\end{aligned}
\end{equation*}
The functions $\rho_{m_{1},m_{2},k}(x)$ and $\tilde{\rho}_{m_{1},m_{2},k}(x)$, which introduce inhomogeneous terms in the differential equations, account for the remaining contributions of the $t$-expansion at each fixed order and can be explicitly computed for any choice of $m_1$ and $m_2$, similarly to the functions $r_{m_1,m_2,k}$ and $\tilde{r}_{m_1,m_2,k}$ in the previous section. The second and third sets of equations have been put in normal form for $\tilde{f}_{k,m_2}$ and $\tilde{g}_{k,m_2}$ respectively. The specific form chosen for the coefficients anticipates the method of solution, which will become evident in the following discussion. The former set can also be written in normal form with respect to $g'_{k,m_2}$, obtaining:
\begin{equation}
\label{eq: odes gk}
g_{k,m_2}''+\alpha_{m_1,m_2,k}\,g_{k,m_2}'+\beta_{m_1,m_2,k}=0,
\end{equation}
where
\begin{equation*}
\begin{aligned}
&\alpha_{m_{1},m_{2},k}= \partial_{x}\left[\log\left[ u_{m_2} \left(\tilde{v}_{m_2}^{m_1}-d_{m_1,m_2} \tilde{v}_{m_2}^{-m_1}\right)\right]+2 \left(m_2+1+\frac{k-1}{m_2}\right) \log x\right],\\
&\beta_{m_{1},m_{2},k}=\frac{1}{m_1}e^{-\int \alpha_{m_{1},m_{2},k}\,dx}\bigg\{\partial_{x}\left[ x^{2 \left(m_2+\frac{k-1}{m_2}\right)} u_{m_2} \left( \tilde{v}_{m_2}^{m_{1}}+d_{m_{1},m_{2}}\tilde{v}_{m_2}^{-m_{1}}\right) \tilde{f}_{k,m_2}\right]\\
&\quad\quad\quad\,\,\,+2\frac{u_{m_2} }{m_2}x^{2 \left(m_2-m_1+\frac{k-1}{m_2}\right)-1} \rho_{m_1,m_2,k}   \bigg\}.\\
\end{aligned}
\end{equation*}
As previously anticipated, the structure of these equations remains unchanged even for $N_f>0$, as can be verified using formulas (\ref{eq: Gexp Nf}) and (\ref{eq: Ginst Nf}). The mass dependence is implicitly introduced through appropriate replacements, such as $d_{m_{1},m_{2}}\to d_{m_{1},m_{2}}(\{\mu_i\})$, $\rho_{m_{1},m_{2},k}(x)\to \rho_{m_{1},m_{2},k}(x,\{\mu_i\})$, $\tilde{\rho}_{m_{1},m_{2},k}(x)\to \tilde{\rho}_{m_{1},m_{2},k}(x,\{\mu_i\})$ and so forth.
\subsubsection{Solvability of the ODEs}
We now examine in more detail the general structure of the equations derived in the previous subsection, showing that, upon providing appropriate boundary conditions, they admit a closed form solution in terms of elementary functions. In practice, they can be solved in the order in which they arise from the expansion. Let us begin by analyzing equations (\ref{eq: 1st ode gen}). These equations are straightforwardly solvable via the method of variation of coefficients, and their general solution is given by:
\begin{equation}
    u_{m_2}=\frac{\kappa_{m_{1},m_{2},1}x^{-2m_{2}}}{\tilde{v}_{m_2}^{m_{1}}+c_{m_{1},m_{2}}\tilde{v}_{m_2}^{-m_{1}}}
\end{equation}
where
\begin{equation}
    \kappa_{m_{1},m_{2},1}(x)= \kappa_{m_{1},m_{2},1}^{(0)}-\frac{4}{m_{2}}\int x^{2(m_{2}-m_{1})-1}\rho_{m_{1},m_{2},1}(x)dx ,
\end{equation}
and $\kappa_{m_{1},m_{2},1}^{(0)}$ is an integration constant. In the case $m_2=1$ and $m_1=1$, this integration constant can be fixed by requiring $u_{1}(0)=1$ and $\tilde{v}_{1}(x)\simeq 1/x^2$, i.e. $\tilde{f}_{1}(0)=0$ and $g'_1(0)=\text{const}$, while already for $m_2=1$ and $m_1=2$ this is not the case. However, in Appendix \ref{app: blow-up example 2}, we show that this second integration constant is in fact fixed by the blow-up equation itself. To determine $\tilde{v}_{m_2}$ we can simply equate two $u_{m_2}$'s for different values of $m_{1}$, obtaining:
\begin{equation*}
    \frac{\kappa_{m_{1},m_{2},1}}{\tilde{v}_{m_2}^{m_{1}}+c_{m_{1},m_{2}}\tilde{v}_{m_2}^{-m_{1}}}=\frac{\kappa_{\tilde{m}_{1},m_{2},1}}{\tilde{v}_{m_2}^{\tilde{m}_{1}}+c_{\tilde{m}_{1},m_{2}}\tilde{v}_{m_2}^{-\tilde{m}_{1}}}.
\end{equation*}
Due to the form of $d_{m_{1},m_{2}}$ in (\ref{eq: dm1m2 pure}), setting $m_1=1$ and $\tilde{m}_1=2$, the function $\tilde{v}_{m_2}$ solves a simple quadratic equation:
\begin{equation}
\label{eq: C2Z2 root branch}
    \tilde{v}_{m_2}^{2}-\frac{\kappa_{2,m_{2},1}}{\kappa_{1,m_{2},1}}\tilde{v}_{m_2}+\frac{1}{m_{2}^{4}[(m_{2}-1)!]^{8}}=0.
\end{equation}
We note that, once again, this is the origin of the square roots that appear in the resummation functions. The same holds also for $N_f > 0$, as can be verified from equation (\ref{eq: constant c Nf}). Of the two solutions of the quadratic equation, we keep the only one compatible with $\tilde{v}_{m_2}(x)\simeq 1/x^{2}$, which is then determined uniquely once the remaining integration constant has been fixed. Moving on to equations (\ref{eq: odes fk}), we observe that the coefficients $a_{m_1, m_2, k}$ are still total derivatives. As a result, the equations can be solved in closed form, with solutions given by:
\begin{equation}
\tilde{f}_{k,m_2}=\frac{\kappa_{m_1,m_2,k}}{\kappa_{m_1,m_2,1}}x^{-2\frac{k-1}{m_2}},
\end{equation}
where:
\begin{equation}
    \kappa_{m_1,m_2,k}(x)=\kappa_{m_1,m_2,k}^{(0)}-\int e^{\,\int a_{m_1,m_2,k}\, dx}\,b_{m_1,m_2,k}\,dx,
\end{equation}
and $\kappa_{m_1,m_2,k}^{(0)}$ is an integration constant, which can be fixed by requiring the vanishing of resummation functions at zero. Once $\tilde{f}_{k,m_2}$ is expressed in terms of $g_{k,m_2}'$, we can substitute this expression into (\ref{eq: odes gk}) for a different value of $m_1$, thereby obtaining a linear differential equation for $g_{k}'$. This equation can again be solved by performing simple integrals, and the two integration constants for $g_{k}$ are then fixed by requiring the function to vanish at zero. Finally, also equations (\ref{eq: odes gktilde}) can be solved in closed form, with their solution reading: 
\begin{equation}
    \tilde{g}_{k,m_2}=\frac{\tilde{\kappa}_{m_1,m_2,k}}{\kappa_{m_1,m_2,1}}x^{-2\frac{k-1}{m_2}-1},
\end{equation}
where:
\begin{equation}
   \tilde{\kappa}_{m_1,m_2,k}(x)=\tilde{\kappa}_{m_1,m_2,k}^{(0)}-\frac{2}{m_2}\int x^{2\left(m_2-m_1+\frac{k-1}{m_2} \right)}\,\tilde{\rho}_{m_1,m_2,k}(x)\,dx,
\end{equation}
and $\tilde{\kappa}_{m_1,m_2,k}^{(0)}$ is again an integration constant, which can be determined by requiring the functions $\tilde{g}_{k,m_2}$ to vanish at zero. Thus, we have seen that expanding the blow-up equation \eqref{eq: blow-up 1 explicit 2} at two distinct values of $m_1$ is also sufficient to determine all the resummation functions. One can then verify that, even in this case, the expansion of the blow-up equation is well-defined, similarly to the discussion in Section~\ref{sec: structure expansion}. Again, similarly to the $\mathbb{C}^2$ case, the Taylor coefficients of the resummation functions for $j > m_2$ introduce certain subtleties, which we discuss in Appendix~\ref{app: blow-up example 2} through the example of the solution of the first few equations.

\subsection{\texorpdfstring{Solving the blow-up equations: $\mathcal{N}=2^{*}$ theory}{Solving the blow-up equations 2}}

In the $\mathcal{N}=2^{*}$ case, denoting again by $\mu$ the mass of the adjoint hypermultiplet, we can write the first-order blow-up equation (\ref{eq: blow-up 1 NS lame 1}) as:
\begin{equation}
\label{eq: blow-up lame expl}
    \sum_{n\in \mathbb{Z}}\mathfrak{q}^{n^{2}/2}\widetilde{\mathcal{G}}_{1-loop}(a,\mu,n)\widetilde{\mathcal{G}}_{exp}(a,\mu,n;\mathfrak{q})\widetilde{\mathcal{G}}_{inst}(a,\mu,n;\mathfrak{q})=0,
\end{equation}
where explicitly the three terms read:
\begin{subequations}
\begin{align}
&\widetilde{\mathcal{G}}_{1\text{-}loop}(a,\mu,n)=
\begin{cases}
\frac{(2a+\mu)(n-\mu+2a)^{2n-1}}{2a(n+2a)^{2n-1}}\frac{\prod_{j=1}^{n-1}(j+\mu+2a)^{2j+1}(j-\mu+2a)^{2j-1}}{\prod_{j=1}^{n-1}(j+2a)^{4j}}, & \text{if } n \geq 0,\\[4pt]
\frac{(2a-\mu)(n+\mu+2a)^{-2n-1}}{2a(n+2a)^{-2n-1}}\frac{\prod_{j=1}^{-n-1}(j+\mu-2a)^{2j+1}(j-\mu-2a)^{2j-1}}{\prod_{j=1}^{-n-1}(j-2a)^{4j}}, & \text{if } n < 0,
\end{cases}\label{eq: G1L C2Z2 *}\\[4pt]
&\widetilde{\mathcal{G}}_{exp}(a,\mu,n;\mathfrak{q})=e^{ 2\mathcal{W}_{1,\, inst}- \left(2 \partial_{\hbar}+n \partial_{a}\right) \mathcal{W}_{0,\, inst}  },\\
&\widetilde{\mathcal{G}}_{inst}(a,\mu,n;\mathfrak{q})=n(a + \frac{n}{2} ) -\frac{1}{2}\gamma_{0}+ \mathfrak{q}\partial_{\mathfrak{q}} \left[\mathcal{W}_{1,\, inst} - \left( \partial_{\hbar}+\frac{n}{2} \partial_{a}+1\right) \mathcal{W}_{0,\, inst} \right],
\end{align}
\end{subequations}
with $\mathcal{W}_{i,\, inst}=\mathcal{W}_{i,\,inst}\left(a + \frac{n}{2},\mu,1; \mathfrak{q}\right)$. To solve equation (\ref{eq: blow-up lame expl}), we consider the ansatz \eqref{ansatzeN2s0}, \eqref{ansatzeN2s1} and expand in $\mathfrak{q}$ the blow-up equation (\ref{eq: blow-up lame expl}), setting the Coulomb branch parameter $a$ to be as in equation \eqref{eq: a pole q}. Then, we can compute the analogues of (\ref{eq: gexp general}) and (\ref{eq: ginst general}), which read:
\begin{equation}
\label{eq: Gexp lame}
    \log \widetilde{\mathcal{G}}_{exp}^{\,\mp}=2\sum_{k\geq 1}\mathfrak{q}^{k-1}\left[\tilde{f}_{k}\mp m_{1}y^{2}\partial_{y}g_{k}\mp \mathfrak{q}^{j/2}(g_{k}-\tilde{g}_{k}+y \partial_{y}g_{k}-\mu\partial_{\mu}g_{k}) \right]+\sigma_{exp}^{*\,\mp},
\end{equation}
\begin{equation}
\label{eq: Ginst lame}
\begin{aligned}
&\widetilde{\mathcal{G}}_{inst}^{\,\mp}=\frac{j}{2}(j\pm m_{1})+\sum_{k\geq 1}\mathfrak{q}^{k-1}\bigg[(k-1)\tilde{f}_{k}+\frac{j}{2}y\partial_{y}\tilde{f}_{k}\mp m_{1}(k+j-1)y^{2}\partial_{y}g_{k}\\
&\mp \frac{j}{2}m_{1}y^{3}\partial_{y}^{2}g_{k}\pm \mathfrak{q}^{j/2}\bigg((k+j/2-1)\tilde{g}_{k}+\frac{j}{2}y \partial_{y}\tilde{g}_{k}-2(k+j/2-1)g_{k}\\
&-(k+2j-1)y\partial_{y}g_{k}-\frac{j}{2}y^{2}\partial_{y}^2g_{k}+\mu(k+j/2-1)\partial_{\mu} g_{k}+\frac{j}{2}\mu y\partial_{\mu}\partial_{y}g_{k}\bigg) \bigg]+\sigma_{inst}^{*\,\mp},
\end{aligned}
\end{equation}
where we recall that $y=x \mathfrak{q}^{(j-m_{2})/2}$. These two expressions allow us to write the explicit form of the equations, up to the usual inhomogeneous part. Indeed, the equations we obtain have precisely the same form as those presented in the previous section, with the appropriate replacements to incorporate the implicit $\mu$-dependence. For this reason, we do not repeat the discussion of their solution. However, in Appendix~\ref{app: blow-up example 3}, we show that the solution of these differential equations closely follows the same steps as those for the analogous equations in the pure gauge theory case. One can, once again, compute explicitly the constant $d_{m_{1},m_{2}}(\mu)$ from \eqref{eq: G1L C2Z2 *}, obtaining:
\begin{equation}
    d_{m_{1},m_{2}}(\mu)=-\left(\frac{\mu(m_{2}-\mu)}{m_{2}[(m_{2}-1)!]^{2}}\prod_{j=1}^{m_{2}-1}(j-\mu)(j+\mu)\right)^{4m_1},
\end{equation}
and show that the resulting $\mathfrak{q}$-expansion of the blow-up equation is well behaved.
\section{The periodic Lamé spectral problem}
\label{sec: spectral problem}
In this section we study the real spectral problem ($\ref{eq: hill}$) on the rectangular torus, $\tau \in i\mathbb{R}$. For convenience, we rescale the periods of the torus by a factor of $\pi$, so that the real section is parametrized by $z=x+\pi\tau/2$, $x\in [0,\pi]$. Then, the Lamé equation reads:
\begin{equation}
\label{eq: lame std}
    \left(\partial_{x}^{2}+\text{\Large $\mathfrak{e}$} -\mu(\mu-1)\wp(x+\pi\tau/2;\pi,\pi\tau)\right)\psi(x,\nu,\mu;\mathfrak{q})=0,
\end{equation}
where the rescaled eigenvalue is given by:
\begin{equation}
\label{eq:ns energy}
   \text{\Large $\mathfrak{e}$}\left(\nu,\mu;\mathfrak{q}\right)=-2\mu(\mu-1)\frac{\eta_1}{\pi^{2}}-4\mathfrak{q}\partial_{\mathfrak{q}}f(\nu, \mu ; \mathfrak{q}).
\end{equation}
From the point of view of the periodic spectral problem, the branch points of $f(\nu,\mu;\mathfrak{q})$ at $\nu \in \mathbb{Z}$ correspond to the edges between gaps and bands in the spectrum, thus to eigenvalues associated to periodic and anti-periodic solutions of the equation. Indeed, the partial resummation of the semi-classical block allows one to obtain a finite answer for these eigenvalues, simply by taking first the limit $\nu \to n^{\pm}$, or equivalently $\nu \to -n^{\mp}$, $n \in \mathbb{N}$, and then expanding around $\mathfrak{q}=0$. In particular, from this perspective, it is clear that the fractional powers of $\mathfrak{q}$ appearing in the expansion of the eigenvalues arise from the square roots present in the resummation functions. Meanwhile, the distinction between eigenvalues corresponding to even and odd solutions stems from the direction in which the limit $\nu \to n$ is taken. As an example, the eigenvalues associated to $\nu^{2}=1,4$ are respectively\footnote{As can be seen from the explicit formulae \eqref{eq: eigenvalues lame} and \eqref{eq: eigenvalues mathieu}, in some cases the coefficients of the fractional powers can be obtained by just swapping the sign, however this is not a general feature.}:
\begin{subequations}
\label{eq: eigenvalues lame}
\begin{align}
&\text{\Large $\mathfrak{e}$}_{1,\pm}\left(\mu;\mathfrak{q}\right)=1-\frac{1}{3}\mu(\mu-1)\pm 4\mu(\mu-1)\mathfrak{q}^{1/2}-2\mu(\mu-1)(\mu^{2}-\mu-4)\mathfrak{q}+O(\mathfrak{q}^{3/2}),\\
&\text{\Large $\mathfrak{e}$}_{2,+}\left(\mu;\mathfrak{q}\right)=4-\frac{1}{3}\mu(\mu-1)+\frac{20}{3} \mu ^2(\mu -1)^2 \mathfrak{q}+\mathcal{O}(\mathfrak{q}^2),\\
&\text{\Large $\mathfrak{e}$}_{2,-}\left(\mu;\mathfrak{q}\right)=4-\frac{1}{3}\mu(\mu-1)-\frac{4}{3} \mu(\mu-1)(\mu -4)(\mu +3) \mathfrak{q}+\mathcal{O}(\mathfrak{q}^2),
\end{align}
\end{subequations}
\begin{figure}[t]
    \centering
\includegraphics[width=0.9\linewidth]{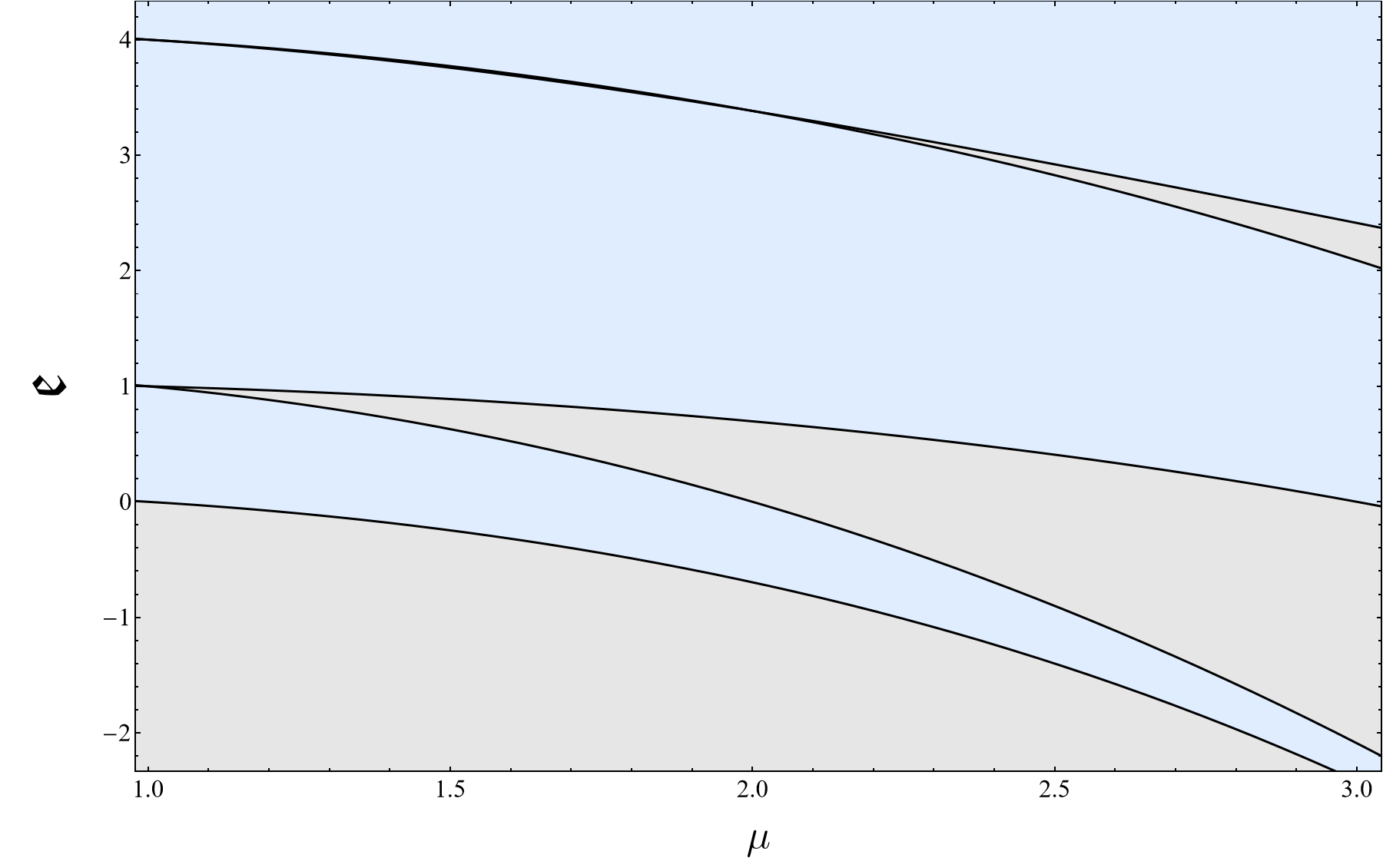}
    \caption{A portion of the stability chart of the $\tau= i$ Lamé equation (\ref{eq: lame std}) is reconstructed by plotting the band-gap edges \eqref{eq: eigenvalues lame} (black lines) as functions of the coupling $\mu$. Bands are shown as light blue regions while gaps as gray regions.}
    \label{fig: stability chart}
\end{figure}\noindent
where $\pm$ refers to the $\nu \rightarrow n^{\pm}$ limit. In the Mathieu limit $\mathfrak{q}=t/\mu^{4}$, $\mu \to \infty$, they reduce to the corresponding eigenvalues\footnote{As mentioned at the end of Section \ref{sec:connection}, in order to compare with the expansion presented in \cite{NIST:DLMF}, Section \href{https://dlmf.nist.gov/28.6.i}{28.6}, one must identify $t=\frac{1}{16}q^2$.}: 
\begin{subequations}
\label{eq: eigenvalues mathieu}
\begin{align}
&\lambda_{1,\,\pm}(t)=1\mp 4t^{1/2}-2t+\mathcal{O}(t^{3/2}),\\
&\lambda_{2,\,+}(t)=4+\frac{20 t}{3}-\frac{763 t^2}{54}+\mathcal{O}(t^{3}),\\
&\lambda_{2,\,-}(t)=4-\frac{4t}{3}+\frac{5 t^2}{54}+\mathcal{O}(t^{3}).
\end{align}
\end{subequations}
The resulting expansions can be checked against the ones obtained from standard methods, as shown in Appendix \ref{app: checks 1}, and allow one to construct the stability chart of the equations. It is interesting to observe that, by choosing an integer value of $\mu \geq 2$, all poles in $\nu$ vanish from the perturbative $\mathfrak{q}$-expansion of the eigenvalue, except for the first $\mu-1$. This occurs due to the finite-gap property of the Lamé potential: for these specific values of $\mu$, only the first $\mu-1$ gaps in the spectrum remain open, while all subsequent ones are closed. This property is even more evident in the resummed semi-classical block, where the functions $g_{k,j}$ as in (\ref{eq: results star}) vanish when $\mu\in\{1-j,\dots,j\}$, for any value of $k$. Moreover, the resummed semi-classical block enables the construction of an infinite set of periodic functions of the Floquet exponent, corresponding to the bands in the spectrum. These can be defined as:
\begin{equation}
\label{eq: bands k}
\text{\Large $\mathfrak{e}$}^{(n)}(\nu,\mu;\mathfrak{q})=
\begin{cases}
\text{\Large $\mathfrak{e}$}(n+\nu,\mu;\mathfrak{q}), & \text{for } n\geq0 \text{ even},\\[4pt]
\text{\Large $\mathfrak{e}$}(n+1-\nu,\mu;\mathfrak{q}),     & \text{for } n>0 \text{ odd},
\end{cases}
\end{equation}
\noindent
for $\nu\in [0,1]$, together with the required symmetry $\text{\Large $\mathfrak{e}$}^{(n)}(-\nu,\mu;\mathfrak{q})=\text{\Large $\mathfrak{e}$}^{(n)}(\nu,\mu;\mathfrak{q})$ and the identification of branch cuts via $\nu \sim -\nu$. Finally, we remark that studying the spectral problem (\ref{eq: lame std}) is equivalent to studying the Jacobi form of the Lamé equation:
\begin{equation}
\label{eq:lame jacobi}
    \left(\partial_{z}^{2}+h-\mu(\mu-1)\kappa^{2} \sn^{2}(z;\kappa^2)\right)\psi(z)=0,
\end{equation}
where $\kappa$ is the elliptic modulus of the elliptic sine function, see Appendix \ref{app: special functions}. Indeed, the eigenvalues of the two forms are related by:
\begin{equation}
h(\nu,\mu;\kappa^{2}) = \left(\frac{\pi}{2K(\kappa^{2})}\right)^{2}\text{\Large $\mathfrak{e}$}\left(\nu,\mu;\mathfrak{q}\right) -\mu(\mu-1)\frac{e_{3}}{e_{1}-e_{3}},
\end{equation}
where the rescaling of \text{\Large $\mathfrak{e}$} is due to the rescaling of the periods of the torus, which are $2K(\kappa^{2})$ and $2iK(1-\kappa^{2})$ for equation (\ref{eq:lame jacobi}). For instance, in the case of a square torus with $\tau= i$ and $\kappa^{2}=1/2$, we have the particularly simple relation:
\begin{equation}
h\left(\nu,\mu;\kappa^{2}\right) = \Gamma\left(\frac{3}{4}\right)^{4}\frac{\text{\Large $\mathfrak{e}$}\left(\nu,\mu;\mathfrak{q}\right)}{\pi} +\frac{\mu(\mu-1)}{2}.
\end{equation}

\subsection{(Anti-)Periodic eigenvalues from isomonodromic deformations}
An alternative approach to the study of the periodic spectral problem ($\ref{eq: lame std}$) comes from the theory of isomonodromic deformations of linear ODEs. In this subsection, we closely follow the discussion in \cite{Bershtein:2021uts}, to which the reader may refer for further details, and extend it to the periodic case. Consider the following Fuchsian linear system on the complex torus:
\begin{equation}
\label{eq: tfuchs}
    \partial_{z}Y(z,\tau)=A(z,\tau)Y(z,\tau),
\end{equation}
with the matrix $A(z,\tau)$ given by:
\begin{equation}
\label{eq: A(z)}
    A(z,\tau)= \begin{pmatrix}
        p & m x(2Q,z)\\
        m x(-2Q,z) & -p
    \end{pmatrix},
\end{equation}
where: 
\begin{equation*}
    x(u,z)=\frac{\theta_{1}(z-u|\tau)\theta_{1}^{'}(0|\tau)}{\theta_{1}(z|\tau)\theta_{1}(u|\tau)},
\end{equation*}
is a Lamé function, while $p$ and $Q$ are complex auxiliary parameters. The matrix (\ref{eq: A(z)}) has a simple pole at $z=0$ and we can consider the associated isomonodromic deformation problem, thus the combined system of (\ref{eq: tfuchs}) and
\begin{equation}
    2\pi i \partial_{\tau}Y(z,\tau)=-MY(z,\tau),\quad M(z,\tau)=m \begin{pmatrix}
        \wp(2Q) & \partial_{Q}x(2Q,z)\\
        \partial_{Q}x(-2Q,z) & \wp(2Q)
    \end{pmatrix}.
\end{equation}
Compatibility between the two linear systems then leads to:
\begin{subequations}
\begin{align}
 &p= 2\pi i \partial_{\tau}Q,\\
 &(2\pi i)^{2}\frac{d^{2}}{d\tau^{2}}Q=m^{2}\wp(2Q),\label{eq: eCM}
\end{align}
\end{subequations}
the latter corresponding to the equation of motion for the classical non-autonomous 2-particle elliptic Calogero-Moser (eCM) system and a specific elliptic form of the Painlevé VI equation. Being a second order ODE, (\ref{eq: eCM}) must be supplemented with two integration constants $(\sigma,\eta)$ which are chosen such that:
\begin{equation}
    Q\raisebox{-1ex}{\big|}_{m=0}=\tau \sigma + \frac{\eta}{4\pi}.
\end{equation}
These two parameters can be seen as coordinates on the space of monodromy data of the linear system (\ref{eq: tfuchs}). The non-autonomous eCM system is a classical integrable system, with Hamiltonian:
\begin{equation}
    \mathcal{H} = p^{2}-m^{2}\left(\wp(2Q)+2\eta_{1}(\tau)\right),
\end{equation}
and corresponding tau-function $\mathcal{T}$ defined by:
\begin{equation}
    \mathcal{H}=2\pi i \partial_{\tau}log \mathcal{T}.
\end{equation}
This tau-function admits an explicit representation in terms of dual Nekrasov partition functions (\ref{eq: zdual}), which in turn are expressed in terms of self-dual Nekrasov partition functions\footnote{In particular, $\mathcal{Z}_{inst}(\sigma,m;\mathfrak{q})=\mathcal{Z}_{inst}(\sigma,m;1,-1;\mathfrak{q})$, where the latter is defined in equation \eqref{eq: zinst 2*}, corresponds to a $c=1$ conformal block via the identification ($\ref{eq: block = inst}$).} \cite{Bonelli:2019boe,Bonelli:2019yjd}. Namely:
\begin{equation}
    \mathcal{T}(\sigma,m,\eta,\tau)=\eta(\tau)\frac{\mathcal{Z}_{1/2}^{D}(\sigma,m,\eta; \mathfrak{q})}{\theta_{2}(2Q|2\tau)}=\eta(\tau)\frac{\mathcal{Z}_{0}^{D}(\sigma,m,\eta;\mathfrak{q})}{\theta_{3}(2Q|2\tau)},
\end{equation}
where $Q=Q(\sigma,m,\eta,\tau)$ is a solution of (\ref{eq: eCM}) and 
\begin{equation}
\label{eq: zdual}
    \mathcal{Z}_{\epsilon}^{D}(\sigma,m,\eta;\mathfrak{q})=\sum_{n\in \mathbb{Z}+\epsilon}e^{in\eta}\frac{\prod_{\epsilon'=\pm}G(1-m+2\epsilon'(\sigma+n))}{\prod_{\epsilon'=\pm}G(1+2\epsilon'(\sigma+n))}\mathfrak{q}^{(\sigma+n)^2-\frac{1}{24}}\mathcal{Z}_{inst}(\sigma+n,m; \mathfrak{q}).
\end{equation}
The Lamé equation is obtained from (\ref{eq: tfuchs}) by turning the linear system into a second order equation for one of the components of $Y(z,\tau)$, provided the $singularity$ $matching$ $condition$:
\begin{equation}
\label{eq: sing match q}
    Q(\sigma,m,\eta,\tau)=0,
\end{equation}
is satisfied, which can be shown to be equivalent to:
\begin{equation}
\label{eq: sing match z}
 \theta_{2}(0|2\tau)\mathcal{Z}_{0}^{D}(\sigma,m,\eta;\mathfrak{q})-\theta_{3}(0|2\tau)\mathcal{Z}_{1/2}^{D}(\sigma,m,\eta;\mathfrak{q})=0.
\end{equation}
In particular, one finds:
\begin{equation}
    \left(\partial_{z}^{2}-m(m\pm 1)\wp(z)-E_{\pm}\right)\psi(z)=0,
\end{equation}
where the eigenvalue $E_{\pm}$, due to the singularity matching condition, is expressed in terms of the Painlevé tau-function $\mathcal{T}(\sigma,m,\eta,\tau)$ evaluated on a solution of (\ref{eq: sing match q}), and is given by:
\begin{equation}
\label{eq: E}
E_{\pm} = \left[2\pi i \partial_{\tau}\left(\log\,\mathcal{Z}_{0}^{D}(\sigma, m, \eta;\mathfrak{q})+\log \frac{\eta(\tau)}{\theta_{3}(0|2\tau)}\right)\pm 2m\frac{\theta_{3}''(0|2\tau)}{\theta_{3}(0|2\tau)}+2m(m\pm 1)\eta_{1}(\tau)\right]{\bigg|}_{Q=0}.
\end{equation}
The sign $\pm$ corresponds to the two possible branches of the solution to equation (\ref{eq: eCM}) near $Q=0$. In order to compute the eigenvalues associated to periodic and anti-periodic solutions to the spectral problem (\ref{eq: lame std}), we interpret the singularity matching condition (\ref{eq: sing match z}) as a quantization condition for the system, which we solve in two non-generic limits\footnote{From the gauge theory viewpoint, this should correspond to the alignment of the central charges of two BPS particles, defining a wall of marginal stability. It would be interesting to further investigate the problem from this perspective.
} of the monodromy data $(\sigma,\eta)$, corresponding to limits to logarithmic Painlevé tau-functions \cite{Oleg-Pasha-Misha,OlegTalk,OlegMovie}. In particular, setting $\mathfrak{t}=\mathfrak{q}^{1/2}=e^{i\pi\tau}$, in order to obtain \enquote{periodic} eigenvalues we will solve (\ref{eq: sing match z}) in the limit $\eta = \Omega \,\sigma$ and $\sigma \rightarrow 0$, as an equation for $\Omega = \Omega(m,\mathfrak{t})$, while to obtain \enquote{anti-periodic} eigenvalues we will solve it in the limit $\eta = \Omega(\sigma-1/2)$ and $\sigma \rightarrow 1/2$. Once a solution is found, this can be substituted in the corresponding limit of the eigenvalue (\ref{eq: E}) to obtain the desired result. In both periodic and anti-periodic cases, we find:
\begin{equation}
    \Omega_{|\nu|}(m,\mathfrak{t})= \omega_{|\nu|}(m,\mathfrak{t})+8i H(-1-m) +4i \log \mathfrak{t}, \quad \nu \in \mathbb{Z},
\end{equation}
where $H(x)$ is the harmonic number function, and $\omega_{|\nu|}(m,\mathfrak{t})=\mathfrak{t}^{-|\nu|}(\omega_{-|\nu|}(m)+\mathcal{O}(\mathfrak{t}))$ depends on the absolute value of the specific integer Floquet exponent considered. In particular, we obtain: 
\begin{equation}
    \Omega_{0}(m,\mathfrak{t}) = -4i \lim_{\nu\to 0}\frac{1}{\nu} \frac{\partial f}{\partial \nu}\left(\nu,\mu=m;\mathfrak{t}^2\right),
\end{equation}
where the semi-classical conformal block $f(\nu,\mu;\mathfrak{q})$ in this formula contains also a 1-loop part, as in equation (\ref{eq: eta old}). Thus, we find a connection between $c=1$ conformal blocks and semi-classical ones, or equivalently between the self-dual limit of Nekrasov functions and the NS one. The corresponding eigenvalues are also observed to match, i.e. we find:
\begin{equation*}
    E_{-}(\nu,m;\mathfrak{t}){\Big|}_{\nu \to 0} = \pi^{2}\text{\Large $\mathfrak{e}$}\left(0,m;\mathfrak{t}^2   \right),
\end{equation*}
where $\text{\Large $\mathfrak{e}$}(\nu,\mu;\mathfrak{q})$ is as in equation (\ref{eq:ns energy}). Meanwhile, for $|\nu| >0$, we find a quadratic equation that determines the first coefficient $\omega_{-|\nu|}(m)$, which, in turn, selects the eigenvalue associated to odd or even solutions. Interestingly, we find that $\omega_{-|\nu|}(m)$ has the form:
\begin{equation}
\omega_{-|\nu|}^{\mp}(m)=\mp(-1)^{|\nu|} \frac{iN}{(m\mp|\nu|) \prod_{j=1-|\nu|}^{|\nu|-1}(m+j)},
\end{equation}
with $N$ some positive integer. Once $\Omega_{|\nu|}(m,\mathfrak{t})$ has been determined up to the desired order, the eigenvalues can be obtained as follows: the solution derived from $\omega_{-|\nu|}^{-}$ can be directly substituted into equation (\ref{eq: E}) for $E_{-}$, whereas the solution obtained from $\omega_{-|\nu|}^{+}$ must be substituted into the equation for $E_{+}$, with the additional step of sending $m \to -m$.

\subsection{(Anti-)Periodic eigenfunctions from orbifold defects}
Another approach to studying the periodic spectral problem emerges directly from the $\mathcal{N}=2^{*}$ $SU(2)$ gauge theory, in which equation (\ref{eq: lame std}) can be interpreted as a quantization of the Seiberg–Witten curve \cite{Seiberg:1994rs,Seiberg:1994aj}, capturing the dynamical content of the low-energy effective theory. In this framework, solutions to the Lamé equation correspond to the NS limit of expectation values of certain BPS surface defect operators \cite{Nekrasov:2017rqy, Nekrasov:2017gzb}, which can be computed via localization in the $\Omega$-background. At least two types of BPS surface defects can be considered; the first is associated with the insertion of a degenerate primary operator in the dual CFT picture \cite{Alday:2009fs, Alday:2010vg}, precisely the one we examined at the beginning of this work. The second corresponds to an orbifold defect \cite{Kanno:2011fw,Jeong:2017pai}, wherein the theory is defined on $\mathbb{C} \times \mathbb{C}/\mathbb{Z}_{2}$ instead of on $\mathbb{C}^{2} \simeq \mathbb{R}^{4}$. To fully characterize this second type of defect, one must specify a coloring function $c: \mathbb{Z}_{2} \to \mathbb{Z}_{2} \subset SU(2)$, which determines the gauge group component of the embedding of $\mathbb{Z}_{2}$ into the symmetry group of the theory. In particular, we focus on regular orbifold defects, characterized by a surjective coloring function. In this case, there are only two possibilities: either $c = id : \mathbb{Z}_{2} \to \mathbb{Z}_{2}$, or $c = (01) : \mathbb{Z}_{2} \to \mathbb{Z}_{2}$, the elementary transposition. Let $\boldsymbol{Y} = (Y_{0}, Y_{1})$ denote a pair of Young diagrams. In the presence of such a defect, the elementary characters appearing in the equivariant localization formulas used to compute the instanton partition functions decompose as:
\begin{subequations}
\begin{align}
N_{\omega} &= \sum_{\alpha = c^{-1}(\omega)} e^{a_{\alpha}},\\
K_{\omega} &= \sum_{\alpha = 0,1} \sum_{\substack{(r, s) \in Y_{\alpha} \\ c(\alpha) + s\, \equiv \,\omega + 1 \!\!\!\!\pmod{2}}} 
    e^{a_{\alpha} + \epsilon_{1}(r - 1) + \epsilon_{2}(s - 1)},
\end{align}
\end{subequations}
where $\omega \in \mathbb{Z}_{2}$. The full $\mathbb{Z}_{2}$-invariant character is then given by:
\begin{equation}
T[\boldsymbol{Y}]^{\mathbb{Z}_{2},c}=\left(1-e^{-\mu}\right)\sum_{\omega\in \mathbb{Z}_{2}}\left[N_{\omega}K_{\omega}^{*}+q_{1}q_{2}N_{\omega}^{*}K_{\omega-1}-(1-q_{1})K_{\omega}K_{\omega}^{*}+q_{2}(1-q_{1})K_{\omega}K_{\omega+1}^{*}\right],
\end{equation}
where $q_{i}=e^{\epsilon_{i}}$ and the $*$ operation reverses the sign of all the weights in the exponents. In particular, for the $id$ case we have $N_{0}=e^{a_{0}},$ $N_{1}=e^{a_{1}}$ and
\begin{equation}
    K_{0}=\sum_{(r,s)\in Y_{0},\,\, s\,\,odd}e^{a_{0}+\epsilon_{1}(r-1)+\epsilon_{2}(s-1)}+\sum_{(r,s)\in Y_{1},\,\, s\,\,even}e^{a_{1}+\epsilon_{1}(r-1)+\epsilon_{2}(s-1)},
\end{equation}
\begin{equation}
    K_{1}=\sum_{(r,s)\in Y_{0},\,\, s\,\,even}e^{a_{0}+\epsilon_{1}(r-1)+\epsilon_{2}(s-1)}+\sum_{(r,s)\in Y_{1},\,\, s\,\,odd}e^{a_{1}+\epsilon_{1}(r-1)+\epsilon_{2}(s-1)},
\end{equation}
while, for the $(01)$ case, we have $N_{0}=e^{a_{1}},$ $N_{1}=e^{a_{0}}$ and
\begin{equation}
    K_{0}=\sum_{(r,s)\in Y_{0},\,\, s\,\,even}e^{a_{0}+\epsilon_{1}(r-1)+\epsilon_{2}(s-1)}+\sum_{(r,s)\in Y_{1},\,\, s\,\,odd}e^{a_{1}+\epsilon_{1}(r-1)+\epsilon_{2}(s-1)},
\end{equation}
\begin{equation}
    K_{1}=\sum_{(r,s)\in Y_{0},\,\, s\,\,odd}e^{a_{0}+\epsilon_{1}(r-1)+\epsilon_{2}(s-1)}+\sum_{(r,s)\in Y_{1},\,\, s\,\,even}e^{a_{1}+\epsilon_{1}(r-1)+\epsilon_{2}(s-1)}.
\end{equation}
Defining the fractional coupling constants $\mathfrak{t}_{0}=\mathfrak{t} z$ and $\mathfrak{t}_{1}=\mathfrak{t}/z$, such that $\mathfrak{t}^{2}=\mathfrak{q}=e^{2\pi i \tau}$, the instanton part of the orbifold defect partition function is given by:
\begin{equation}
\Psi_{c}^{inst}\left(z,\nu,\mu;\epsilon_{1},\epsilon_{2};\mathfrak{t} \right)=\sum_{\boldsymbol{Y}}\prod_{\omega\in \mathbb{Z}_{2}}\mathfrak{t}^{k_{\omega}}_{\omega}\mathcal{E}(-T[\boldsymbol{Y}]^{\mathbb{Z}_{2},c}),
\end{equation}
where $k_{\omega}=\dim K_{\omega}$, $\nu=2(a_{0}-a_{1})$, and $\mathcal{E}$ acts by transforming sums of exponentials of weights into products of weights:
\begin{equation*}
    \mathcal{E}\left(\sum_{\alpha} e^{w_{\alpha}} - \sum_{\beta} e^{w_{\beta}} \right) = \frac{\prod_{\alpha} w_{\alpha}}{\prod_{\beta} w_{\beta}}.
\end{equation*}
Thus, setting $z=e^{2ix}$, we define the full defect partition functions as:
\begin{subequations}
\label{eq: orbifolds}
\begin{align}
\Psi_{id}\left(x,\nu,\mu;\epsilon_{1},\epsilon_{2};\mathfrak{t} \right)&=e^{i\nu x}\theta_{4}(x|\tau)^{1-\mu}\,\Psi_{id}^{inst}\left(e^{2ix},\nu,\mu;\epsilon_{1},\epsilon_{2};\mathfrak{t} \right)\label{eq: psi id},\\
\Psi_{(01)}\left(x,\nu,\mu;\epsilon_{1},\epsilon_{2};\mathfrak{t} \right)&=e^{-i\nu x}\theta_{4}(x|\tau)^{1-\mu}\,\Psi_{(01)}^{inst}\left(e^{2ix},\nu,\mu;\epsilon_{1},\epsilon_{2};\mathfrak{t} \right)\label{eq: psi 01}.
\end{align}
\end{subequations}
Employing the qq-characters formalism \cite{Nekrasov:2017gzb}, one can show that the NS limit of the two defect partition functions: 
\begin{equation}
\psi_c(x,\nu,\mu;\mathfrak{t})=\lim_{\epsilon_2 \to 0}\frac{\Psi_c(x,\nu,\mu;1,\epsilon_2;\mathfrak{t})}{\sqrt{\mathcal{Z}_{inst}^{U(2)}(\nu/2,\mu;1,\epsilon_2;\mathfrak{t})}},
\end{equation}
where $\mathcal{Z}_{inst}^{U(2)}$ is as in equation (\ref{eq: U(2) inst star}), constitute a pair of linearly independent solutions to the Lamé equation (\ref{eq: lame std}). Indeed, defining the effective twisted superpotential $\mathcal{W}_0$ as:
\begin{equation}
\mathcal{W}_0\left(\nu,\mu,\epsilon_{1};\mathfrak{t} \right) = \lim_{\epsilon_2 \to 0} \epsilon_2 \log \Psi_{c}^{inst}\left(z,\nu,\mu;\epsilon_{1},\epsilon_{2};\mathfrak{t} \right),
\end{equation}
the analogue of formula (\ref{eq:ns energy}) takes the form:
\begin{equation}
    \text{\Large $\mathfrak{e}$}\left(\nu,\mu;\mathfrak{t}\right) = \nu^{2}-4\mathfrak{t} \partial_{\mathfrak{t}}\mathcal{W}_0\left(\nu,\mu,1;\mathfrak{t} \right) +\frac{1}{3}\mu(\mu-1)(1-2E_{2}(\tau)),
\end{equation}
which yields the standard perturbative expansion of the eigenvalue. By instead considering the even and odd combinations:
\begin{equation}
\Psi_{\pm}\left(x,\nu,\mu;\epsilon_{1},\epsilon_{2};\mathfrak{t}\right)=\Psi_{id}\left(x,\nu,\mu;\epsilon_{1},\epsilon_{2};\mathfrak{t}\right)\pm\Psi_{(01)}\left(x,\nu,\mu;\epsilon_{1},\epsilon_{2};\mathfrak{t} \right),
\end{equation}
the effective twisted superpotential computed using these functions remains regular even for integer values of the Floquet exponent. Namely, if we define:
\begin{equation}
\mathcal{W}_0^{\,n,\,\pm}\left(\mu;\mathfrak{t} \right)=\lim_{\epsilon_2 \to 0}\epsilon_2\, \log\Psi_{\pm}\left(x,\nu=n,\mu;\epsilon_{1}=1,\epsilon_{2};\mathfrak{t} \right),
\end{equation}
we can compute the periodic and anti-periodic eigenvalues simply through\footnote{We note that the $\pm$ symbol here indicates the parity of the (anti-)periodic eigenfunction, in contrast to its use in equation (\ref{eq:ns energy}), where it refers to the direction of the limit $\nu \to n^{\pm}$. These two conventions do not necessarily agree.}:
\begin{equation}
    \text{\Large $\mathfrak{e}$}_{n,\,\pm}\left(\mu;\mathfrak{t}\right)=n^{2}-4\mathfrak{t}\partial_{\mathfrak{t}}\mathcal{W}_0^{\,n,\,\pm}\left(\mu;\mathfrak{t} \right)+\frac{1}{3}\mu(\mu-1)(1-2E_{2}(\tau)).
\end{equation}
Finally, also periodic and anti-periodic eigenfunctions can be obtained from $\Psi_{\pm}$, by considering:
\begin{equation}
    \psi_{n,\,\pm}(x,\mu;\mathfrak{t})=\lim_{\epsilon_2 \to 0}e^{-\mathcal{W}_0^{\,n,\,\pm}(\mu;\mathfrak{t})/\epsilon_2}\Psi_{\pm}\left(x,\nu=n,\mu;\epsilon_{1}=1,\epsilon_{2};\mathfrak{t} \right).
\end{equation}

\subsection{B-cycle periodic spectral problem}
Until now, we have studied the periodic spectral problem on the A-cycle (\ref{eq: lame std}), however this is not the only possibility. Indeed, we can also examine the periodic spectral problem on the B-cycle:
\begin{equation}
\label{eq: lame B cycle}
\left(\partial_{x}^{2}-\text{\Large $\mathfrak{e}$}+\mu(\mu-1)\wp(ix+\pi/2;\pi,\pi\tau)\right)\psi=0,
\end{equation}
where, once again, $\tau\in i\mathbb{R}$, and the eigenvalue $\text{\Large $\mathfrak{e}$}$ is given by equation (\ref{eq:ns energy}). Due to the modular properties of the $\wp$-function, this problem is actually equivalent to the previous one under the transformation $\tau \to -1/\tau$. Indeed, since: 
\begin{equation*}
    \wp(z;\tau)=\frac{1}{\tau^{2}}\wp(z/\tau;-1/\tau),
\end{equation*}
by making the change of variable $y=ix/\tau$, the equation becomes:
\begin{equation*}
    \left(-\partial_{y}^{2}-\text{\Large $\mathfrak{e}$}\tau^{2}+\mu(\mu-1)\wp(y+\pi/2\tau;\pi,-\pi/\tau)\right)\psi=0.
\end{equation*}
In particular, this implies that studying the B-cycle spectral problem (\ref{eq: lame B cycle}) in the limit $\tau \to i\infty$ is equivalent to studying the A-cycle spectral problem (\ref{eq: lame std}) in the opposite limit $\tau \to 0$. To this end, we can leverage the knowledge of the monodromy matrices to gain insights into the spectrum in this regime. Recalling that $\nu = 2a = 2\alpha - 1$, the trace of the B-cycle monodromy matrix $M_{B}^{(n)}$ in equation (\ref{eq: B-cycle mm}) can be written as:
\begin{equation}
\label{eq: trmb}
    \Tr M_{B}=\frac{\sin\pi(2a+\mu)}{\sin2\pi a}e^{\eta/2}+\frac{\sin\pi(2a-\mu)}{\sin2\pi a}e^{- \eta/2}.
\end{equation}
Here, a term that coincides with the 1-loop contribution to the NS prepotential of the $\mathcal{N}=2^{*}$ gauge theory has been absorbed into the definition of the conformal block $f(a, \mu; \mathfrak{q})$, so that the dual period $\eta$ is given by:
\begin{equation}
\label{eq: eta old}
    \eta(a,\mu;\mathfrak{q}) =\partial_{a}f(a,\mu;\mathfrak{q})= -2a\log \mathfrak{q} +2\log\frac{\Gamma(1+2a)\Gamma(1-2a-\mu)}{\Gamma(1-2a)\Gamma(1+2a-\mu)}+\partial_{a}f_{inst}(a,\mu;\mathfrak{q}).
\end{equation}
Since we are interested in the spectral problem for values of the coupling $\mu \geq 1$, it is more convenient to rewrite (\ref{eq: trmb}) as:
\begin{equation}
\label{eq: trmb final}
    \Tr M_{B}=-\frac{\sin\pi(2a-\mu)}{\sin2\pi a}e^{\eta/2}-\frac{\sin\pi(2a+\mu)}{\sin2\pi a}e^{- \eta/2},
\end{equation}
where the dual period now reads:
\begin{equation}
\label{eq: eta}
    \eta(a,\mu;\mathfrak{q}) =\partial_{a}f(a,\mu;\mathfrak{q})= -2a\log \mathfrak{q} +2\log\frac{\Gamma(1+2a)\Gamma(\mu-2a)}{\Gamma(1-2a)\Gamma(\mu+2a)}+\partial_{a}f_{inst}(a,\mu;\mathfrak{q}),
\end{equation}
From the general theory of periodic differential equations, we can write:
\begin{equation}
\label{eq: trmb=nudual}
    \Tr M_{B}=2\cos(\pi\nu_{D}),
\end{equation}
where $\nu_{D}$ is the B-cycle Floquet exponent, such that when it is an integer, the corresponding solutions are either periodic or anti-periodic along the B-cycle. Thus, we can solve this equation for $a$ as a function of $\nu_{D}$ and the other variables, and then use the explicit expression for the eigenvalue (\ref{eq:ns energy}) to determine the energy of the system as a function of these parameters.

\subsubsection{Scattering states}
For generic values of $\mu$, a first set of solutions can be obtained by considering the following double expansion of the Coulomb branch parameter:
\begin{equation*}
    a_{n}(\nu_{D},\mu,\tau)=\frac{n}{2\tau}+\sum_{k\geq 0}c_{k}^{(n)}(\nu_{D},\mu,\tau)\mathfrak{q}^{k},\quad c_{k}^{(n)}(\nu_{D},\mu,\tau)=\sum_{j\geq 2}c_{k,j}^{(n)}(\nu_{D},\mu)\tau^{-j},
\end{equation*}
with $n \in \mathbb{Z}$, which implies a corresponding double expansion for the eigenvalue. For physical values of the parameters, these solutions are purely imaginary and the very first terms read:
\begin{equation}
    a_{n}(\nu_{D},\mu,\tau)=\frac{n}{2\tau}+\frac{in\left[\pi\left(\cos{\pi\mu}+\cos{\pi(\nu_{D}+n)}\right) \csc{\pi\mu} +2H(\mu-1)\right]}{2\pi \tau^{2}}+\mathcal{O}\left(\tau^{-3}\right),
\end{equation}
while for the corresponding eigenvalues we find:
\begin{equation}
\begin{aligned}
\text{\Large $\mathfrak{e}$}_{n}(\nu_{D},\mu,\tau)&=-\frac{1}{3}\mu(\mu-1)+\frac{n^{2}}{\tau^{2}}\\
&+\frac{2in^{2}\left[\pi\left(\cos{\pi(\nu_{D}+n)}+\cos\pi\mu\right)\csc\pi\mu+2H(\mu-1))\right]}{\pi \tau^{3}}+\mathcal{O}\left(\tau^{-4}\right).
\end{aligned}
\end{equation}
In particular, one can notice that the dual Floquet exponent $\nu_{D}$ always appears in combination with the integer $n$ as $\cos{\pi(\nu_D + n)}$. When $\nu_{D}=0$, even $n$ corresponds to periodic solutions, while odd $n$ corresponds to anti-periodic ones. Conversely, when $\nu_{D}=\pm 1$, the roles are reversed. These expansions exhibit poles for integer values of $\mu$, due to terms proportional to $\csc{\pi\mu}$, except if:
\begin{equation*}
    \cos{\pi(\nu_{D}+n)}+\cos{\pi\mu}=0 \iff \cos{\pi(\nu_{D}+n)}=\cos{\pi(\mu-1)},
\end{equation*}
which implies that the two monodromy exponents $\mu$ and $\nu_D$ are essentially identified, and in particular, $\nu_{D}$ is constrained to be an integer when $\mu$ is an integer. We will now show that, in the case where $\mu$ is an integer, this term is precisely what is missing from the expansion, without imposing any restriction on $\nu_{D}$. Indeed, for $\mu=k\in \mathbb{Z}$, equation (\ref{eq: trmb=nudual}) simplifies to:
\begin{equation}
\label{eq:qcinteger}
    (-1)^{k-1}\cosh{\frac{\eta}{2}}= \cos{\pi\nu_{D}},
\end{equation}
and $\eta$ also simplifies significantly; in particular, the $\Gamma$-functions disappear leaving behind only simple rational functions. In this case, we can solve directly for the B-cycle Floquet exponent $\nu_{D}=\nu_{D}(a,k,\tau)$ as:
\begin{equation}
    \nu_{D}(a,k,\tau)=n\pm \frac{i}{2\pi}\eta(a,k;\mathfrak{q}).
\end{equation}
This relation can then be inverted to express $a$ as $a=a(\nu_{D},k,\tau)$, leading to an expansion analogous to (\ref{eq: expansion ns lame}) for the eigenvalue as a function of the B-cycle Floquet exponent. The first few terms of the resulting expression read:
\begin{equation}
\label{eq: en scattering}
\begin{aligned}
&\text{\Large $\mathfrak{e}$}_{n}(\nu_D,k,\tau)=-\frac{1}{3}k(k-1)+\frac{\nu_D^{2}}{\tau^{2}}+\frac{4i\nu_D^{2}H(k-1)}{\pi \tau^{3}}-\frac{12\nu_D^{2}H(k-1)^{2}}{\pi^{2} \tau^{4}}+\mathcal{O}\left(\tau^{-5}\right)\\
&+8k(k-1)\left((1+k-k^{2})-k(k-1)\nu_D^{2}\left(\frac{1}{\tau^{2}}-\frac{i(1+4H(k-1))}{\pi \tau^{3}}\right)+\mathcal{O}\left(\tau^{-4}\right)\right)\mathfrak{q}+\mathcal{O}\left(\mathfrak{q}^{2}\right).
\end{aligned}
\end{equation}
Interestingly, when also $\nu_D \in \mathbb{Z}$, equation (\ref{eq:qcinteger}) is equivalent to the usual NS quantization condition:
\begin{equation}
    \partial_{a}f(a,\mu;\mathfrak{q}) = 2\pi i n,\quad n \in \mathbb{Z}.
\end{equation}
The solutions are then characterized by the integer $n$ and the first few terms of their expansion reads:
\begin{equation}
\begin{aligned}
a_{n}(\mu,\tau)&=\frac{n}{2\tau}+\frac{inH(\mu-1)}{\pi\tau^{2}}-\frac{2nH(\mu-1)^{2}}{\pi^{2}\tau^{3}}+\mathcal{O}\left(\tau^{-4}\right)\\
&+\left(-\frac{2in\mu^{2}(\mu-1)^{2}}{\pi\tau^{2}}+\frac{8n\mu^{2}(\mu-1)^{2}H(\mu-1)}{\pi^{2}\tau^{3}}+\mathcal{O}\left(\tau^{-4}\right)\right)\mathfrak{q} + \mathcal{O}\left(\mathfrak{q}^{2}\right),
\end{aligned}
\end{equation}
while the corresponding eigenvalues are given by equation (\ref{eq: en scattering}) where $\nu_D$ is replaced by $n$. Finally, the quantization condition simplifies also in the case $\mu = k + 1/2$, $k\in \mathbb{Z}$, resulting in:
\begin{equation}
    (-1)^{k} \cot(2\pi a)\sinh\left(\frac{\eta}{2}\right)=\cos(\pi\nu_{D}).
\end{equation}

\subsubsection{Quasi-bound states}
A second set of qualitatively different solutions can instead be obtained by expanding the quantization condition (\ref{eq: trmb=nudual}) for $a$ near the poles of the $\Gamma$-functions in ($\ref{eq: eta}$). Assuming $\mu \geq 1$, we can consider $a \in \mathbb{R}$. In particular, the trace of the monodromy matrix (\ref{eq: trmb final}) is invariant under $a \to -a$. Consequently, we can restrict to $a > 0$, with $a=0$ being shared with the previous set of solutions. Then, to ensure that $\nu_{D}$ remains finite in the $\tau\to i\infty$ limit, the only possibility is:
\begin{equation}
    \hat{a}_{n}(\nu_{D},\mu;\mathfrak{q})=\frac{\mu-n}{2}+\sum_{i\geq 1}\sum_{j\geq 0}a_{i,j}(\nu_{D},\mu)\mathfrak{q}^{i(\mu-n)/2+j}= \frac{\mu-n}{2} + \mathcal{O}\left(\mathfrak{q}^{(\mu-n)/2}\right),
\end{equation}
where $n= [\mu]-1,-\,,1$. This second set of solutions consists again of double expansions, this time in $\mathfrak{q}^{\frac{\mu-n}{2}}$ and $\mathfrak{q}$, which correspond to the bound states of the Poschl-Teller potential in the $\tau = i\infty$ limit, see Appendix \ref{app: checks 2}.  The first terms of the corresponding expansion of the eigenvalue read: 
\begin{equation}
\begin{aligned}
\hat{\text{\Large $\mathfrak{e}$}}_{n}(\nu_{D},\mu;\mathfrak{q})&=(\mu-n)^{2}-\frac{1}{3}\mu(\mu-1) -\frac{4\cos\pi\nu_{D}}{\pi}\frac{\Gamma(1+n-\mu)\Gamma(2\mu-n)\sin\pi\mu}{\Gamma(n)\Gamma(\mu-n)}\mathfrak{q}^{\frac{\mu-n}{2}}\\
&+\frac{8\mu(\mu-1)\left(\mu(\mu-1)-1+(\mu-n)^{2}\right)}{(\mu-n)^{2}-1}\mathfrak{q} +\mathcal{O}\left(\mathfrak{q}^{\mu-n}\right).
\end{aligned}
\end{equation}
Expanding for $a$ near the remaining poles of the $\Gamma$-functions results in a divergent $\Tr M_{B}$, except in the case where $\mu$ is an integer. In this situation, the poles at $a = n/2$, with $n\in \mathbb{N}$, coincide with those associated with quasi-bound states. Indeed, due to cancellations between the $\Gamma$-functions, equation (\ref{eq:qcinteger}) can be explicitly written as: 
\begin{equation}
e^{2\pi a t+\frac{1}{2}\partial_{a}f_{inst}}\frac{(\mu-1-2a)\cdots (1-2a)}{(\mu-1+2a)\cdots(1+2a)}+(a\to-a)=2(-1)^{\mu-1}\cos{\pi\nu_{D}}.
\end{equation}
Assuming $\mu\geq 1$ and $a>0$, the remaining poles of the $\Gamma$-functions are now located at $a = 1/2,-,(\mu-1)/2$. This implies that analyzing these points requires the resummation of $f_{inst}$ around these specific values of $a$. For instance, in the case $\mu = 2$, there is a unique solution near $a \sim \tfrac{1}{2}$. Using the methods introduced in the previous sections to compute $\partial_{a} f_{inst}$ around this singular point, we find:
\begin{equation}
    \hat{\text{\Large $\mathfrak{e}$}}_{1}(\nu_{D};\mathfrak{q})=\frac{1}{3}+8\cos\pi\nu_{D}\sqrt{\mathfrak{q}}+8\big(1-4(1+\log \sqrt{\mathfrak{q}})\sin^{2}\pi\nu_{D}\big)\mathfrak{q}+\mathcal{O}(\mathfrak{q}^{3/2}).
\end{equation}
Meanwhile, for $\mu = 3$, there are two solutions, $a \sim \frac{1}{2}$ and $a \sim 1$, for which the corresponding eigenvalues are given by:
\begin{equation}
\begin{aligned}
\hat{\text{\Large $\mathfrak{e}$}}_{1}(\nu_{D};\mathfrak{q})&=2+48(3+2\cos\pi\nu_{D})\mathfrak{q}\\
&-48\big(127+128\cos\pi\nu_{D}-8(1-6\log \mathfrak{q})\sin^{2}\pi\nu_{D}\big)\mathfrak{q}^{2}+\mathcal{O}(\mathfrak{q}^{3}),
\end{aligned}
\end{equation}
\begin{equation}
    \hat{\text{\Large $\mathfrak{e}$}}_{2}(\nu_{D};\mathfrak{q})=-1-24\cos\pi\nu_{D}\sqrt{\mathfrak{q}}-24\big(1+(28+6\log \mathfrak{q})\sin^{2}\pi\nu_{D}\big)\mathfrak{q}+\mathcal{O}(\mathfrak{q}^{3/2}).
\end{equation}

\newpage
\appendix
\section{Special Functions}
\label{app: special functions}
\subsection{Classical special functions}
In this sub-appendix, we summarize the conventions used for the various special functions appearing throughout the main text. The Weierstrass $\wp$-function, defined by:
\begin{equation}
\wp(z;2\omega_1,2\omega_2)=\frac{1}{z^{2}}+\sum_{(m,n)\neq (0,0)}\left(\frac{1}{(z-2m\omega_1-2n \omega_2)^{2}}-\frac{1}{(2m\omega_1+2n \omega_2)^{2}}\right),
\end{equation}
admits the following alternative representations:
\begin{subequations}
\begin{align}
&\wp(z)=\frac{\pi^{2}}{\sin^{2}\pi z}-\frac{\pi^{2}}{3}+\pi^{2}\sum_{n\neq 0}\left(\frac{1}{\sin^{2}\pi(z-n\tau)}-\frac{1}{\sin^{2}\pi n\tau} \right),\\
&\wp(z)=\frac{\pi^{2}/\tau^{2}}{\sin^{2}\frac{\pi z}{\tau}}-\frac{\pi^{2}}{3\tau^{2}}+\frac{\pi^{2}}{\tau^{2}}\sum_{m\neq 0}\left(\frac{1}{\sin^{2}\frac{\pi(z-m)}{\tau}}-\frac{1}{\sin^{2}\frac{\pi m}{\tau}} \right),
\end{align}
\end{subequations}
where $\wp(z)$ always denotes $\wp(z;1,\tau)$, $\tau=\omega_2/\omega_1$. In particular, in the limit $\tau\to +i \infty$, we find:
\begin{equation}
\label{eq: wp q-exp}
    \wp(z)=\frac{\pi^{2}}{\sin^{2}\pi z}-\frac{\pi^{2}}{3}+16\pi^{2}\mathfrak{q}\sin^{2}\pi z+\mathcal{O}\left(\mathfrak{q}^{2}\right),
\end{equation}
where $\mathfrak{q}=e^{2\pi i \tau}$. Moreover, the values of the $\wp$-function at the half-periods are denoted respectively by: $e_1=\wp(\omega_1;2\omega_1,2\omega_2)$, $e_2=\wp(\omega_2;2\omega_1,2\omega_2)$  and $e_3=\wp(\omega_1+\omega_2;2\omega_1,2\omega_2)$. The Weierstrass $\zeta$-function, defined as:
\begin{equation}
    \zeta(z; 2\omega_1, 2\omega_2)=\frac{1}{z}+\sum_{(m,n)\neq (0,0)}\left(\frac{1}{z-2m\omega_1-2n\omega_2}+\frac{1}{2m\omega_1+2n\omega_2}+\frac{z}{(2m\omega_1+2n\omega_2)^2} \right),
\end{equation}
is related to the Weierstrass $\wp$-function by:
\begin{equation}
    \wp(z;2\omega_1,2\omega_2)=-\zeta'(z;2\omega_1,2\omega_2),
\end{equation}
and, once again, $\zeta(z)$ always denotes $\zeta(z;1,\tau)$. For the Jacobi theta functions, we use the following conventions:
\begin{subequations}
\begin{align}
&\theta_{1}(z|\tau)=-i\sum_{n\in \mathbb{Z}}(-1)^{n}e^{i\pi\tau\left(n+\frac{1}{2}\right)^{2}}e^{2\pi i z\left(n+\frac{1}{2}\right)},\\
&\theta_{2}(z|\tau)=\sum_{n\in \mathbb{Z}}e^{i\pi\tau\left(n+\frac{1}{2}\right)^{2}}e^{2\pi i z\left(n+\frac{1}{2}\right)},\\
&\theta_{3}(z|\tau)=\sum_{n\in \mathbb{Z}}e^{i\pi\tau n^{2}}e^{2\pi i zn},\\
&\theta_{4}(z|\tau)=\sum_{n\in \mathbb{Z}}(-1)^{n}e^{i\pi\tau n^{2}}e^{2\pi i zn}.
\end{align}
\end{subequations}
The Euler $\varphi$-function is defined by:
\begin{equation}
\varphi(\mathfrak{q})=\prod_{n\geq 1}(1-\mathfrak{q}^n),
\end{equation}
while the Dedekind $\eta$-function as:
\begin{equation}
\eta(\tau)=\mathfrak{q}^{1/24}\varphi(\mathfrak{q}).
\end{equation}
In terms of Jacobi theta functions, the Weierstrass elliptic functions can be represented as follows:
\begin{subequations}
\begin{align}
&\wp(z)=-\partial_{z}^{2}\log\theta_1(z|\tau)-2\eta_1(\tau),\\[1ex]
&\zeta(z)=\partial_{z}\log\theta_1(z|\tau)+2z\eta_1(\tau),
\end{align}
\end{subequations}
where
\begin{equation}
    \eta_{1}(\tau)=-2\pi i \partial_{\tau}\log\eta(\tau)=4\pi^2 \left(\frac{1}{24}-\sum_{n=1}^{\infty}\frac{n\mathfrak{q}^{n}}{1-\mathfrak{q}^{n}}\right).
\end{equation}
Moreover, we have the relation:
\begin{equation}
    E_{2}(\tau)=\frac{6}{\pi^2}\eta_{1}(\tau).
\end{equation}
Finally, the Jacobi elliptic sine function is given by:
\begin{equation}
    \sn(z;\kappa^{2})=\frac{\theta_{3}(0|\tau)\theta_{1}(u(z)|\tau)}{\theta_{2}(0|\tau)\theta_{4}(u(z)|\tau)},\quad u(z)=\frac{\ z}{2K(\kappa^{2})},
\end{equation}
where
\begin{equation}
K(\kappa^{2})=\int_{0}^{1}\frac{dt}{\sqrt{(1-t^{2})(1-\kappa^{2}t^{2})}},
\end{equation}
is the complete elliptic integral of the first kind. The function $\sn(z;\kappa^2)$ is an odd, doubly periodic function of periods $4K(\kappa^2)$ and $2iK(1-\kappa^2)$, related to the Weierstrass $\wp$-function by:
\begin{equation}
    \wp(z)=e_{3}+\frac{e_{1}-e_{3}}{\sn^{2}(z\sqrt{e_{1}-e_{3}};\kappa^2)},
\end{equation}
with the following formulas relating modular parameters:
\begin{subequations}
\begin{align}
&\kappa^2=\frac{e_{2}-e_{3}}{e_{1}-e_{3}}=\frac{\theta_{2}(0|\tau)^{4}}{\theta_{3}(0|\tau)^{4}},\\[1ex]
&\tau = i\,\frac{K(1-\kappa^{2})}{K(\kappa^{2})}.
\end{align}
\end{subequations}
\subsection{Nekrasov partition functions, conformal blocks \& AGT dictionary}
Starting from the Nekrasov partition function of the $SU(2)$ gauge theory with $N_f$ fundamental hypermultiplets, we adopt the conventions of \cite{Bonelli:2025owb} and write:
\begin{equation}
    \mathcal{Z}^{[N_{f}]}(a,\{\mu_{i}\};\epsilon_{1},\epsilon_{2}; t)=\mathcal{Z}^{[N_{f}]}_{cl}(a;\epsilon_{1},\epsilon_{2}; t)\mathcal{Z}^{[N_{f}]}_{1\text{-}loop}(a,\{\mu_{i}\};\epsilon_{1},\epsilon_{2})\mathcal{Z}^{[N_{f}]}_{inst}(a,\{\mu_{i}\};\epsilon_{1},\epsilon_{2}; t),
\end{equation}
where:
\begin{subequations}
\begin{align}
&\mathcal{Z}^{[N_{f}]}_{cl}(a;\epsilon_{1},\epsilon_{2}; t)=t^{\frac{(\epsilon_1+\epsilon_2)^2/4-a^2}{\epsilon_1 \epsilon_2}},\label{eq: classical part Nf}\\[1ex]
&\mathcal{Z}^{[N_{f}]}_{1\text{-}loop}(a,\{\mu_{i}\};\epsilon_{1},\epsilon_{2})=\prod_{\pm}\frac{\prod_{i=1}^{N_f}\exp\left(\gamma_{\epsilon_1,\epsilon_2}(\mu_i\pm a-(\epsilon_1+\epsilon_2)/2) \right)}{\exp(\gamma_{\epsilon_1,\epsilon_2}(\pm2 a)) },\label{eq: 1-loop part Nf}\\[1ex]
&\mathcal{Z}^{[N_{f}]}_{inst}(a,\{\mu_{i}\};\epsilon_{1},\epsilon_{2}; t)=\sum_{\boldsymbol{Y}=(Y_+,Y_-)}\frac{\prod_{i=1}^{N_f}z_{fund}(a,\mu_i,\boldsymbol{Y})}{z_{vec}(a,\boldsymbol{Y})}t^{|\boldsymbol{Y}|},
\end{align}
\end{subequations}
and $\boldsymbol{Y}=(Y_+,Y_-)$ denotes a couple of Young tableaux, $|\boldsymbol{Y}|$ being the total number of boxes of the pair. The double gamma function $\gamma_{\epsilon_1,\epsilon_2}$ is defined as:
\begin{equation}
    \gamma_{\epsilon_1,\epsilon_2}(x)=\frac{d}{ds}\frac{1}{\Gamma(s)}\int_{0}^{\infty}\frac{dz}{z}z^{s} \frac{e^{-xz}}{(e^{\epsilon_1 z}-1)(e^{\epsilon_2 z}-1)}\bigg{|}_{s=0}, \quad \Re(x)>0, \quad \Re(\epsilon_{1/2})\neq 0,
\end{equation}
and satisfies the difference equation:
\begin{equation}
   \gamma_{\epsilon_1,\epsilon_2}(x-\epsilon_1)+\gamma_{\epsilon_1,\epsilon_2}(x-\epsilon_2)-\gamma_{\epsilon_1,\epsilon_2}(x-\epsilon_1-\epsilon_2)-\gamma_{\epsilon_1,\epsilon_2}(x)=\log x.
\end{equation}
Many useful properties of this function, including its shift relations, can be found e.g. in the appendices of \cite{Bonelli:2025owb} and \cite{Bershtein:2018zcz}. Coming back to the instanton part of the partition function, we have \cite{Bruzzo:2002xf,Flume:2002az}:
\begin{equation}
    z_{fund}(a,\mu,\boldsymbol{Y})=\prod_{\pm}\prod_{(i,j)\in Y_{\pm}}(\mu\pm a+(i-1/2)\epsilon_1 +(j-1/2)\epsilon_2),
\end{equation}
while, in order to define $z_{vec}$, we first define:
\begin{equation}
\begin{aligned}
z_{bif}(\vec{a},\vec{b},m,\boldsymbol{Y},\boldsymbol{W})=&\prod_{\alpha,\beta=\pm}\prod_{s_1\in Y_{\alpha}}\left[a_{\alpha}-b_{\beta}+\epsilon_{2}\left(A_{Y_{\alpha}}(s_1)+1\right)-\epsilon_{1}L_{W_{\beta}}(s_1)-m\right]\\
&\prod_{s_2\in W_{\beta}}\left[a_{\alpha}-b_{\beta}+\epsilon_{1}\left(L_{Y_{\alpha}}(s_2)+1\right)-\epsilon_{2}A_{W_{\beta}}(s_2)-m\right],
\end{aligned}
\end{equation}
where, following \cite{Alday:2009aq}, $\vec{a}=(a_+,a_-)$ and $\vec{b}=(b_+,b_-)$, $s_{1}$ and $s_{2}$ label a box in the corresponding Young tableau, $A_Y $ and $L_Y$ are defined as: 
\begin{equation}
    A_{Y}(s)= \lambda_{i}-j\,, \quad L_{Y}(s)=\lambda^{'}_{j}-i, \quad s=(i,j)\in Y,
\end{equation}
and the Young tableau $Y$ is identified with the corresponding partition $Y=\{\lambda_{1}\geq...\geq\lambda_{n}>0\}$, $Y^{T}=\{\lambda^{'}_{1}\geq...\geq\lambda^{'}_{m}>0\}$ being its transpose. Then, we have:
\begin{equation}
    z_{adj}(a,m,\boldsymbol{Y})=z_{bif}(\vec{a},\vec{a},m,\boldsymbol{Y},\boldsymbol{Y}),
\end{equation}
where $\vec{a}=(a,-a)$, and finally:
\begin{equation}
    z_{vec}(a,\boldsymbol{Y})=z_{adj}(a,0,\boldsymbol{Y}).
\end{equation}
Moving on to the partition function of the $\mathcal{N}=2^{*}$ $SU(2)$ gauge theory, we have:
\begin{equation}
\label{eq: SU(2) Z *}
    \mathcal{Z}(a,m;\epsilon_1,\epsilon_2;\mathfrak{q})=\mathcal{Z}_{cl}(a;\epsilon_{1},\epsilon_{2};\mathfrak{q})\mathcal{Z}_{1\text{-}loop}(a,m;\epsilon_1,\epsilon_2)\mathcal{Z}_{inst}(a,m;\epsilon_1,\epsilon_2;\mathfrak{q}),
\end{equation}
where:
\begin{subequations}
\begin{align}
&\mathcal{Z}_{cl}(a;\epsilon_{1},\epsilon_{2};\mathfrak{q})=\mathfrak{q}^{-\frac{a^2}{\epsilon_1 \epsilon_2}},\\[1ex]
&\mathcal{Z}_{1\text{-}loop}(a,m;\epsilon_1,\epsilon_2)=\prod_{\pm}\frac{\exp(\gamma_{\epsilon_1,\epsilon_2}(\pm2 a-m))}{\exp(\gamma_{\epsilon_1,\epsilon_2}(\pm2 a))}\label{eq: SU(2) 1-loop star},\\[1ex]
&\mathcal{Z}_{inst}(a,m;\epsilon_1,\epsilon_2;\mathfrak{q})=\varphi(\mathfrak{q})^{1-\frac{2m(Q-m)}{\epsilon_1 \epsilon_2}}\mathcal{Z}_{inst}^{U(2)}(a,m;\epsilon_1,\epsilon_2;\mathfrak{q})\label{eq: zinst 2*},
\end{align}
\end{subequations}
and finally:
\begin{equation}
\label{eq: U(2) inst star}
   \mathcal{Z}_{inst}^{U(2)}(a,m;\epsilon_1,\epsilon_2;\mathfrak{q})= \sum_{\boldsymbol{Y}}\mathfrak{q}^{|\boldsymbol{Y}|}\frac{z_{adj}(a,m,\boldsymbol{Y})}{z_{vec}(\vec{a},\boldsymbol{Y})}.
\end{equation}
In conclusion, we describe how the semi-classical limit $\psi_{s}$ of the degenerate torus blocks in the $necklace$ basis $\Phi^{(n)}_{s}$, as in equation \eqref{eq: 2pt classical}, can be expressed in terms of surface defect partition functions of the $\mathcal{N}=2^{*}$ $SU(2)$ gauge theory. We consider the quiver gauge theory \cite{Katz:1997eq,Gaiotto:2009we, Nekrasov:2012xe} associated to the circular quiver with two $SU(2)$ gauge nodes, one of which is degenerate \cite{Nekrasov:2017rqy}, and compute its instanton partition function as:
\begin{equation}
\begin{aligned}
\mathcal{Z}_{s,\, inst}^{defect}(z,\sigma,m;\epsilon_1,\epsilon_2;\mathfrak{q})=&\sum_{\boldsymbol{Y},\boldsymbol{W}}z^{|\boldsymbol{Y}|-|\boldsymbol{W}|}\mathfrak{q}^{\frac{|\boldsymbol{Y}|+|\boldsymbol{W}|}{2}}\frac{z_{bif}(\vec{\sigma}_1,\vec{\sigma}_2,m,\boldsymbol{Y},\boldsymbol{W})}{z_{vec}(\sigma,\boldsymbol{Y})}\\
&\cdot\frac{z_{bif}(\vec{\sigma}_2,\vec{\sigma}_1,-\epsilon_2/2,\boldsymbol{W},\boldsymbol{Y})}{z_{vec}(\sigma-s\epsilon_2/2,\boldsymbol{W})},
\end{aligned}
\end{equation}
where $\vec{\sigma}_1=(\sigma,-\sigma)$ and $\vec{\sigma}_2=(\sigma-s\epsilon_2/2,-\sigma+s\epsilon_2/2)$, $s=\pm$. Up to a simple overall prefactor, the AGT correspondence implies the relation:
\begin{equation}
    \Phi^{(n)}_{s}(x+i\tau/2,a,m;\mathfrak{q})\propto \mathcal{Z}^{U(1)}(z,m;\mathfrak{q})\mathcal{Z}_{s,\, inst}^{defect}(e^{2i\pi x},a-Q/2,m;b,b^{-1};\mathfrak{q}),
\end{equation}
where $Q=\epsilon_1+\epsilon_2=b+b^{-1}$, while:
\begin{equation}
\begin{aligned}
\mathcal{Z}^{U(1)}(z,m;\mathfrak{q})=&\prod_{l\geq 0}\left(1-\mathfrak{q}^{l+1}\right)^{2m(Q-m)-3b^{2}/2-2}\left(1-z\mathfrak{q}^{l+1/2}\right)^{-b(Q-m)}\\
&\cdot\left(1-z^{-1}\mathfrak{q}^{l+1/2}\right)^{2m(Q+b/2)}.
\end{aligned}
\end{equation}
Thus, taking the semi-classical limit, we obtain the following representation for a pair of linearly independent Floquet solutions of the Lamé equation (\ref{eq: hill}):
\begin{equation}
\label{eq: classical defect}
\psi_s(x,\nu,\mu;\mathfrak{q})=e^{-2\pi isx}\lim_{b\to 0}\frac{\mathcal{Z}_{s,\, inst}^{defect}(e^{2i\pi x},\frac{\nu}{2b},\frac{\mu}{b};b,b^{-1};\mathfrak{q})}{\mathcal{Z}_{ inst}(\frac{\nu}{2b},\frac{\mu}{b};b,b^{-1};\mathfrak{q})}.
\end{equation}

\section{Checks}
\label{app: checks}

\subsection{Comparison with Hill determinant method}
\label{app: checks 1}
In this sub-appendix, we describe how the expansions of the eigenvalues and eigenfunctions of the Lamé equation can be computed using standard techniques. The first method we present concerns the computation of eigenvalues and is known as the Hill determinant method \cite{Wang:1989}. Consider the following complex Hill equation:
\begin{equation}
\label{eq: hill appendix}
    -\partial_{x}^{2}\psi(x)+V(x)\psi(x)=0,
\end{equation}
with $V(x)$ an even, periodic potential of period $\pi$. Its Fourier expansion reads:
\begin{equation}
    V(x)=a_{0}+\sum_{k\neq 0}a_{k}e^{2ikx},
\end{equation}
where $a_{-k}=a_{k}$, due to the parity of the potential, and $a_{k}\in \mathbb{C}$ for any $ k\in\mathbb{Z}$. We search for a Floquet solution $\psi(x)$, meaning a quasi-periodic function such that $\psi(x+\pi)=e^{i \nu \pi}\psi(x)$, where $\nu \in \mathbb{C}$ is the Floquet exponent. We can then write $\psi(x)$ in the form $\psi(x)=e^{i \nu x}\phi(x)$, where $\phi(x)$ is a periodic function of period $\pi$, and expand $\phi(x)$ in Fourier series, obtaining:
\begin{equation}
    \psi(x)=e^{i\nu x}\sum_{n=-\infty}^{\infty}b_{n}e^{2inx},
\end{equation}
where $b_{n}\in \mathbb{C}$ for any $n \in \mathbb{Z}$. Inserting this expansion into equation (\ref{eq: hill appendix}), we find:
\begin{equation}
    \sum_{n=-\infty}^{\infty}b_{n}(\nu+2n)^{2}e^{i(\nu+2n)x}+\sum_{k=-\infty}^{\infty}a_{k}e^{2ikx}\sum_{n=-\infty}^{\infty}b_{n}e^{i(\nu+2n)x}=0,
\end{equation}
and thus an infinite number of linear equations of the form:
\begin{equation}
    (\nu+2n)^{2}b_{n}+\sum_{k=-\infty}^{\infty}a_{k}b_{n-k}=0, \quad\forall n \in \mathbb{Z}.
\end{equation}
This system can be reformulated as the following matrix equation:
\begin{equation}
    \sum_{m\in \mathbb{Z}}B_{nm}b_{m}=0\,, \quad B_{nn}=a_{0}+(\nu+2n)^{2}\,,\quad B_{nm}=a_{n-m},
\end{equation}
which admits non-trivial solutions if and only if the determinant of the matrix $B$ vanishes. We can then use this last condition as an equation for determining the eigenvalues. In the case of equation (\ref{eq: lame std}), the relevant Fourier expansion of the potential is given by:
\begin{equation}
\label{eq: p fourier}
\wp(x+\pi\tau/2;\pi,\pi\tau)=-\frac{2}{\pi^2}\eta_{1}(\tau)-4\sum_{k\geq 1}2k\,\cos2k x\, \frac{\mathfrak{q}^{k/2}}{1-\mathfrak{q}^{k}},
\end{equation}
which leads to the identification:
\begin{equation*}
a_{0}=\text{\Large $\mathfrak{e}$}\,,\quad a_{k}=4\mu(\mu-1)k\frac{\mathfrak{q}^{k/2}}{1-\mathfrak{q}^{k}}.
\end{equation*}
We can then solve the condition $\det B=0$ as an equation for $\text{\Large $\mathfrak{e}$}=\nu^{2}+\cdots$, obtaining a power series expansion in $\mathfrak{q}^{1/2}$. In particular, solving the equation for generic values of $\nu$ we recover the expansion (\ref{eq: expansion ns lame}). Meanwhile, fixing $\nu$ to a specific integer value from the outset yields the expansion for the corresponding periodic or anti-periodic eigenvalue. The two possible eigenvalues corresponding to the same value of $|\nu|$ arise from the two distinct solutions of a quadratic equation satisfied by the first non-trivial coefficient in the expansion. An equivalent approach for computing the eigenvalues is provided by a continued fraction expansion, which is also available in the literature: \cite{doi:10.1142/S0219530504000023}. Alternatively, one can directly solve equation (\ref{eq: lame std}) as a power series in $\mathfrak{q}^{1/2}$ by expanding both the eigenfunction and the eigenvalue accordingly. Indeed, considering the Fourier expansion (\ref{eq: p fourier}), the equation takes the form:
\begin{equation}
\label{eq: lamé fourier}
    \Big(\partial_{x}^{2}+4\mu(\mu-1)\sum_{k\geq 1}2k\,\cos2kx\,\frac{\mathfrak{q}^{k/2}}{1-\mathfrak{q}^{k}}+\text{\Large $\mathfrak{e}$}\Big)\psi=0,
\end{equation}
where the eigenvalue is expanded as:
\begin{equation}
\text{\Large $\mathfrak{e}$}(\nu,\mu;\mathfrak{q}) = \sum_{n\geq 0}e_{n}(\nu,\mu)\mathfrak{q}^{n/2}.
\end{equation}
To proceed, we must specify the form of the eigenfunction $\psi(x)$, distinguishing between the case of non-integer and integer $\nu$, which correspond respectively to quasi-periodic and (anti-)periodic solutions. In the first case, we can write:
\begin{equation}
    \psi_{\pm}(x,\nu,\mu;\mathfrak{q})=e^{\pm i \nu x}\sum_{n\geq 0}\psi_{n}(x,\nu,\mu)\mathfrak{q}^{n/2},
\end{equation}
where the coefficients $\psi_{n}(x,\nu,\mu)$ are periodic functions of period $\pi$, more specifically Laurent polynomials in $e^{ 2 i x}$. Equivalently, we can consider the even and odd combinations:
\begin{subequations}
\begin{align}
&\psi_{even}(x,\nu,\mu;\mathfrak{q})=\psi_{+}(x,\nu,\mu;\mathfrak{q})+\psi_{-}(x,\nu,\mu;\mathfrak{q}),\\[1ex]
&\psi_{odd}(x,\nu,\mu;\mathfrak{q})=\psi_{+}(x,\nu,\mu;\mathfrak{q})-\psi_{-}(x,\nu,\mu;\mathfrak{q}).
\end{align}
\end{subequations}
Meanwhile, for a fixed integer value of $\nu$, we can write:
\begin{subequations}
\begin{align}
&\psi_{even}^{(\nu)}(x,\mu;\mathfrak{q})\propto \cos(|\nu|x)+\sum_{n\geq 1}\psi_{even,\,n}^{(\nu)}(x,\mu)\mathfrak{q}^{n/2},\\[1ex]
&\psi_{odd}^{(\nu)}(x,\mu;\mathfrak{q})\propto \sin(|\nu|x)+\sum_{n\geq 1}\psi_{odd,\,n}^{(\nu)}(x,\mu)\mathfrak{q}^{n/2},
\end{align}
\end{subequations}
where each coefficient $\psi_{even,\,n}^{(\nu)}$ and $\psi_{odd,\,n}^{(\nu)}$ is a linear combination of elementary trigonometric functions. For example, the solution associated with a vanishing Floquet exponent $\nu=0$ takes the form:
\begin{equation}
    \psi^{(0)}(x)\propto 1 + 2 \mu(\mu-1)\cos\,2x \,\sqrt{\mathfrak{q}} + 
 \frac{1}{2} \mu (-2 + 3 \mu - 2 \mu^2 + \mu^3) \cos\,4x\,\mathfrak{q} +O(\mathfrak{q}^{3/2}),
\end{equation}
up to a choice of initial condition.
\subsection{\texorpdfstring{Consistency check at $\tau\to +i\infty$}{Check at tau to infinity}}
\label{app: checks 2}
In this sub-appendix we perform a consistency check of the quantization condition (\ref{eq: trmb=nudual}) in the $\tau\to +i \infty$ limit, where the problem reduces to the analysis of the hyperbolic Pöschl–Teller potential \cite{practicalQM}. Indeed, consider the Lamé equation:
\begin{equation}
    -\partial_{x}^{2}\psi+\mu(\mu-1)\wp(x;1,\tau)\psi=E\psi,
\end{equation}
Using the transformation properties of the Weierstrass $\wp$-function, we can write:
\begin{equation}
    \wp(x;1,\tau)=t^{-2}\wp(x/t;1/t,i),
\end{equation}
where $t=\Im(\tau)$, so that, performing the change of variable $z=x/t$, the equations becomes: 
\begin{equation}
     -\partial_{z}^{2}\psi+\mu(\mu-1)\wp(z;1/t,i)\psi=t^{2}E\psi.
\end{equation}
The Weierstrass $\wp$-function can be represented as:
\begin{equation}
    \wp(z;1/t,i)=t^{2}\left(\frac{\pi^{2}}{\sin^{2}\pi z t}-\frac{\pi^{2}}{3}+\sum_{k\neq 0}\frac{\pi^{2}}{\sin^{2}\pi t(z-k i)}-\frac{\pi^{2}}{\sin^{2}\pi tk i}\right),
\end{equation}
and in the $t \to +\infty$ limit we can approximate it by:
\begin{equation}
    \wp(z;1/t,i)\simeq t^{2}\left(\frac{\pi^{2}}{\sin^{2}\pi z t}-\frac{\pi^{2}}{3}\right).
\end{equation}
Setting now $z=i v +1/2t$, so as to lie along the vertical line at the center of the torus, we find:
\begin{equation}
    \left[-\partial_{v}^{2}-\mu(\mu-1)\frac{t^{2}\pi^{2}}{\cosh^{2}\pi t v}+t^{2}\left(E+\frac{\pi^{2}}{3}\mu(\mu-1)\right)\right]\psi=0.
\end{equation}
Up to exponentially small corrections, we can approximate the potential on the B-cycle as an infinite sum of such localized contributions centered at $v \in \mathbb{Z}$, with the previous equation serving as the effective description in the vicinity of $v=0$. To solve it, we make the substitutions:
\begin{equation}
    y = \cosh^{2}\pi t v\,, \quad E=-\kappa^{2}\pi^{2}-\frac{\pi^{2}}{3}\mu(\mu-1),
\end{equation}
obtaining the hypergeometric equation:
\begin{equation*}
y\left(1-y\right)\psi''+\left(\frac{1}{2}-y\right)\psi'-\frac{1}{4}\left(\frac{\mu(\mu-1)}{y}+\kappa^{2}\right)\psi=0 .
\end{equation*}
Setting then $\psi=y^{\mu/2}\phi$, we find:
\begin{equation*}
    y(1-y)\phi''+\left(\frac{1}{2}+\mu-y(1+\mu)\right)\phi'-\frac{1}{4}\left(\kappa^{2}+\mu^{2}\right)\phi=0,
\end{equation*}
which we can easily solve near $y=1$, or $v=0$, in terms of hypergeometric functions. Expressing the solutions in terms of the variable $v$, we find:
\begin{subequations}
\begin{align}
&\psi_{even}(v)=\cosh^{\mu}\pi t v 
\ _2F_1\left(\frac{\mu+i\kappa}{2},\frac{\mu-i\kappa}{2},\frac{1}{2};-\sinh^{2}\pi tv\right),\\
&\psi_{odd}(v)=\sinh\pi t v\,\cosh^{\mu}\pi t v \ _2F_1\left(\frac{1+\mu+i\kappa}{2},\frac{1+\mu-i\kappa}{2},\frac{3}{2};-\sinh^{2}\pi tv\right),
\end{align}
\end{subequations}
distinguishing the two linearly independent solutions by parity. Then, in the vicinity of the other integer $v$ points, we will have similar local solutions. For example, near $v=1$, we would find: $\Tilde{\psi}_{even/odd}(v)=\psi_{even/odd}(1-v)$. To obtain the desired quantization condition, we must impose the periodicity of the solutions, which can be achieved by matching their asymptotic behaviors. Indeed, assuming $t>>1$, for $0<v<1$, we have:
\begin{equation}
\begin{aligned}
\psi_{even}(v) &\simeq\ 
\frac{\Gamma\left(\frac{i\kappa}{2}\right) \Gamma\left(\frac{1}{2} + \frac{i\kappa}{2}\right)}{\Gamma\left(\frac{i\kappa}{2} + \frac{\mu}{2}\right) \Gamma\left(\frac{1}{2} - \frac{\mu}{2} + \frac{i\kappa}{2}\right)} e^{i \kappa \pi t v} +\frac{\Gamma\left(-\frac{i\kappa}{2}\right) \Gamma\left(\frac{1}{2} - \frac{i\kappa}{2}\right)}{\Gamma\left(-\frac{i\kappa}{2} + \frac{\mu}{2}\right) \Gamma\left(\frac{1}{2} - \frac{\mu}{2} - \frac{i\kappa}{2}\right)} e^{-i \kappa \pi t v}\\[4pt]
&=\Omega_{even}e^{i\kappa \pi t v}+\Bar{\Omega}_{even}e^{-i\kappa \pi t v},
\end{aligned}
\end{equation}
and imposing (anti-)periodicity, namely $\psi_{even}(v)=\pm\psi_{even}(v+1)$, we obtain:
\begin{equation}
\label{eq: qc limit even}
    e^{i\kappa \pi t}\frac{\Omega_{even}}{\Bar{\Omega}_{even}}=\pm 1.
\end{equation}
Similarly, for the odd solution we have:
\begin{equation}
\begin{aligned}
\psi_{\text{odd}}(v) &\simeq \frac{1}{4}  \frac{\Gamma\left(\frac{i\kappa}{2}\right) \Gamma\left(\frac{1}{2} + \frac{i\kappa}{2}\right)}{\Gamma\left(\frac{1}{2} + \frac{i\kappa}{2} + \frac{\mu}{2}\right) \Gamma\left(1 - \frac{\mu}{2} + \frac{i\kappa}{2}\right)} e^{i \kappa \pi t v}
+\frac{1}{4}  \frac{\Gamma\left(-\frac{i\kappa}{2}\right) \Gamma\left(\frac{1}{2} - \frac{i\kappa}{2}\right)}{\Gamma\left(\frac{1}{2} - \frac{i\kappa}{2} + \frac{\mu}{2}\right) \Gamma\left(1 - \frac{\mu}{2} - \frac{i\kappa}{2}\right)} e^{-i \kappa \pi t v} \\[4pt]
&= \Omega_{\text{odd}} e^{i \kappa \pi t v} + \Bar{\Omega}_{\text{odd}} e^{-i \kappa \pi t v},
\end{aligned}
\end{equation}
and imposing $\psi_{odd}(x)=\pm\psi_{odd}(x+1)$, we find the quantization condition:
\begin{equation}
\label{eq: qc limit odd}
    e^{i\kappa \pi t}\frac{\Omega_{odd}}{\Bar{\Omega}_{odd}}=\mp 1 .
\end{equation}
By solving equation (\ref{eq: trmb=nudual}) for $e^{\eta/2}$ in the case of periodic or anti-periodic solutions, corresponding to $\cos\nu_D=\pm1$, we obtain:
\begin{subequations}
\begin{align}
& e^{\eta/2}= \pm \frac{\sin2\pi a - \sin\pi \mu}{\sin\pi(2a-\mu)},\label{eq: qc limit 1}\\
& e^{\eta/2}= \pm \frac{\sin2\pi a + \sin\pi \mu}{\sin\pi(2a-\mu)},\label{eq: qc limit 2}
\end{align}
\end{subequations}
where 
\begin{equation}
    \eta = 4\pi a t+2\log\frac{\Gamma(\mu-2a)\Gamma(1+2a)}{\Gamma(\mu+2a)\Gamma(1-2a)},
\end{equation}
neglecting corrections that are exponentially suppressed in the limit $t \to +\infty$. Upon identifying $\kappa = 2a/i$, we find that the quantization condition (\ref{eq: qc limit even}) matches with (\ref{eq: qc limit 1}), while the quantization condition (\ref{eq: qc limit odd}) corresponds to (\ref{eq: qc limit 2}).

\section{Blow-up equations}
\label{app: blow-up}
\subsection{\texorpdfstring{Useful formulae for $N_f>0$}{Formulas for Nf>0.}}
\label{app: blow-up formulas}
In this subsection, we collect some lengthy formulas relevant to the $\mathbb{C}^2$ and $\mathbb{C}^2/\mathbb{Z}_2$ blow-up equations for generic $N_f\leq 4$. First, the regular parts of the resummation ansatz for $N_f = 4$ are given by \cite{Poghossian:2009mk}:
\begin{equation}
\label{eq: reg0}
\begin{aligned}
&R_{0}^{[4]}(a,\{\mu_i \},\hbar;t)=\frac{a^2}{\hbar}\left(\frac{\pi  K(1-t)}{K(t)}+\log \left(\frac{t}{16}\right)\right)+\frac{1}{4\hbar}\left(\sum_{i=1}^{4}\mu_i\right)^{2}\log (1-t)\\
&+\frac{1}{\hbar}\left(\sum_{i=1}^{4}\mu_{i}^{2}\right)\log\left(\frac{2}{\pi}K(t) \right)+\frac{1}{2}\left(\sum_{i=1}^{4}\mu_i\right)\log(1-t)+\frac{\hbar}{4}\left(\log (1-t)-2 \log \left(\frac{2 K(t)}{\pi }\right)\right)\\
\end{aligned}
\end{equation}
\begin{equation}
\label{eq: reg1}
R_{1}^{[4]}(\{\mu_i \},\hbar;t)=\frac{1}{2} \left(\log (1-t)-2 \log \left(\frac{2 K(t)}{\pi }\right)\right)+\frac{1}{2\hbar}\left(\sum_{i=1}^{4}\mu_i\right)\log(1-t).
\end{equation}
From these expressions, it is possible to derive all other regular parts for $N_f \leq 3$ by decoupling the appropriate number of hypermultiplets. The function $R_{1}^{[4]}(\{\mu_i \},\hbar;t)$ is determined in terms of $R_{0}^{[4]}(\{\mu_i \},\hbar;t)$, for example by taking the large $a$ limit of the blow-up equation (\ref{eq: blow-up 1 pure ns}). Indeed, in this limit, the equation reduces to:
\begin{equation}
    t\partial_{t}\left(\mathcal{W}_{1,\, inst}(a,\{\mu_i\},\hbar;t)-\left(\partial_{\hbar}+\frac{1}{\hbar}\right)\mathcal{W}_{0,\, inst}(a,\{\mu_i\},\hbar;t)\right)\simeq 0
\end{equation}
thus fixing $R_{1}^{[4]}$ up to a $t$-independent term, which can then be seen to vanish. Then, writing: 
\begin{equation*}
\widetilde{\mathcal{G}}_{exp/inst}^{\,\mp,\,[N_f]} = \widetilde{\mathcal{G}}_{exp/inst}^{[N_f]}\left(\frac{m_{1}x+t^{m_{2}/2}}{2x},\{\mu_{i}\},\mp j -m_{1},t\right)
\end{equation*}
the expressions necessary to determine the differential equations satisfied by the resummation functions read:
\begin{equation}
\label{eq: Gexp Nf}
\begin{aligned}
&\log \widetilde{\mathcal{G}}_{exp}^{\,\mp,\,[N_f]}=2t^{k-1} \bigg\{
\tilde{f}_{k,j} 
\mp m_1 y^2\, \partial_{y}g_{k,j}
\mp t^{j/2} \bigg(\big[1-(4-N_f)(k+j/2-1)\big]g_{k,j} \\
&+\big[1-(4-N_f)j/2\big]y\,  \partial_{y}g_{k,j}-\tilde{g}_{k,j}- \sum_{i=1}^{N_{f}}\mu_i\, \partial_{\mu_i} g_{k,j}
\bigg)\bigg\}    + \sigma_{exp}^{\,\mp,\,[N_f]}(x,\{\mu_{i}\},t),
\end{aligned}
\end{equation}
\begin{equation}
\label{eq: Ginst Nf}
\begin{aligned}
&\widetilde{\mathcal{G}}_{inst}^{\,\mp,\,[N_f]}=\frac{j}{2} (j \pm m_1) 
+ t^{k-1} \bigg\{ (k - 1)\, \tilde{f}_{k,j}
+j \frac{y}{2} \,  \partial_{y}\tilde{f}_{k,j} \\
&\mp m_1 y \left[
(j + k - 1)\,  \partial_{y}g_{k,j}
+ j \frac{y}{2}\,  \partial_{y}^2g_{k,j}
\right]
\pm t^{j/2} \Big[
(k+j/2 - 1)\, \tilde{g}_{k,j}\\
&+ j\frac{y}{2}\,  \partial_{y}\tilde{g}_{k,j}-\big[2-(4-N_f)(k+j/2-1)\big](k+j/2 - 1)\, g_{k,j}\\
&- \big[k+2j-1-j(4-N_f)(k+3j/4-1)\big]y\,  \partial_{y}g_{k,j}
- j\big(1-\frac{j}{2}(4-N_f)\big)\frac{y^2}{2}\,  \partial_{y}^2g_{k,j}\\
& + \sum_{i=1}^{N_{f}} \mu_i \left(
(k+j/2 - 1)\, \partial_{\mu_i} g_{k,j}+
+ j\frac{y}{2}\,  \partial_{\mu_i} \partial_{y}g_{k,j}
\right)
\Big]\bigg\}+ \sigma_{inst}^{\,\mp,\,[N_f]}(x,\{\mu_{i}\},t),
\end{aligned}
\end{equation}
where $y=x t^{(j-m_{2})/2}$.  Assuming the asymptotic behavior of the resummation functions as in Section \ref{sec: results}, we find again that the expansions of the blow-up equations under consideration are proper series expansions. Finally, if $\varsigma_{m_{2}}=\frac{1-(-1)^{m_2}}{2}$ and we define:
\begin{equation}
\label{eq: c tilde}
    \tilde{c}_{m_{1},m_{2}}(\mu)=\mu^{m_{1}\varsigma_{m_{2}}}\prod_{j=\frac{1}{2}+\frac{\varsigma_{m_{2}}}{2}}^{\frac{m_2 -1}{2}}(\mu-j)^{m_1}(\mu+j)^{m_1},
\end{equation}
we have:
\begin{equation}
\label{eq: constant c Nf}
    d_{m_1,m_2}(\{\mu_{i}\})=-\frac{\prod_{i=1}^{N_f}\tilde{c}^{\,2}_{m_1, m_2}(\mu_i)}{m_{2}^{4m_1}[(m_{2}-1)!]^{8m_1}}.
\end{equation}
\subsection{\texorpdfstring{Example: $\mathbb{C}^2$ blow-up equations, $\mathcal{N}=2^{*}$ theory}{C2, N2star theory}}
\label{app: blow-up example 1}
In this section, we present and solve the first few equations derived from the expansion of the $\mathbb{C}^2$ blow-up equation \eqref{eq: C2 blow-up lame expl}. We focus on this case because the presence of both integer and half-integer shifts allows us to work with lower values of \( m_1 \), thereby reducing the complexity introduced by the Taylor coefficients of the resummation functions appearing in the equations. Specifically, we analyze the equations for $m_2=1$ and \( m_1 = 1, 2 \). The first two of these read, respectively:
\begin{subequations}
\begin{align}
& e^{2 \tilde{f}_{1,1}(x, \mu) + x^2 \partial_{x}g_{1,1}(x, \mu)} + \mu^2(\mu - 1)^2 x^2 \, e^{2 \tilde{f}_{1,1}(x, \mu) - x^2 \partial_{x}g_{1,1}(x, \mu)} - 1 = 0, \\
& e^{2 \tilde{f}_{1,1}(x, \mu) + 2x^2 \partial_x g_{1,1}(x, \mu) } 
- \mu^4(\mu - 1)^4 x^4 \, e^{2 \tilde{f}_{1,1}(x, \mu) - 2 x^2 \partial_x g_{1,1}(x, \mu)} 
- 1 = 0.
\end{align}
\end{subequations}
Performing the substitutions \(\tilde{f}_{1,1} = \frac{1}{2} \log u_1\) and \(g_{1,1}' = \frac{1}{x^2} \log(x v_1)\), and solving for \(u_1\), we obtain:
\begin{subequations}
\begin{align}
&u_1(x,\mu)=\frac{1}{ x\left(v_1(x, \mu) +\mu^2(-1 + \mu)^2 \,  v_1(x, \mu)^{-1}\right)},\\
&u_1(x,\mu) =\frac{1}{x^2 \left(v_1(x, \mu)^2- \mu^4(-1 + \mu)^4  \, v_1(x, \mu)^{-2}  \right)}
\end{align}
\end{subequations}
so that, by equating the two expressions for $ u_1 $, we find:
\begin{equation}
    v_1(x,\mu)=\frac{1\pm\sqrt{1+4 \mu ^2(\mu -1)^2 x^2}}{2 x},
\end{equation}
where only the choice of the $+$ sign in front of the square root is compatible with the boundary condition $g_{1,1}(0, \mu) = 0$. Thus, we find the following expressions for the first two resummation functions:
\begin{subequations}
\begin{align}
&\tilde{f}_{1,1}(x,\mu)=-\frac{1}{4} \log \left(1+4\mu ^2 (\mu -1)^2 x^2\right)\\
&g_{1,1}(x,\mu) = \frac{
\sqrt{1 + 4 \mu^2 (\mu - 1)^2 x^2} 
- \log\left( \tfrac{1}{2} + \tfrac{1}{2}\sqrt{1 + 4 \mu^2 (\mu - 1)^2 x^2} \right) - 1
}{x}.
\end{align}
\end{subequations}
Using the explicit forms of $\tilde{f}_{1,1}$ and $g_{1,1}$, and after some simplification, the second equation for $m_1 = 1$ takes the form:
\begin{equation}
    \left(1+ \sqrt{1+4\mu^2 (\mu - 1)^2 x^2} \right) \tilde{g}_{1,1}(x, \mu)
+ 2 \mu \left( \mu^3 - 2\mu + 1 \right) =0
\end{equation}
which can easily be solved to obtain:
\begin{equation}
\tilde{g}_{1,1}(x,\mu) = \frac{(-1 + \mu + \mu^2)\left(1- \sqrt{1 + 4 \mu^2 (\mu - 1)^2 x^2}  \right)}{2 \mu (\mu - 1) x}.
\end{equation}
Then, the second equation for $m_1=2$ is identically satisfied. Moving on to the third equation for $m_1 = 1$, and expanding it to the first few orders around $x = 0$, one obtains the following relation:
\begin{equation}
\begin{aligned}
\partial_x^2 \tilde{f}_{1,2}(\mu, 0)&=\frac{1}{4(\mu + 1)} \left[
- \mu^2 (\mu - 1)^2 ( -20 + 56\mu - 27\mu^2 - 45\mu^3 + 19\mu^4 + \mu^5 )\right.\\
&\left.
+\, 8 \, \partial_x^2 \tilde{f}_{2,1}(\mu, 0)
- 4 (\mu + 1) \, \partial_x g_{1,2}(\mu, 0)
+ 8 \, \partial_x g_{2,1}(\mu, 0)
\right]
\end{aligned}
\end{equation}
which simplifies the equation and allows it to be easily solved for $\tilde{f}_{2,1}$, yielding:
\begin{equation}
\tilde{f}_{2,1}(x, \mu)=-\frac{x^2 \, }{2 \sqrt{1+4 \mu^2 (\mu - 1)^2 x^2 }}\partial_x g_{2,1}(x, \mu)+ \dots
\end{equation}
where the remaining terms form a lengthy expression in $x$ and $\mu$, involving only $\partial_x^2 \tilde{f}_{2,1}(\mu, 0)$ and $\partial_x g_{2,1}(\mu, 0)$. Similarly, expanding the third equation for $m_1=2$ around $x=0$, we find the following relation:
\begin{equation}
\begin{aligned}
\partial_x g_{1,2}(\mu,0)=&\frac{1}{4(\mu + 1)} \left[
(\mu - 1)^2 \left(2 \mu^6 - 5 \mu^5 + 17 \mu^4 - 31 \mu^3 - 29 \mu^2 + 50 \mu - 16\right) \mu^2
\right.\\
&\left.
+ \left(-\mu^2 + \mu - 6\right) \partial_x^2 \tilde{f}_{2,1}(\mu, 0)
- 2 \left(\mu^2 - \mu + 2\right) \partial_x g_{2,1}(\mu, 0)
\right]
\end{aligned}
\end{equation}
which can be used to simplify it. Substituting the expression of $\tilde{f}_{2,1}$ as a function of $g_{2,1}$, one can easily solve the equation to find:
\begin{equation}
\begin{aligned}
&\partial_x g_{2,1}(\mu,x)=\frac{1}{2}\bigg[
\mu^4(\mu - 1)^4 \left(5 \mu^4 - 10 \mu^3 - 7 \mu^2 + 12 \mu - 6\right)+ 2\, \partial_x g_{2,1}(\mu,0)
\\
&
+ x^2 \left(6 \mu^2(\mu - 1)^2 \, \partial_x g_{2,1}(\mu,0) \right)
+ \bigg(2\, \partial_x g_{2,1}(\mu,0)+ 2 (\mu - 1)^2 \mu^2 \, \partial_x g_{2,1}(\mu,0) x^2\\
& +(\mu - 1)^4 \mu^4 \left(\mu^4 - 2 \mu^3 - 3 \mu^2 + 4 \mu - 2\right) x^2\bigg) \sqrt{1 + 4 \mu^2 (\mu - 1)^2 x^2} \bigg]\cdot \\
&
\left[
1 + 5 \mu^2 (\mu - 1)^2 x^2
+ 4 \mu^4 (\mu - 1)^4 x^4
+ \left(1 + 3 \mu^2 (\mu - 1)^2 x^2\right) \sqrt{1 + 4 \mu^2 (\mu - 1)^2 x^2}
\right]^{-1}.
\end{aligned}
\end{equation}
Expanding the previous function for $x \to +\infty$, we find at leading order:
\begin{equation}
\partial_x g_{2,1}(\mu, x) \simeq \frac{1}{4 \mu (\mu - 1)} \bigg(
2\, \partial_x g_{2,1}(\mu, 0)- \mu^2 (\mu - 1)^2 \left(2 - 4\mu + 3\mu^2 + 2\mu^3 + \mu^4 \right)
\bigg)x^{-1}
\end{equation}
which is clearly incompatible with the expansion \eqref{eq: C2 Gexp gen}, thereby forcing $\partial_x g_{2,1}(\mu, 0)$ to take the form:
\begin{equation}
\partial_x g_{2,1}(\mu, 0)
= -\frac{1}{2} (\mu - 1)^2 \mu^2 \left( \mu^4 - 2\mu^3 - 3\mu^2 + 4\mu - 2 \right).
\end{equation}
Thus, we obtain:
\begin{equation}
\begin{aligned}
g_{2,1}(\mu,x)&=-\frac{2 (\mu -2) (\mu -1) (\mu +1) \mu-2 +3 (\mu -1)^4 \mu ^4 x^2}{12 (\mu -1)^2 \mu ^2 x^3}\\
&-\frac{\sqrt{1+4 (\mu -1)^2 \mu ^2 x^2}}{12 (\mu -1)^2 \mu ^2 x^3}\Big[2-2 \mu ^4+4 \mu ^3+2 \mu ^2-4 \mu\\
&+\left(\mu ^8-4 \mu ^7-2 \mu ^6+20 \mu ^5-27 \mu ^4+16 \mu ^3-4 \mu ^2\right) x^2\Big],
\end{aligned}
\end{equation}
\begin{equation}
    \tilde{f}_{2,1}(\mu,x)= \frac{\left(3 \mu ^8-12 \mu ^7+10 \mu ^6+12 \mu ^5-25 \mu ^4+16 \mu ^3-4 \mu ^2\right) x^2}{2\big(1+4 (\mu -1)^2 \mu ^2 x^2\big)}.
\end{equation}
Moving on to the next equation for $m_1 = 1$, and expanding around $x = 0$ to the first few orders, we find the relations:
\begin{subequations}
\begin{align}
\partial_x^2 \tilde{f}_{1,2}(\mu, 0) &= -\frac{1}{2} \mu^2 (\mu + 1)^2 \left(\mu^2 - 3\mu + 2\right)^2, \\
\partial_x g_{1,2}(\mu, 0) &= \frac{1}{4} \mu^2 (\mu + 1)^2 \left(\mu^2 - 3\mu + 2\right)^2,
\end{align}
\end{subequations}
and
\begin{equation}
\begin{aligned}
\partial_\mu \partial_x g_{1,2}(\mu, 0)
&= \frac{-8 (\mu + 1) \partial_x \tilde{g}_{1,2}(\mu, 0)
+ 16 \partial_x \tilde{g}_{2,1}(\mu, 0)}{4 \mu (\mu + 1)} \\
&+ \frac{(\mu - 1) \left(7 \mu^7 - 42 \mu^6 + 28 \mu^5 + 118 \mu^4 - 67 \mu^3 - 68 \mu^2 + 48 \mu - 8\right)}{4 (\mu + 1)},
\end{aligned}
\end{equation}
which allow us to simplify the equation. Then, the equation, now involving only the remaining Taylor coefficient $\partial_x \tilde{g}_{2,1}(\mu, 0)$, can be readily solved for $\tilde{g}_{2,1}$, which at leading order as $x \to +\infty$ reads:
\begin{equation}
\begin{aligned}
    \tilde{g}_{2,1}(\mu, x) \simeq\; & 
- \mu(\mu - 1) \Big( \mu(\mu -1)  \left(2 - 6\mu + 4\mu^2 + 9\mu^3 - 9\mu^4 - 5\mu^5 + 3\mu^6\right)  \\
& - 2\, \partial_x \tilde{g}_{2,1}(\mu, 0)
\Big)x^2 + \mathcal{O}(1).
\end{aligned}
\end{equation}
Again, this is clearly incompatible with the expansion \eqref{eq: C2 Gexp gen}, and thus we find:
\begin{equation}
   \partial_x \tilde{g}_{2,1}(\mu, 0) =\frac{1}{2} \mu(\mu -1)  \left(2 - 6\mu + 4\mu^2 + 9\mu^3 - 9\mu^4 - 5\mu^5 + 3\mu^6\right).
\end{equation}
which, when substituted into $\tilde{g}_{2,1}(\mu, x)$, yields:
\begin{equation}
\begin{aligned}
&\tilde{g}_{2,1}(\mu,x)= \big(12 (\mu - 1)^3 \mu^3 x^3\big)^{-1}\Big[
  2 \mu^6 - 6 \mu^5 + 6 \mu^4 + 2 \mu^3 - 6 \mu^2 + 6 \mu
  - 2\\
&+ \left(6 \mu^{10} - 24 \mu^9 + 33 \mu^8 - 12 \mu^7 - 12 \mu^6 + 12 \mu^5 - 3 \mu^4\right) x^2\Big]\\
&+\left(12 (\mu - 1)^3 \mu^3 x^3 \sqrt{1+4 (\mu - 1)^2 \mu^2 x^2} \right)^{-1}\Big[ -2 \mu^6 + 6 \mu^5 - 6 \mu^4 - 2 \mu^3\\
&+ 6 \mu^2 - 6 \mu + 2+ \big(
    10 \mu^{14} - 58 \mu^{13} + 84 \mu^{12} + 90 \mu^{11}
    - 372 \mu^{10} + 354 \mu^9\\
&- 20 \mu^8 - 202 \mu^7 + 162 \mu^6 - 56 \mu^5 + 8 \mu^4
  \big) x^4\\
&+ \left(
    -10 \mu^{10} + 44 \mu^9 - 73 \mu^8 + 44 \mu^7 + 20 \mu^6
    - 52 \mu^5 + 43 \mu^4 - 20 \mu^3 + 4 \mu^2
  \right) x^2\Big],
\end{aligned}
\end{equation}
while the corresponding equation for $m_1 = 2$ vanishes identically. This procedure can be straightforwardly extended to higher orders in the expansion, and to different values of $m_1$ and $m_2$, with no essential change in the steps involved.

\subsection{\texorpdfstring{Example: $\mathbb{C}^2/\mathbb{Z}_2$ blow-up equations, $N_f=0$ theory}{C2Z2, pure gauge theory}}
\label{app: blow-up example 2}
In this section we show and solve the first few differential equations obtained from the expansion of the blow-up equation (\ref{eq: blow-up 1 explicit 2}). This is instructive in order to discuss a couple of subtleties which are not manifest from the general form of the equations presented in the main part of this work. In particular, we consider the expansions for $m_2=1$ and $m_{1}=1,2$, up to the first six equations. The very first equation for $m_{1}=1$ reads:
\begin{equation}
\begin{aligned}
& e^{- 2 x^2  g_{1,1}'(x)}
   x^3 \left(-2 - x \tilde{f}_{1,1}'(x) + 
    2 x^2 g_{1,1}'(x) + x^3 g_{1,1}''(x) \right)\\
&+ e^{2 x^2 g_{1,1}'(x)}
    \left( \tilde{f}_{1,1}'(x) + 
    2x   g_{1,1}'(x) + x^{2}  g_{1,1}''(x)  \right)=0,
\end{aligned}
\end{equation}
and can be solved algebraically for $ \tilde{f}_{1,1}'$. Integrating the solution and imposing the vanishing of the function at $x=0$, we find: 
\begin{equation}
\label{eq: f11}
    \tilde{f}_{1,1}(x)=-\frac{1}{2}\log (e^{2x^{2} g_{1,1}'(x)}-x^{4}e^{-2x^{2} g_{1,1}'(x)}),
\end{equation}
or equivalently, expressing the solution in terms of $u_1(x)$ and $\tilde{v}_1(x)$, where $\tilde{f}_{1,1}=\frac{1}{2}\log u_1$ and $g_{1,1}'=\frac{1}{2x^{2}}\log(x^{2} \tilde{v}_1)$, we can write:
\begin{equation}
\label{eq: u1 1}
    u_1(x)=\frac{1}{x^{2}}\frac{1}{\tilde{v}_1-\tilde{v}_1^{-1}}\,.
\end{equation}
Considering now the first equation for $m_{1}=2$, we find:
\begin{equation}
\begin{aligned}
&1 + e^{2 \tilde{f}_{1,1}(x) - 4x^2 g_{1,1}'(x)} \cdot x^8 \left( 
    -3 - x \tilde{f}_{1,1}'(x) + 4x^2 g_{1,1}'(x) + 2x^3 g_{1,1}''(x) 
\right) \\
&+e^{2 \tilde{f}_{1,1}(x) + 4x^2 g_{1,1}'(x)} \cdot \left( 
    -1 + x \tilde{f}_{1,1}'(x) + 4x^2 g_{1,1}'(x) + 2x^3 g_{1,1}''(x) 
\right)=0.
\end{aligned}
\end{equation}
Writing again everything in terms of $u_1$, $\tilde{v}_1$ and solving for the former, we find:
\begin{equation}
\label{eq: u1 2}
    u_1(x)=\frac{1-c_{1}x^{2}}{x^{4}}\frac{1}{\tilde{v}_1^{2}-\tilde{v}_1^{-2}}\,,
\end{equation}
where now the integration constant $c_{1}$ cannot be fixed by requiring $u_1(0)=1$ and $\tilde{v}_1(x)\simeq 1/x^{2}$, i.e. $\tilde{f}_{1,1}(0)=g_{1,1}(0)=0$. Solving the quadratic equation obtained from (\ref{eq: u1 1}) and (\ref{eq: u1 2}), we find:
\begin{equation}
\label{eq: v(x) 1}
    \tilde{v}_1(x)=\frac{1}{2x^{2}} \left( 1 - x^2 c_1 \pm \sqrt{\left(x^2 c_1-1\right)^2-4x^4 } \right).
\end{equation}
In order to satisfy $\tilde{v}_1(x)\simeq 1/x^{2}$, we must chose the plus sign in front of the square root. Then, $c_{1}$ is related to the value of $g_{1,1}'$ at zero by $c_{1}=-2g_{1,1}'(0)$, and by consequence of $\tilde{f}_{1,1}''(0)=-2g_{1}'(0)$, also to the value of $\tilde{f}_{1,1}''$ at zero. To determine $c_{1}$, we can consider the second equation for $m_{1}=1$. Writing $\tilde{f}_{1,1}(x)$ as in equation (\ref{eq: f11}) and $g_{1,1}'(x)$ in terms of $\tilde{v}_1(x)$, using (\ref{eq: v(x) 1}), the resulting equation reads:
\begin{equation}
\label{eq: g1tilde}
\begin{aligned}
&\left(1 - 2 x^2 c_1 + x^4 (c_{1}^{2}-4)\right)^{-2} \bigg\{ x c_1 
+ x^3(4 - 3c_1^2) 
+ 3x^5 c_1 (c_1^2-4) 
- x^7 (c_1^2-4)^2 \\
& + \sqrt{1 - 2x^2 c_1 + x^4 (c_1^2-4)} \cdot \Bigg[
  -4x^3 \left(1 + \log \frac{1}{2} \left( 1 - x^2 c_1 + \sqrt{ (x^2 c_1-1)^2-4x^4 } \right) \right)  \\
& - 8x^4 g_{1,1}(x)+ \left( -1 + 3x^2 c_1 + x^6 c_1 (c_1^2-4) - x^4 (4 + 3c_1^2) \right) \tilde{g}_{1,1}(x) \\
& + x(-1 + x^2(c_1-2))(x^2 c_1-1)(x^2(2 + c_1)-1)  \tilde{g}_{1,1}'(x)
\Bigg]\bigg\}=0.
\end{aligned}
\end{equation}
For simplicity, we do not substitute $g_{1,1}(x)$ yet. We now look at the equation near $x=+\infty$, setting $\tilde{g}_{1,1}(x)=\tilde{k}+O(x^{-1})$, $g_{1,1}(x)=k+O(x^{-1})$. The asymptotic behavior of $g_{1,1}(x)$ is dictated by that of its first derivative, which we just determined up to $c_{1}$, while that of $\tilde{g}_{1,1}(x)$ is dictated by the equation itself. We obtain:
\begin{equation}
\label{eq: integration constant c1}
    \frac{c_{1}\tilde{k}}{\sqrt{c_{1}^{2}-4}}-x^{-1} +\mathcal{O}(x^{-2})=0.
\end{equation}
Clearly, any choice of $c_{1}\neq \pm 2$ will not allow us to solve the equation. Assuming instead $c_{1}^{2}=4$, we see that the proper choice is $c_{1}=-2$, since $c_{1}=2$ would imply that $\tilde{v}_1(x)$ as in (\ref{eq: v(x) 1}) becomes imaginary when $x^{2}>1/4$. In view of the next section, we note here that the correct choice is the one for which the coefficient of $x^{4}$ inside the square root in (\ref{eq: v(x) 1}) vanishes, while also maintaining the function real on the real axis. With this choice, we find:
\begin{equation}
    \tilde{f}_{1,1}(x)=-\frac{1}{4}\log(1+4x^{2}) ,\quad g_{1,1}(x)=- \frac{\log\left( 1/2 + \sqrt{1/4 + x^2} \right) + 1 - \sqrt{1 + 4x^2}}{x},
\end{equation}
and equation (\ref{eq: g1tilde}) for $\tilde{g}_{1,1}(x)$ simplifies to:
\begin{equation}
4x^3 
- 2\sqrt{1 + 4x^2}(x + 4x^3) 
- (1 + 6x^2 + 16x^4)\, \tilde{g}_{1,1}(x) 
- (x + 6x^3 + 8x^5)\, \tilde{g}_{1,1}'(x)=0,
\end{equation}
which we can solve, to obtain:
\begin{equation}
    \tilde{g}_{1,1}(x)=\frac{1-\sqrt{1+4x^{2}}}{2x},
\end{equation}
where the integration constant has been fixed by enforcing $\tilde{g}_{1,1}(0)=0$. Then, having already found $\tilde{g}_{1,1}$, the second equation for $m_{1}=2$ becomes a simple algebraic relation between the Taylor coefficients of the functions for $m_{2}=2$, namely:
\begin{equation}
8 g_{1,2}'(0)+2\tilde{f}_{1,2}''(0)-1=0 \implies\tilde{f}_{1,2}''(0)=\frac{1}{2}-4g_{1,2}'(0),
\end{equation}
which we can use to simplify the next equations. The third equation for $m_{1}=1$ reads:
\begin{equation}
    \begin{aligned}
& 0=\sqrt{1 + 4x^2}-1 
- 2x^2 \left(7 + 6x^2 - 2\sqrt{1 + 4x^2} \right) + 4(1 + 4x^2)^{3/2}  \tilde{f}_{2,1}(x) \\
& + 2x(1 + 4x^2)^{3/2} \tilde{f}_{2,1}'(x) + 8x^2 (1 + 6x^2 + 10x^4)  g_{2,1}'(x) + 2x^2 (1 + 6x^2 + 8x^4)  g_{2,1}''(x),
\end{aligned}
\end{equation}
which we can solve to find:
\begin{equation}
\tilde{f}_{2,1}(x)=- \frac{
  (2 + x^2)\left(-1 - 2x^2 + \sqrt{1 + 4x^2} \right)
  + 4x^4(1 + 2x^2)  g_{2,1}'(x)
}{
  4x^2 \sqrt{1 + 4x^2}
},
\end{equation}
where the integration constant has been fixed by requiring $\tilde{f}_{2,1}(0)=0$. This solution can then be substituted into the third equation for $m_{1}=2$, which simplifies to:
\begin{equation}
    \begin{aligned}
& -1 - 2x^2 + x^4 
- 2(x^4 + 6x^6) g_{2,1}'(x) 
- (x^5 + 4x^7) g_{2,1}''(x) \\
& \quad + \sqrt{1 + 4x^2} \left( 1 + x^3 \left(9 - 40g_{1,2}'(0) - 8\tilde{g}_{1,2}'(0) \right) \right)=0.
\end{aligned}
\end{equation}
Expanding this equation at leading order around $x=0$, we find:
\begin{equation}
(9-40g_{1,2}'(0)-8\tilde{g}_{1,2}'(0))x^{3}+\mathcal{O}(x^{4})\implies \tilde{g}_{1,2}'(0)= \frac{9}{8}-5g_{1,2}'(0),
\end{equation}
which gives us another relation between Taylor coefficients. Using this relation and solving for $g_{2,1}(x)$, we obtain:
\begin{equation}
    g_{2,1}(x)=\frac{
\left(2 - x^2\right) \sqrt{1 + 4x^2} - \left(2 + 3x^2\right)}{12x^3} \quad \implies \quad \tilde{f}_{2,1}(x)=\frac{3}{2}\frac{x^{2}}{1+4x^{2}},
\end{equation}
where now both integration constants of $g_{2,1}$ are fixed by enforcing its vanishing at zero. Then, the next equation for $m_{1}=1$ reads:
\begin{equation}
\begin{aligned}
&3 
+ x^2 \bigg(
10 + x \big(
3 - 10x + 24x^2 - 80x^3 + 48x^4 
- 12(1 + 4x^2)^2 g_{1,2}'(0)
\big)
\bigg)+ \sqrt{1 + 4x^2} \\
&\cdot \bigg(3x^2(1 + 6x^2 + 8x^4)\, \tilde{g}_{2,1}'(x)+ x(9 + 54x^2 + 96x^4)\, \tilde{g}_{2,1}(x) 
-3 - 22x^2 - 36x^4 
\bigg)=0.
\end{aligned}
\end{equation}
Expanding this equation near $x=0$, we find:
\begin{equation}
    \left(18 - 12\, \tilde{g}_{2,1}'(0)\right)x^2 
+ \frac{3}{2} \left(-2 + 8\, g_{1,2}'(0)\right)x^3+\mathcal{O}(x^{4})=0,
\end{equation}
obtaining:
\begin{equation}
    g_{1,2}'(0)=1/4 \quad\implies \quad\tilde{f}_{1,2}''(0)=-1/2, \quad \tilde{g}_{1,2}'(0)=-1/8.
\end{equation}
We notice here that the blow-up equation already determines at this stage the analogue of the previous integration constant $c_{1}$, for the $m_{2}=2$ functions. Using the value of $g_{1,2}'(0)$ to simplify the equation, we can solve it for $\tilde{g}_{2,1}(x)$ obtaining:
\begin{equation}
    \tilde{g}_{2,1}(x)=\frac{(5x^4 - 5x^2-1 )(1 + 4x^2)^{-1/2} + 3x^2+1}{6x^3}.
\end{equation}
Then, the fourth equation for $m_{1}=2$ reads:
\begin{equation}
    2\left(8 + 36\,  g_{2,2}'(0) + 9\, \tilde{f}_{2,2}''(0)\right) 
+ x^2\left(3 + 48\, g_{1,2}^{(3)}(0) + 2\, \tilde{f}_{1,2}^{(4)}(0)\right)=0,
\end{equation}
from which we find the following relations:
\begin{equation}
    \tilde{f}_{2,2}''(0)=-\frac{4}{9}\big(2+9 g_{2,2}'(0)\big), \quad \tilde{f}_{1,2}^{(4)}(0)=-\frac{3}{2}\big(1+16g_{1,2}^{(3)}(0)\big).
\end{equation}
Carrying on, the fifth equation for $m_{1}=1$ reads:
\begin{equation}
    \begin{aligned}
&27 + 341 x^2 + 2796 x^4 + 9888 x^6 + 12320 x^8 
+ 72 x^2 (1 + 4x^2)^3 \left( 4 \tilde{f}_{3,1}(x) + x \tilde{f}_{3,1}'(x) \right) \\
&+ (1 + 4x^2)^{3/2} \Big( -27 - 179 x^2 + 438 x^4 + 1270 x^6 \\
& + 72 x^4 \left[ \left(6 + 36x^2 + 56x^4\right) g_{3,1}'(x) + x(1 + 6x^2 + 8x^4) g_{3,1}''(x) \right] \Big)=0,
\end{aligned}
\end{equation}
which can be solved to find:
\begin{equation}
\begin{aligned}
\tilde{f}_{3,1}(x)&=  \frac{
-72 - x^2 \left(630 + 1601 x^2 + 1864 x^4 + 3080 x^6\right)
}{
288 x^4 (1 + 4x^2)^2
}\\
&+  \frac{
(1 + 2x^2)\left(72 + 54x^2 - 127x^4 - 288x^6 g_{3,1}'(x)\right)
}{
288 x^4 (1 + 4x^2)^{1/2}
}.
\end{aligned}
\end{equation}
Using this expression, the fifth equation for $m_{1}=2$ simplifies to:
\begin{equation}
    \begin{aligned}
&488 
- \frac{162}{(1 + 4x^2)^2}
- \frac{809}{\sqrt{1 + 4x^2}}
- 381 \sqrt{1 + 4x^2}
- \frac{1152 (x^2 + 5x^4) g_{3,1}'(x)}{\sqrt{1 + 4x^2}} \\
&- 288 x^3 \sqrt{1 + 4x^2} \, g_{3,1}''(x)
+ \frac{8}{x^{4}} \Bigg[
18 - 18 \sqrt{1 + 4x^2} + 36x^2 \sqrt{1 + 4x^2} \\
& - 16x^5 \left(118 + 243\, g_{2,2}'(0) + 27\, \tilde{g}_{2,2}'(0) \right)
+ x^6 \left(1 - 288\, g_{1,3}'(0) - 72\, \tilde{f}_{1,3}''(0)\right) \\
& - 12x^7 \left(15 + 88\, g_{1,2}^{(3)}(0) + 8\, \tilde{g}_{1,2}^{(3)}(0) \right)
\Bigg]=0.
\end{aligned}
\end{equation}
Expanding the equation around $x=0$, we have:
\begin{equation}
\begin{aligned}
&\left(-\frac{236}{9} - 54\, g_{2,2}'(0) - 6\, \tilde{g}_{2,2}'(0) \right)x + \left(
\frac{7}{4} - 4\, g_{1,3}'(0) - 2\, g_{3,1}'(0) - \tilde{f}_{1,3}''(0)
\right)x^2 \\
&+ \frac{1}{6} \left(
-15 - 88\, g_{1,2}^{(3)}(0) - 8\, \tilde{g}_{1,2}^{(3)}(0)
\right)x^3 + \mathcal{O}(x^{4})=0,
\end{aligned}
\end{equation}
from which we find:
\begin{equation}
    \tilde{g}_{2,2}'(0)=-\frac{118}{27}-9g_{2,2}'(0), \quad \tilde{g}_{1,2}^{(3)}(0)=-\frac{15}{8}-11g_{1,2}^{(3)}(0).
\end{equation}
Using these relations in the previous equation and solving for $g_{3,1}(x)$, we obtain:
\begin{equation}
\label{eq: g31}
\begin{aligned}
g_{3,1}(x)&= \frac{
-72 - 90x^2 + 635x^4 + \dfrac{72 + 234x^2 - 599x^4 + 88x^6}{\sqrt{1 + 4x^2}}
}{
1440x^5
}\\
&-\frac{1}{288}\left( -1 + 288\, g_{1,3}'(0) + 72\, \tilde{f}_{1,3}''(0) \right)\arcsinh(2x)
\end{aligned}
\end{equation}
where the $\arcsinh(2x)$ term is clearly incompatible with the asymptotics of the function near infinity. Indeed, (\ref{eq: g31}) behaves as:
\begin{equation}
    g_{3,1}(x)\simeq \pm \frac{11}{360}+\left(\frac{1}{576}-\frac{1}{2}g_{1,3}'(0)-\frac{1}{8}\tilde{f}_{1,3}''(0)\right)\log16x^{2}+\mathcal{O}\left(x^{-1}\right).
\end{equation}
Requiring the vanishing of this term leads to another relation between Taylor coefficients of resummation functions:
\begin{equation}
    \tilde{f}_{1,3}''(0)=\frac{1}{72}-4g_{1,3}'(0).
\end{equation}
Finally, the sixth equation for $m_{1}=1$ reads:
\begin{equation}
\begin{aligned}
&-810 + 3932x^2 - 2(1 + 4x^2)^{3/2} \Big(
    -405 + 4396x^2 + 41280x^4 + 76900x^6 \\
& + 1080x^3 \left[
    (5 + 30x^2 + 48x^4)\, \tilde{g}_{3,1}(x)
    + x(1 + 6x^2 + 8x^4)\, \tilde{g}_{3,1}'(x)
\right] \Big) \\
&x^4 \Big[ 74622 + 280152x^2 + 324992x^4 - 87168x^6+ 1200x \left(4 + 9\, g_{2,2}'(0)\right) \\
&+ 45x^3 \left(1307 + 2880\, g_{2,2}'(0) + 144\, g_{1,2}^{(3)}(0)\right) + 180x^5 \left(1361 + 2880\, g_{2,2}'(0) + 432\, g_{1,2}^{(3)}(0)\right) \\
&+ 240x^7 \left(1523 + 2880\, g_{2,2}'(0) + 1296\, g_{1,2}^{(3)}(0)\right) + 25920x^9 \left(3 + 16\, g_{1,2}^{(3)}(0)\right)\Big]=0.
\end{aligned}
\end{equation}
Considering its expansion around $x=0$, the first few orders are:
\begin{equation}
\begin{aligned}
&-30x^4 \left(1861 + 432\, \tilde{g}_{3,1}'(0) \right) 
+ 1200x^5 \left(4 + 9\, g_{2,2}'(0) \right) \\
& - 40x^6 \left(10625 + 3888\, \tilde{g}_{3,1}'(0) + 72\, \tilde{g}_{3,1}^{(3)}(0) \right) \\
& + 45x^7 \left(1307 + 2880\, g_{2,2}'(0) + 144\, g_{1,2}^{(3)}(0) \right)+\mathcal{O}(x^{8})=0,
\end{aligned}
\end{equation}
from which we obtain:
\begin{equation}
    g_{2,2}'(0)=-\frac{4}{9}, \quad g_{1,2}^{(3)}(0)=-\frac{3}{16}.
\end{equation}
Using these values to simplify the previous equation, we find:
\begin{equation}
    \begin{aligned}
&-810 + 3932x^2 + 74622x^4 + 280152x^6 + 324992x^8 - 87168x^{10} \\
& - 2(1 + 4x^2)^{3/2} \Big(
    -405 + 4396x^2 + 41280x^4 + 76900x^6 \\
& + 1080x^3 \left[
    (5 + 30x^2 + 48x^4)\, \tilde{g}_{3,1}(x)
    + x(1 + 6x^2 + 8x^4)\, \tilde{g}_{3,1}'(x)
\right] \Big)=0,
\end{aligned}
\end{equation}
which we can easily solve, obtaining:
\begin{equation}
    \tilde{g}_{3,1}(x)=\frac{216 + 405x^2 - 3845x^4}{2160x^5}
+ \frac{-216 - 1701x^2 + 119x^4 + 12199x^6 - 1816x^8}{2160x^5 (1 + 4x^2)^{3/2}}.
\end{equation}
Then, the sixth equation for $m_{1}=2$ reads:
\begin{equation}
\begin{aligned}
&\frac{1}{300} \left(-143 + 21312\, g_{1,3}'(0) + 2160\, \tilde{g}_{1,3}'(0) \right) + \left(-\frac{55}{432} + 19\, g_{1,3}'(0) + 2\, \tilde{g}_{1,3}'(0) \right) x \\
&+ \left(-\frac{14161}{1296} + 16\, g_{3,2}'(0) + 4\, \tilde{f}_{3,2}''(0) \right) x^2 + \frac{5}{36} \left(-16 + 72\, g_{2,2}^{(3)}(0) + 3\, \tilde{f}_{2,2}^{(4)}(0) \right) x^4 \\
&+ \left( g_{1,2}^{(5)}(0) + \frac{1}{60} \left( -15 + \tilde{f}_{1,2}^{(6)}(0) \right) \right) x^6=0,
\end{aligned}
\end{equation}
from which we find:
\begin{equation}
\begin{aligned}
&g_{1,3}'(0) = \frac{1}{144},\quad
\tilde{g}_{1,3}'(0) = -\frac{1}{432},\quad 
\tilde{f}_{3,2}''(0) = \frac{14161}{5184} - 4\, g_{3,2}'(0), \\
&\tilde{f}_{2,2}^{(4)}(0) = \frac{16}{3} - 24\, g_{2,2}^{(3)}(0), \quad
\tilde{f}_{1,2}^{(6)}(0) = 15 - 60\, g_{1,2}^{(5)}(0).
\end{aligned}
\end{equation}
As a final summary, the relations we found up until this order read:
\begin{equation}
    \begin{aligned}
& g_{1,2}'(0) = \frac{1}{4},\quad \tilde{g}_{1,2}'(0) = -\frac{1}{8}, \quad \tilde{f}_{1,2}''(0) = -\frac{1}{2},\\
& g_{1,2}^{(3)}(0) = -\frac{3}{16}, \quad \tilde{g}_{1,2}^{(3)}(0) = \frac{3}{16},\quad \tilde{f}_{1,2}^{(4)}(0) = 3,\\
& g_{1,3}'(0) = \frac{1}{144}, \quad \tilde{g}_{1,3}'(0) = -\frac{1}{432},\quad \tilde{f}_{1,3}''(0) = -\frac{1}{72}, \\
& g_{2,2}'(0) = -\frac{4}{9},\quad \tilde{g}_{2,2}'(0) = -\frac{10}{27}, \quad \tilde{f}_{2,2}''(0) = \frac{8}{9}, \\
&\tilde{f}_{1,2}^{(6)}(0) = 15 - 60\, g_{1,2}^{(5)}(0), \quad \tilde{f}_{2,2}^{(4)}(0) = \frac{16}{3} - 24\, g_{2,2}^{(3)}(0), \\
&\tilde{f}_{3,2}''(0) = \frac{14161}{5184} - 4\, g_{3,2}'(0).
\end{aligned}
\end{equation}
Going to higher order in the expansion or to higher values of $m_{1}$ and $m_{2}$ does not introduce any new feature in the equations, which can be solved straightforwardly. As we have seen, the basic strategy to obtain the desired solution is to impose the proper parity of the functions and the correct behavior at zero and at infinity. Expanding the equations around zero or infinity before solving them often helps identify combinations of Taylor coefficients of the resummation functions that vanish, which in turn makes the integration of the equations easier.

\subsection{\texorpdfstring{Example: $\mathbb{C}^2/\mathbb{Z}_2$ blow-up equations, $\mathcal{N}=2^{*}$ theory}{C2Z2, N2star theory}}
\label{app: blow-up example 3}
In this section we follow the same steps as in the previous one, showing and solving the differential equations for the resummation functions in the $\mathcal{N}=2^{*}$ case. We suppress the arguments of the functions and denote by $\tilde{f}_{1,1}'$ the derivative of $\tilde{f}_{1,1}(\mu,x)$ with respect to $x$, unless the functions are valued at some particular point. The first equation for $m_{1}=1$ reads:
\begin{equation}
\begin{aligned}
& (\mu - 1)^4 \mu^4 x^4 e^{-2 x^2 g_{1,1}'} \left( -x \tilde{f}_{1,1}' + x^3 g_{1,1}'' + 2 x^2 g_{1,1}' - 2 \right)\\ 
& +x e^{2 x^2 g_{1,1}'} \left( \tilde{f}_{1,1}' + 2x g_{1,1}' + x^{2} g_{1,1}''  \right)=0,
\end{aligned}
\end{equation}
and depends algebraically on $\tilde{f}_{1,1}'$. Solving it and integrating the result, we find:
\begin{equation}
 \tilde{f}_{1,1}(\mu,x)= - \frac{1}{2} \log\left(e^{2 x^2 g_{1,1}'} - (\mu - 1)^4 \mu^4 x^4 e^{-2 x^2 g_{1,1}'} \right),
\end{equation}
where the integration constant has been determined by imposing the vanishing of the function $\tilde{f}_{1,1}$ at $x=0$. Writing $\tilde{f}_{1,1}=\frac{1}{2}\log u_1$ and $g_{1,1}'=\frac{1}{2x^{2}}\log(x^{2} \tilde{v}_1)$, we have:
\begin{equation}
\label{eq: u1 1 2}
    u_1(\mu,x)=\frac{1}{x^{2}}\frac{1}{\tilde{v}_1-\mu^{4}(\mu-1)^{4}\tilde{v}_1^{-1}}\,.
\end{equation}
Meanwhile, the first equation for $m_{1}=2$ reads:
\begin{equation}
\begin{aligned}
0&=(\mu - 1)^8 \mu^8 x^8 \left( x \tilde{f}_{1,1}' - 2 x^3 g_{1,1}'' - 4 x^2 g_{1,1}' + 3 \right)
e^{2 \tilde{f}_{1,1} - 4 x^2 g_{1,1}'}\\
&+\left(1 - x\tilde{f}_{1,1}' - 4x^{2} g_{1,1}' - 2x^{3} g_{1,1}'' \right)
e^{2 \tilde{f}_{1,1} + 4 x^2 g_{1,1}'}-1,
\end{aligned}
\end{equation}
which we can solve for $u_{1}$, to obtain:
\begin{equation}
\label{eq: u1 2 2}
    u_1(\mu,x)=\frac{1-c_{1}x^{2}}{x^{4}}\frac{1}{\tilde{v}_1^{2}-\mu^{8}(\mu-1)^{8}\tilde{v}_1^{-2}}\,.
\end{equation}
Then, equating (\ref{eq: u1 1 2}) and (\ref{eq: u1 2 2}) leads to:
\begin{equation}
    \tilde{v}_{1}(\mu,x)=\frac{1-c_1 x^2\pm\sqrt{1-2 c_1 x^2+x^4 \left(c_1^2-4 \mu ^4 (\mu -1)^4\right)}}{2 x^2},
\end{equation}
where, by looking at the behavior near $x=0$, we have to select the positive relative sign and, by looking at the behavior near infinity, we have to choose $c_{1}=-2\mu^{2}(\mu-1)^{2}$ to cancel the coefficient of $x^{4}$ inside the square root, as discussed in the previous section around equation \eqref{eq: integration constant c1}. We thus find:
\begin{equation}
\begin{aligned}
& \tilde{f}_{1,1}(\mu,x)=-\frac{1}{4}\log \big(1+4\mu^{2}(\mu-1)^{2}x^{2}\big),\\
&  g_{1,1}(\mu,x)=\frac{\sqrt{1+4 \mu ^2(\mu -1)^2 x^2}+\log2-\log \left(1+\sqrt{1+4 \mu ^2(\mu -1)^2 x^2}\right)-1}{x},
\end{aligned}
\end{equation}
where the remaining integration constant for $g_{1,1}$ is fixed by $g_{1,1}(\mu,0)=0$. We can then look at the second equation for $m_1=1$, which reads:
\begin{equation}
\begin{aligned}
&\left[1+8 (\mu -1)^4 \mu^4 x^4 + 6 (\mu -1)^2 \mu^2 x^2 \right]x \tilde{g}_{1,1}' \\
& +\left[1+16 (\mu -1)^4 \mu^4 x^4 + 6 (\mu -1)^2 \mu^2 x^2 \right] \tilde{g}_{1,1}\\
&     - 4 (\mu -1)^3 \mu^3 (\mu^2 + \mu - 1) x^3
+ 2 \mu (\mu^3 - 2\mu + 1) x \left[ 1+4\mu^2 (\mu -1)^2 x^2\right]^{3/2}=0.
\end{aligned}
\end{equation}
The equation can be easily solved, obtaining:
\begin{equation}
    \tilde{g}_{1,1}(\mu,x)=\frac{\left(\mu ^2+\mu -1\right) \left(1-\sqrt{1+4 \mu ^2(\mu -1)^2 x^2}\right)}{2\mu (\mu -1)  x},
\end{equation}
where the integration constant has been fixed by $ \tilde{g}_{1,1}(\mu,0)=0$. Then, the second equation for $m_1=2$ simplifies to:
\begin{equation}
    2 \tilde{f}_{1,2}''(\mu,0) + 8 g_{1,2}'(\mu,0) - (\mu -2)^2 (\mu -1)^2 \mu^2 (\mu +1)^2 =0,
\end{equation}
from which we find:
\begin{equation}
\label{eq: f12 mu}
    \tilde{f}_{1,2}''(\mu,0) = \frac{1}{2} (\mu -2)^2 (\mu -1)^2 \mu^2 (\mu +1)^2 - 4 g_{1,2}'(\mu,0).
\end{equation}
Moving on, the third equation for $m_{1}=1$ reads:
\begin{equation}
\begin{aligned}
&\left[ 1+4 (\mu -1)^2 \mu^2 x^2\right]^{3/2}
\left[4 \tilde{f}_{2,1} + 2x \tilde{f}_{2,1}' + (\mu -1)^2 \mu^2\right]\\
&+ 8 \left[10 (\mu -1)^4 \mu^4 x^6 + 6 (\mu -1)^2 \mu^2 x^4 + x^2\right] g_{2,1}'\\
& + 2 \left[8 (\mu -1)^4 \mu^4 x^7 + 6 (\mu -1)^2 \mu^2 x^5 + x^3\right) g_{2,1}''\\
&    - (\mu -1)^2 \mu^2 \left[12 (\mu -1)^4 \mu^4 x^4 + 2 \left(7 \mu ^4-14 \mu ^3-\mu ^2+8 \mu -4\right) x^2 + 1 \right]=0,
\end{aligned}
\end{equation}
which can be solved to obtain:
\begin{equation}
\begin{aligned}
&\tilde{f}_{2,1}(\mu,x)= \frac{-2 \mu ^4+4 \mu ^3+2 \mu ^2-4 \mu+2 +\left(-\mu ^8+4 \mu ^7-6 \mu ^6+4 \mu ^5-\mu ^4\right) x^2}
{4 (\mu -1)^2 \mu^2 x^2}\\
&+\frac{1+2 (\mu -1)^2 \mu^2 x^2}{4 (\mu -1)^2 \mu^2 x^2 \sqrt{1+4 (\mu -1)^2 \mu^2 x^2}}\Big[2 \mu ^4-4 \mu ^3-2 \mu ^2+4 \mu-2 \\
&+\left(\mu ^8-4 \mu ^7+6 \mu ^6-4 \mu ^5+\mu ^4\right) x^2-4 (\mu -1)^2 \mu^2 x^4 g_{2,1}' \Big],
\end{aligned}
\end{equation}
where the integration constant is determined by requiring $\tilde{f}_{2,1}(\mu\,,0)=0$. Then, the third equation for $m_{1}=2$ reads:
\begin{equation*}
\begin{aligned}
&\sqrt{1+4 (\mu -1)^2 \mu^2 x^2}\Bigg\{
  (\mu -2)(\mu +1)
  \Big[\mu ^4-2 \mu ^3-\mu ^2+2 \mu-1 \\
&+\left(9 \mu ^{10}-45 \mu ^9+92 \mu ^8-98 \mu ^7+53 \mu ^6-5 \mu ^5-10 \mu ^4+4 \mu ^3\right) x^3  \Big]\\
& - 8 (\mu -1)^2 \mu^2 x^3
  \left(-3 g_{1,2}'(\mu,0) + \mu \partial_{\mu}g_{1,2}'(\mu,0) + \tilde{g}_{1,2}'(\mu,0)\right)
\Bigg\}\\
&+ (\mu -2)(\mu +1) \Bigg\{
  (\mu -1)^2 \mu^2 x^4  \Big[
    -2\left(1+6 (\mu -1)^2 \mu^2 x^2\right) g_{2,1}'\\
&- x \left(1+4 (\mu -1)^2 \mu^2 x^2\right) g_{2,1}''
  \Big]-\mu ^4+2 \mu ^3+\mu ^2-2 \mu+1\\
&+\left(\mu ^{12}-6 \mu ^{11}+15 \mu ^{10}-20 \mu ^9+15 \mu ^8-6 \mu ^7+\mu ^6\right) x^4\\
&+\left(-2 \mu ^8+8 \mu ^7-8 \mu ^6-4 \mu ^5+12 \mu ^4-8 \mu ^3+2 \mu ^2\right) x^2\Bigg\}=0.
\end{aligned}
\end{equation*}
Expanding the equation around $x=0$, the first non-zero coefficient gives us:
\begin{equation}
\begin{aligned}
&\partial_{\mu}g_{1,2}'(\mu,0)=\frac{1}{8\mu}\big( 24 g_{1,2}'(\mu,0) - 8 \tilde{g}_{1,2}'(\mu,0)\\
& + 9\mu^8 - 36\mu^7 + 38\mu^6 + 12\mu^5 - 47\mu^4 + 32\mu^3 - 8\mu\big),
\end{aligned}
\end{equation}
which we can use to simplify and solve the equation, obtaining:
\begin{equation}
\begin{aligned}
g_{2,1}(\mu,x)&=-\frac{2 (\mu -2) (\mu -1) (\mu +1) \mu-2 +3 (\mu -1)^4 \mu ^4 x^2}{12 (\mu -1)^2 \mu ^2 x^3}\\
&-\frac{\sqrt{1+4 (\mu -1)^2 \mu ^2 x^2}}{12 (\mu -1)^2 \mu ^2 x^3}\Big[2-2 \mu ^4+4 \mu ^3+2 \mu ^2-4 \mu\\
&+\left(\mu ^8-4 \mu ^7-2 \mu ^6+20 \mu ^5-27 \mu ^4+16 \mu ^3-4 \mu ^2\right) x^2\Big],
\end{aligned}
\end{equation}
\begin{equation}
    \tilde{f}_{2,1}(\mu,x)= \frac{\left(3 \mu ^8-12 \mu ^7+10 \mu ^6+12 \mu ^5-25 \mu ^4+16 \mu ^3-4 \mu ^2\right) x^2}{2\big(1+4 (\mu -1)^2 \mu ^2 x^2\big)}.
\end{equation}
fixing the integration constants via $g_{2,1}(\mu,0)=0$. Then, the fourth equation for $m_1=1$ reads:
\begin{equation}
\begin{aligned}
&24 (\mu -2)(\mu +1) x^3 \left[ 1+4 (\mu -1)^2 \mu^2 x^2\right]^2 g_{1,2}'(\mu,0) \\
&+ \sqrt{4 (\mu -1)^2 \mu^2 x^2 + 1} \Bigg\{
  -6x \left[32 (\mu -1)^4 \mu^4 x^4 + 18 (\mu -1)^2 \mu^2 x^2 + 3\right] \tilde{g}_{2,1} \\
& -6 \left[8 (\mu -1)^4 \mu^4 x^6 + 6 (\mu -1)^2 \mu^2 x^4 + x^2\right] \tilde{g}_{2,1}' \\
&+ (\mu -1)\mu \Big[
    6 \mu^2-3 + 36 (\mu -1)^4 (2\mu^2 - 1) \mu^4 x^4 \\
&+\left(44 \mu ^6-96 \mu ^5+42 \mu ^4+44 \mu ^3-42 \mu ^2+24 \mu -8\right) x^2
  \Big]
\Bigg\} \\
&+ (\mu -1)\mu \cdot \Bigg[
  3-6 \mu^2 - 96 (\mu -2)^3 (\mu -1)^5 \mu^5 (\mu +1)^3 x^7 \\
&+ 32 (\mu -1)^4 \mu^4 \left(5 \mu ^6-9 \mu ^5-24 \mu ^4+23 \mu ^3+9 \mu ^2-12 \mu +4\right) x^6 \\
& - 48 (\mu -2)^3 (\mu -1)^3 \mu^3 (\mu +1)^3 x^5 \\
& + 4 (\mu -1)^2 \mu^2 \left(5 \mu ^6+3 \mu ^5-150 \mu ^4+95 \mu ^3+75 \mu ^2-84 \mu +28\right) x^4 \\
& - 6 (\mu -2)^3 (\mu -1) \mu (\mu +1)^3 x^3 \\
& + 4 \left(-5 \mu ^6+15 \mu ^5-39 \mu ^4+13 \mu ^3+24 \mu ^2-24 \mu +8\right) x^2
\Bigg]=0.
\end{aligned}
\end{equation}
Expanding it around $x=0$, we find:
\begin{equation}
    g_{1,2}'(\mu,0)=\frac{1}{4} (\mu -2)^2 (\mu -1)^2 \mu ^2 (\mu +1)^2,
\end{equation}
which, due to (\ref{eq: f12 mu}), implies: 
\begin{equation}
    \tilde{f}_{1,2}''(\mu,0)=-\frac{1}{2} (\mu -2)^2 (\mu -1)^2 \mu ^2 (\mu +1)^2,
\end{equation}
thus determining the equivalent of the integration constant $c_{1}$ for the $m_2 = 2$ resummation functions. Then, solving the equation we obtain:
\begin{equation}
\begin{aligned}
&\tilde{g}_{2,1}(\mu,x)= \big(12 (\mu - 1)^3 \mu^3 x^3\big)^{-1}\Big[
  2 \mu^6 - 6 \mu^5 + 6 \mu^4 + 2 \mu^3 - 6 \mu^2 + 6 \mu
  - 2\\
&+ \left(6 \mu^{10} - 24 \mu^9 + 33 \mu^8 - 12 \mu^7 - 12 \mu^6 + 12 \mu^5 - 3 \mu^4\right) x^2\Big]\\
&+\left(12 (\mu - 1)^3 \mu^3 x^3 \sqrt{1+4 (\mu - 1)^2 \mu^2 x^2} \right)^{-1}\Big[ -2 \mu^6 + 6 \mu^5 - 6 \mu^4 - 2 \mu^3\\
&+ 6 \mu^2 - 6 \mu + 2+ \big(
    10 \mu^{14} - 58 \mu^{13} + 84 \mu^{12} + 90 \mu^{11}
    - 372 \mu^{10} + 354 \mu^9\\
&- 20 \mu^8 - 202 \mu^7 + 162 \mu^6 - 56 \mu^5 + 8 \mu^4
  \big) x^4\\
&+ \left(
    -10 \mu^{10} + 44 \mu^9 - 73 \mu^8 + 44 \mu^7 + 20 \mu^6
    - 52 \mu^5 + 43 \mu^4 - 20 \mu^3 + 4 \mu^2
  \right) x^2\Big],
\end{aligned}
\end{equation}
so that the fourth equation for $m_1=2$ becomes:
\begin{equation}
\begin{aligned}
&\frac{1}{6} x^2 \left(
  9 \tilde{f}_{2,2}''(\mu,0)
  + 36 g_{2,2}'(\mu,0)
  + 8 (\mu -1)^4 (-\mu^2 + \mu + 2)^2 \mu^4
\right)\\
&x^4 \left(
  \frac{1}{6} \tilde{f}_{1,2}^{(4)}(\mu,0)
  + 4 g_{1,2}^{(3)}(\mu,0)
  + \frac{1}{4} (\mu -2)^4 (\mu -1)^4 \mu^4 (\mu +1)^4
\right)=0.
\end{aligned}
\end{equation}
Solving it, we find two relations between Taylor coefficients at $x=0$ of the $m_2=2$ resummation functions:
\begin{equation}
\begin{aligned}
& \tilde{f}_{2,2}''(\mu,0) = -\frac{8}{9} (\mu -1)^4 \mu ^4 (\mu -2)^2 (\mu +1)^2 - 4 g_{2,2}'(\mu,0),\\
&\tilde{f}_{1,2}^{(4)}(\mu,0) = -\frac{3}{2}(\mu -2)^4 (\mu -1)^4 \mu^4 (\mu +1)^4-24 g_{1,2}^{(3)}(\mu,0).
\end{aligned}
\end{equation}
In summary, it is clear that the only difference with respect to the pure gauge theory case is the cumbersomeness of the equations and of the resummation functions. Thus, in principle one can proceed to higher orders in the expansion, and to different values of $m_1$ and $m_2$, without any qualitative difference with what we have seen so far.

% Bibliography

%% [A] Recommended: using JHEP.bst file
%% \bibliographystyle{JHEP}
\newpage
\bibliographystyle{JHEP}
\bibliography{refs.bib}

\providecommand{\href}[2]{#2}\begingroup\raggedright\begin{thebibliography}{100}

\bibitem{lame1837}
G.~Lam\'e, \emph{{Mémoire sur les surfaces isothermes dans les corps solides homogènes en équilibre de température}}, {\emph{J. Math. Pures Appl.} {\bfseries 2} (1837) 147}.

\bibitem{PhysRevLett.50.873}
Y.~Alhassid, F.~G\"ursey and F.~Iachello, \emph{Potential scattering, transfer matrix, and group theory}, \href{https://doi.org/10.1103/PhysRevLett.50.873}{\emph{Phys. Rev. Lett.} {\bfseries 50} (1983) 873}.

\bibitem{Forini:2010ek}
V.~Forini, \emph{{Quark-antiquark potential in AdS at one loop}}, \href{https://doi.org/10.1007/JHEP11(2010)079}{\emph{JHEP} {\bfseries 11} (2010) 079} [\href{https://arxiv.org/abs/1009.3939}{{\ttfamily 1009.3939}}].

\bibitem{Giombi:2021zfb}
S.~Giombi, S.~Komatsu and B.~Offertaler, \emph{{Large charges on the Wilson loop in $ \mathcal{N} $ = 4 SYM: matrix model and classical string}}, \href{https://doi.org/10.1007/JHEP03(2022)020}{\emph{JHEP} {\bfseries 03} (2022) 020} [\href{https://arxiv.org/abs/2110.13126}{{\ttfamily 2110.13126}}].

\bibitem{Giombi:2022anm}
S.~Giombi, S.~Komatsu and B.~Offertaler, \emph{{Large charges on the Wilson loop in $ \mathcal{N} $ = 4 SYM. Part II. Quantum fluctuations, OPE, and spectral curve}}, \href{https://doi.org/10.1007/JHEP08(2022)011}{\emph{JHEP} {\bfseries 08} (2022) 011} [\href{https://arxiv.org/abs/2202.07627}{{\ttfamily 2202.07627}}].

\bibitem{Liang:1992mw}
J.-Q.~Liang, H.J.W.~Muller-Kirsten and D.H.~Tchrakian, \emph{{Solitons, bounces and sphalerons on a circle}}, \href{https://doi.org/10.1016/0370-2693(92)90486-N}{\emph{Phys. Lett. B} {\bfseries 282} (1992) 105}.

\bibitem{Belavin:1984vu}
A.A.~Belavin, A.M.~Polyakov and A.B.~Zamolodchikov, \emph{{Infinite Conformal Symmetry in Two-Dimensional Quantum Field Theory}}, \href{https://doi.org/10.1016/0550-3213(84)90052-X}{\emph{Nucl. Phys. B} {\bfseries 241} (1984) 333}.

\bibitem{Bonelli:2022ten}
G.~Bonelli, C.~Iossa, D.P.~Lichtig and A.~Tanzini, \emph{Irregular liouville correlators and connection formulae for heun functions}, \href{https://doi.org/10.1007/s00220-022-04497-5}{\emph{Communications in Mathematical Physics} (2022) } [\href{https://arxiv.org/abs/2201.04491}{{\ttfamily 2201.04491}}].

\bibitem{NIST:DLMF}
``{\it NIST Digital Library of Mathematical Functions}.'' \url{https://dlmf.nist.gov/}, Release 1.2.4 of 2025-03-15.

\bibitem{Litvinov:2013sxa}
A.~Litvinov, S.~Lukyanov, N.~Nekrasov and A.~Zamolodchikov, \emph{{Classical Conformal Blocks and Painleve VI}}, \href{https://doi.org/10.1007/JHEP07(2014)144}{\emph{JHEP} {\bfseries 07} (2014) 144} [\href{https://arxiv.org/abs/1309.4700}{{\ttfamily 1309.4700}}].

\bibitem{Jeong:2018qpc}
S.~Jeong and N.~Nekrasov, \emph{{Opers, surface defects, and Yang-Yang functional}}, \href{https://doi.org/10.4310/ATMP.2020.v24.n7.a4}{\emph{Adv. Theor. Math. Phys.} {\bfseries 24} (2020) 1789} [\href{https://arxiv.org/abs/1806.08270}{{\ttfamily 1806.08270}}].

\bibitem{Consoli:2022eey}
D.~Consoli, F.~Fucito, J.F.~Morales and R.~Poghossian, \emph{{CFT description of BH\textquoteright{}s and ECO\textquoteright{}s: QNMs, superradiance, echoes and tidal responses}}, \href{https://doi.org/10.1007/JHEP12(2022)115}{\emph{JHEP} {\bfseries 12} (2022) 115} [\href{https://arxiv.org/abs/2206.09437}{{\ttfamily 2206.09437}}].

\bibitem{Lisovyy:2022flm}
O.~Lisovyy and A.~Naidiuk, \emph{{Perturbative connection formulas for Heun equations}}, \href{https://doi.org/10.1088/1751-8121/ac9ba7}{\emph{J. Phys. A} {\bfseries 55} (2022) 434005} [\href{https://arxiv.org/abs/2208.01604}{{\ttfamily 2208.01604}}].

\bibitem{Alday:2009aq}
L.F.~Alday, D.~Gaiotto and Y.~Tachikawa, \emph{{Liouville Correlation Functions from Four-dimensional Gauge Theories}}, \href{https://doi.org/10.1007/s11005-010-0369-5}{\emph{Lett. Math. Phys.} {\bfseries 91} (2010) 167} [\href{https://arxiv.org/abs/0906.3219}{{\ttfamily 0906.3219}}].

\bibitem{LeFloch:2020uop}
B.~Le~Floch, \emph{{A slow review of the AGT correspondence}}, \href{https://doi.org/10.1088/1751-8121/ac5945}{\emph{J. Phys. A} {\bfseries 55} (2022) 353002} [\href{https://arxiv.org/abs/2006.14025}{{\ttfamily 2006.14025}}].

\bibitem{Alday:2010vg}
L.F.~Alday and Y.~Tachikawa, \emph{{Affine SL(2) conformal blocks from 4d gauge theories}}, \href{https://doi.org/10.1007/s11005-010-0422-4}{\emph{Lett. Math. Phys.} {\bfseries 94} (2010) 87} [\href{https://arxiv.org/abs/1005.4469}{{\ttfamily 1005.4469}}].

\bibitem{Nekrasov:2002qd}
N.A.~Nekrasov, \emph{{Seiberg-Witten prepotential from instanton counting}}, \href{https://doi.org/10.4310/ATMP.2003.v7.n5.a4}{\emph{Adv. Theor. Math. Phys.} {\bfseries 7} (2003) 831} [\href{https://arxiv.org/abs/hep-th/0206161}{{\ttfamily hep-th/0206161}}].

\bibitem{Nekrasov:2009rc}
N.A.~Nekrasov and S.L.~Shatashvili, \emph{{Quantization of Integrable Systems and Four Dimensional Gauge Theories}},  in \emph{{16th International Congress on Mathematical Physics}}, pp.~265--289, 8, 2009, \href{https://doi.org/10.1142/9789814304634_0015}{DOI} [\href{https://arxiv.org/abs/0908.4052}{{\ttfamily 0908.4052}}].

\bibitem{Bonelli:2010gk}
G.~Bonelli, K.~Maruyoshi, A.~Tanzini and F.~Yagi, \emph{{Generalized matrix models and AGT correspondence at all genera}}, \href{https://doi.org/10.1007/JHEP07(2011)055}{\emph{JHEP} {\bfseries 07} (2011) 055} [\href{https://arxiv.org/abs/1011.5417}{{\ttfamily 1011.5417}}].

\bibitem{Bonelli:2011na}
G.~Bonelli, K.~Maruyoshi and A.~Tanzini, \emph{{Quantum Hitchin Systems via ${\beta}$ -Deformed Matrix Models}}, \href{https://doi.org/10.1007/s00220-017-3053-0}{\emph{Commun. Math. Phys.} {\bfseries 358} (2018) 1041} [\href{https://arxiv.org/abs/1104.4016}{{\ttfamily 1104.4016}}].

\bibitem{Piatek:2013ifa}
M.~Piatek, \emph{{Classical torus conformal block, $N = 2^*$ twisted superpotential and the accessory parameter of Lam\'e equation}}, \href{https://doi.org/10.1007/JHEP03(2014)124}{\emph{JHEP} {\bfseries 03} (2014) 124} [\href{https://arxiv.org/abs/1309.7672}{{\ttfamily 1309.7672}}].

\bibitem{Piatek:2015jva}
M.~Piatek and A.R.~Pietrykowski, \emph{{Classical limit of irregular blocks and Mathieu functions}}, \href{https://doi.org/10.1007/JHEP01(2016)115}{\emph{JHEP} {\bfseries 01} (2016) 115} [\href{https://arxiv.org/abs/1509.08164}{{\ttfamily 1509.08164}}].

\bibitem{Basar:2015xna}
G.~Ba\c{s}ar and G.V.~Dunne, \emph{{Resurgence and the Nekrasov-Shatashvili limit: connecting weak and strong coupling in the Mathieu and Lam\'e systems}}, \href{https://doi.org/10.1007/JHEP02(2015)160}{\emph{JHEP} {\bfseries 02} (2015) 160} [\href{https://arxiv.org/abs/1501.05671}{{\ttfamily 1501.05671}}].

\bibitem{Desiraju:2024fmo}
H.~Desiraju, P.~Ghosal and A.~Prokhorov, \emph{{Proof of Zamolodchikov conjecture for semi-classical conformal blocks on the torus}},  \href{https://arxiv.org/abs/2407.05839}{{\ttfamily 2407.05839}}.

\bibitem{Beccaria:2016nnb}
M.~Beccaria and G.~Macorini, \emph{{Exact partition functions for the \ensuremath{\Omega}-deformed $ \mathcal{N}={2}^{\ast } $ SU(2) gauge theory}}, \href{https://doi.org/10.1007/JHEP07(2016)066}{\emph{JHEP} {\bfseries 07} (2016) 066} [\href{https://arxiv.org/abs/1606.00179}{{\ttfamily 1606.00179}}].

\bibitem{Beccaria:2016vxq}
M.~Beccaria, A.~Fachechi, G.~Macorini and L.~Martina, \emph{{Exact partition functions for deformed $\mathcal{N}=2$ theories with $N_{f}=4$ flavours}}, \href{https://doi.org/10.1007/JHEP12(2016)029}{\emph{JHEP} {\bfseries 12} (2016) 029} [\href{https://arxiv.org/abs/1609.01189}{{\ttfamily 1609.01189}}].

\bibitem{Beccaria:2016wop}
M.~Beccaria, \emph{{On the large $\Omega$-deformations in the Nekrasov-Shatashvili limit of $\mathcal N=2^{*}$ SYM}}, \href{https://doi.org/10.1007/JHEP07(2016)055}{\emph{JHEP} {\bfseries 07} (2016) 055} [\href{https://arxiv.org/abs/1605.00077}{{\ttfamily 1605.00077}}].

\bibitem{Gorsky:2017ndg}
A.~Gorsky, A.~Milekhin and N.~Sopenko, \emph{{Bands and gaps in Nekrasov partition function}}, \href{https://doi.org/10.1007/JHEP01(2018)133}{\emph{JHEP} {\bfseries 01} (2018) 133} [\href{https://arxiv.org/abs/1712.02936}{{\ttfamily 1712.02936}}].

\bibitem{nakajima2003lectures}
H.~Nakajima and K.~Yoshioka, \emph{{Lectures on instanton counting}},  in \emph{{CRM Workshop on Algebraic Structures and Moduli Spaces}}, 11, 2003 [\href{https://arxiv.org/abs/math/0311058}{{\ttfamily math/0311058}}].

\bibitem{Bonelli:2011jx}
G.~Bonelli, K.~Maruyoshi and A.~Tanzini, \emph{{Instantons on ALE spaces and Super Liouville Conformal Field Theories}}, \href{https://doi.org/10.1007/JHEP08(2011)056}{\emph{JHEP} {\bfseries 08} (2011) 056} [\href{https://arxiv.org/abs/1106.2505}{{\ttfamily 1106.2505}}].

\bibitem{Bonelli:2011kv}
G.~Bonelli, K.~Maruyoshi and A.~Tanzini, \emph{{Gauge Theories on ALE Space and Super Liouville Correlation Functions}}, \href{https://doi.org/10.1007/s11005-012-0553-x}{\emph{Lett. Math. Phys.} {\bfseries 101} (2012) 103} [\href{https://arxiv.org/abs/1107.4609}{{\ttfamily 1107.4609}}].

\bibitem{Bershtein:2021uts}
M.~Bershtein, P.~Gavrylenko and A.~Grassi, \emph{{Quantum Spectral Problems and Isomonodromic Deformations}}, \href{https://doi.org/10.1007/s00220-022-04369-y}{\emph{Commun. Math. Phys.} {\bfseries 393} (2022) 347} [\href{https://arxiv.org/abs/2105.00985}{{\ttfamily 2105.00985}}].

\bibitem{Jeong:2017pai}
S.~Jeong, \emph{{Splitting of surface defect partition functions and integrable systems}}, \href{https://doi.org/10.1016/j.nuclphysb.2018.12.007}{\emph{Nucl. Phys. B} {\bfseries 938} (2019) 775} [\href{https://arxiv.org/abs/1709.04926}{{\ttfamily 1709.04926}}].

\bibitem{Belavin:2011pp}
V.~Belavin and B.~Feigin, \emph{{Super Liouville conformal blocks from N=2 SU(2) quiver gauge theories}}, \href{https://doi.org/10.1007/JHEP07(2011)079}{\emph{JHEP} {\bfseries 07} (2011) 079} [\href{https://arxiv.org/abs/1105.5800}{{\ttfamily 1105.5800}}].

\bibitem{Belavin:2012eg}
A.A.~Belavin, M.A.~Bershtein and G.M.~Tarnopolsky, \emph{{Bases in coset conformal field theory from AGT correspondence and Macdonald polynomials at the roots of unity}}, \href{https://doi.org/10.1007/JHEP03(2013)019}{\emph{JHEP} {\bfseries 03} (2013) 019} [\href{https://arxiv.org/abs/1211.2788}{{\ttfamily 1211.2788}}].

\bibitem{Bonelli:2021rrg}
G.~Bonelli, F.~Globlek and A.~Tanzini, \emph{{Counting Yang-Mills Instantons by Surface Operator Renormalization Group Flow}}, \href{https://doi.org/10.1103/PhysRevLett.126.231602}{\emph{Phys. Rev. Lett.} {\bfseries 126} (2021) 231602} [\href{https://arxiv.org/abs/2102.01627}{{\ttfamily 2102.01627}}].

\bibitem{Bonelli:2022iob}
G.~Bonelli, F.~Globlek and A.~Tanzini, \emph{{Toda equations for surface defects in SYM and instanton counting for classical Lie groups}}, \href{https://doi.org/10.1088/1751-8121/ac9e2a}{\emph{J. Phys. A} {\bfseries 55} (2022) 454004} [\href{https://arxiv.org/abs/2206.13212}{{\ttfamily 2206.13212}}].

\bibitem{Nagoya:2015cja}
H.~Nagoya, \emph{{Irregular conformal blocks, with an application to the fifth and fourth Painlev\'e equations}}, \href{https://doi.org/10.1063/1.4937760}{\emph{J. Math. Phys.} {\bfseries 56} (2015) 123505} [\href{https://arxiv.org/abs/1505.02398}{{\ttfamily 1505.02398}}].

\bibitem{Bonelli:2016qwg}
G.~Bonelli, O.~Lisovyy, K.~Maruyoshi, A.~Sciarappa and A.~Tanzini, \emph{{On Painlev\'e/gauge theory correspondence}}, \href{https://doi.org/10.1007/s11005-017-0983-6}{\emph{Lett. Math. Phys.} {\bfseries 107} (2017) pages 2359} [\href{https://arxiv.org/abs/1612.06235}{{\ttfamily 1612.06235}}].

\bibitem{Nagoya:2016mlj}
H.~Nagoya, \emph{{Conformal blocks and Painlev\'e functions}},  \href{https://arxiv.org/abs/1611.08971}{{\ttfamily 1611.08971}}.

\bibitem{Nagoya:2018pgp}
H.~Nagoya, \emph{{Remarks on irregular conformal blocks and Painlev\'e III and II tau functions}},  \href{https://arxiv.org/abs/1804.04782}{{\ttfamily 1804.04782}}.

\bibitem{Gavrylenko:2020gjb}
P.~Gavrylenko, A.~Marshakov and A.~Stoyan, \emph{{Irregular conformal blocks, Painlev\'e III and the blow-up equations}}, \href{https://doi.org/10.1007/JHEP12(2020)125}{\emph{JHEP} {\bfseries 12} (2020) 125} [\href{https://arxiv.org/abs/2006.15652}{{\ttfamily 2006.15652}}].

\bibitem{Fucito:2023plp}
F.~Fucito, J.F.~Morales and R.~Poghossian, \emph{{On irregular states and Argyres-Douglas theories}}, \href{https://doi.org/10.1007/JHEP08(2023)123}{\emph{JHEP} {\bfseries 08} (2023) 123} [\href{https://arxiv.org/abs/2306.05127}{{\ttfamily 2306.05127}}].

\bibitem{Poghosyan:2023zvy}
H.~Poghosyan and R.~Poghossian, \emph{{A note on rank 5/2 Liouville irregular block, Painlev\'e I and the $ \mathcal{H} _{0}$ Argyres-Douglas theory}}, \href{https://doi.org/10.1007/JHEP11(2023)198}{\emph{JHEP} {\bfseries 11} (2023) 198} [\href{https://arxiv.org/abs/2308.09623}{{\ttfamily 2308.09623}}].

\bibitem{Bonelli:2024wha}
G.~Bonelli, P.~Gavrylenko, I.~Majtara and A.~Tanzini, \emph{{Surface observables in gauge theories, modular Painlev\'e tau functions and non-perturbative topological strings}},  \href{https://arxiv.org/abs/2410.17868}{{\ttfamily 2410.17868}}.

\bibitem{Bonelli:2025owb}
G.~Bonelli, A.~Shchechkin and A.~Tanzini, \emph{{Refined Painlev\'e/gauge theory correspondence and quantum tau functions}},  \href{https://arxiv.org/abs/2502.01499}{{\ttfamily 2502.01499}}.

\bibitem{Poghossian:2025nef}
R.~Poghossian and H.~Poghosyan, \emph{{A note on rank $\frac{3}{2}$ Liouville irregular block}},  \href{https://arxiv.org/abs/2502.10169}{{\ttfamily 2502.10169}}.

\bibitem{Iorgov:2025hxt}
N.~Iorgov, K.~Iwaki, O.~Lisovyy and Y.~Zhuravlov, \emph{{Many-faced Painlev\'e I: irregular conformal blocks, topological recursion, and holomorphic anomaly approaches}},  \href{https://arxiv.org/abs/2505.16803}{{\ttfamily 2505.16803}}.

\bibitem{Fucito:2005wc}
F.~Fucito, J.F.~Morales, R.~Poghossian and A.~Tanzini, \emph{{N=1 superpotentials from multi-instanton calculus}}, \href{https://doi.org/10.1088/1126-6708/2006/01/031}{\emph{JHEP} {\bfseries 01} (2006) 031} [\href{https://arxiv.org/abs/hep-th/0510173}{{\ttfamily hep-th/0510173}}].

\bibitem{Bonelli:2013pva}
G.~Bonelli, S.~Giacomelli, K.~Maruyoshi and A.~Tanzini, \emph{{N=1 Geometries via M-theory}}, \href{https://doi.org/10.1007/JHEP10(2013)227}{\emph{JHEP} {\bfseries 10} (2013) 227} [\href{https://arxiv.org/abs/1307.7703}{{\ttfamily 1307.7703}}].

\bibitem{Nakajima:2005fg}
H.~Nakajima and K.~Yoshioka, \emph{{Instanton counting on blowup. II. K-theoretic partition function}},  \href{https://arxiv.org/abs/math/0505553}{{\ttfamily math/0505553}}.

\bibitem{Bershtein:2016aef}
M.A.~Bershtein and A.I.~Shchechkin, \emph{{q-deformed Painlev\'e $\tau$ function and q-deformed conformal blocks}}, \href{https://doi.org/10.1088/1751-8121/aa5572}{\emph{J. Phys. A} {\bfseries 50} (2017) 085202} [\href{https://arxiv.org/abs/1608.02566}{{\ttfamily 1608.02566}}].

\bibitem{Bonelli:2017gdk}
G.~Bonelli, A.~Grassi and A.~Tanzini, \emph{{Quantum curves and $q$-deformed Painlev\'e equations}}, \href{https://doi.org/10.1007/s11005-019-01174-y}{\emph{Lett. Math. Phys.} {\bfseries 109} (2019) 1961} [\href{https://arxiv.org/abs/1710.11603}{{\ttfamily 1710.11603}}].

\bibitem{Bershtein:2017swf}
M.~Bershtein, P.~Gavrylenko and A.~Marshakov, \emph{{Cluster integrable systems, $q$-Painlev\'e equations and their quantization}}, \href{https://doi.org/10.1007/JHEP02(2018)077}{\emph{JHEP} {\bfseries 02} (2018) 077} [\href{https://arxiv.org/abs/1711.02063}{{\ttfamily 1711.02063}}].

\bibitem{Bonelli:2020dcp}
G.~Bonelli, F.~Del~Monte and A.~Tanzini, \emph{{BPS Quivers of Five-Dimensional SCFTs, Topological Strings and q-Painlev\'e Equations}}, \href{https://doi.org/10.1007/s00023-021-01034-3}{\emph{Annales Henri Poincare} {\bfseries 22} (2021) 2721} [\href{https://arxiv.org/abs/2007.11596}{{\ttfamily 2007.11596}}].

\bibitem{Bershtein:2018srt}
M.~Bershtein, P.~Gavrylenko and A.~Marshakov, \emph{{Cluster Toda chains and Nekrasov functions}}, \href{https://doi.org/10.1134/S0040577919020016}{\emph{Theor. Math. Phys.} {\bfseries 198} (2019) 157} [\href{https://arxiv.org/abs/1804.10145}{{\ttfamily 1804.10145}}].

\bibitem{Gavrylenko:2025nuo}
P.~Gavrylenko, \emph{{Riemann-Hilbert problems, Fredholm determinants, explicit combinatorial expansions, and connection formulas for the general $q$-Painlev\'e III$_3$ tau functions}},  \href{https://arxiv.org/abs/2501.01419}{{\ttfamily 2501.01419}}.

\bibitem{Sciarappa:2016ctj}
A.~Sciarappa, \emph{{Bethe/Gauge correspondence in odd dimension: modular double, non-perturbative corrections and open topological strings}}, \href{https://doi.org/10.1007/JHEP10(2016)014}{\emph{JHEP} {\bfseries 10} (2016) 014} [\href{https://arxiv.org/abs/1606.01000}{{\ttfamily 1606.01000}}].

\bibitem{Grassi:2017qee}
A.~Grassi and M.~Marino, \emph{{The complex side of the TS/ST correspondence}}, \href{https://doi.org/10.1088/1751-8121/aaec4b}{\emph{J. Phys. A} {\bfseries 52} (2019) 055402} [\href{https://arxiv.org/abs/1708.08642}{{\ttfamily 1708.08642}}].

\bibitem{Wang:2023zcb}
X.~Wang, \emph{{Wilson loops, holomorphic anomaly equations and blowup equations}},  \href{https://arxiv.org/abs/2305.09171}{{\ttfamily 2305.09171}}.

\bibitem{Lukyanov:2011wd}
S.L.~Lukyanov, \emph{{Critical values of the Yang-Yang functional in the quantum sine-Gordon model}}, \href{https://doi.org/10.1016/j.nuclphysb.2011.07.028}{\emph{Nucl. Phys. B} {\bfseries 853} (2011) 475} [\href{https://arxiv.org/abs/1105.2836}{{\ttfamily 1105.2836}}].

\bibitem{Fioravanti:2019vxi}
D.~Fioravanti and D.~Gregori, \emph{{Integrability and cycles of deformed ${\cal N}=2$ gauge theory}}, \href{https://doi.org/10.1016/j.physletb.2020.135376}{\emph{Phys. Lett. B} {\bfseries 804} (2020) 135376} [\href{https://arxiv.org/abs/1908.08030}{{\ttfamily 1908.08030}}].

\bibitem{Alkalaev:2017bzx}
K.B.~Alkalaev and V.A.~Belavin, \emph{{Holographic duals of large-c torus conformal blocks}}, \href{https://doi.org/10.1007/JHEP10(2017)140}{\emph{JHEP} {\bfseries 10} (2017) 140} [\href{https://arxiv.org/abs/1707.09311}{{\ttfamily 1707.09311}}].

\bibitem{Alkalaev:2016ptm}
K.B.~Alkalaev and V.A.~Belavin, \emph{{Holographic interpretation of 1-point toroidal block in the semiclassical limit}}, \href{https://doi.org/10.1007/JHEP06(2016)183}{\emph{JHEP} {\bfseries 06} (2016) 183} [\href{https://arxiv.org/abs/1603.08440}{{\ttfamily 1603.08440}}].

\bibitem{Alekseev:2019gkl}
S.~Alekseev, A.~Gorsky and M.~Litvinov, \emph{{Toward the Pole}}, \href{https://doi.org/10.1007/JHEP03(2020)157}{\emph{JHEP} {\bfseries 03} (2020) 157} [\href{https://arxiv.org/abs/1911.01334}{{\ttfamily 1911.01334}}].

\bibitem{Greene:1997fu}
P.B.~Greene, L.~Kofman, A.D.~Linde and A.A.~Starobinsky, \emph{{Structure of resonance in preheating after inflation}}, \href{https://doi.org/10.1103/PhysRevD.56.6175}{\emph{Phys. Rev. D} {\bfseries 56} (1997) 6175} [\href{https://arxiv.org/abs/hep-ph/9705347}{{\ttfamily hep-ph/9705347}}].

\bibitem{Amin:2011hj}
M.A.~Amin, R.~Easther, H.~Finkel, R.~Flauger and M.P.~Hertzberg, \emph{{Oscillons After Inflation}}, \href{https://doi.org/10.1103/PhysRevLett.108.241302}{\emph{Phys. Rev. Lett.} {\bfseries 108} (2012) 241302} [\href{https://arxiv.org/abs/1106.3335}{{\ttfamily 1106.3335}}].

\bibitem{Iorgov:2014vla}
N.~Iorgov, O.~Lisovyy and J.~Teschner, \emph{{Isomonodromic tau-functions from Liouville conformal blocks}}, \href{https://doi.org/10.1007/s00220-014-2245-0}{\emph{Commun. Math. Phys.} {\bfseries 336} (2015) 671} [\href{https://arxiv.org/abs/1401.6104}{{\ttfamily 1401.6104}}].

\bibitem{Itzykson:1986pj}
C.~Itzykson and J.B.~Zuber, \emph{{Two-Dimensional Conformal Invariant Theories on a Torus}}, \href{https://doi.org/10.1016/0550-3213(86)90576-6}{\emph{Nucl. Phys. B} {\bfseries 275} (1986) 580}.

\bibitem{Bonelli:2019boe}
G.~Bonelli, F.~Del~Monte, P.~Gavrylenko and A.~Tanzini, \emph{{${\mathcal {N}}$ = $2^*$ Gauge Theory, Free Fermions on the Torus and Painlev\'e VI}}, \href{https://doi.org/10.1007/s00220-020-03743-y}{\emph{Commun. Math. Phys.} {\bfseries 377} (2020) 1381} [\href{https://arxiv.org/abs/1901.10497}{{\ttfamily 1901.10497}}].

\bibitem{Hadasz:2009db}
L.~Hadasz, Z.~Jaskolski and P.~Suchanek, \emph{{Recursive representation of the torus 1-point conformal block}}, \href{https://doi.org/10.1007/JHEP01(2010)063}{\emph{JHEP} {\bfseries 01} (2010) 063} [\href{https://arxiv.org/abs/0911.2353}{{\ttfamily 0911.2353}}].

\bibitem{Cho:2017oxl}
M.~Cho, S.~Collier and X.~Yin, \emph{{Recursive Representations of Arbitrary Virasoro Conformal Blocks}}, \href{https://doi.org/10.1007/JHEP04(2019)018}{\emph{JHEP} {\bfseries 04} (2019) 018} [\href{https://arxiv.org/abs/1703.09805}{{\ttfamily 1703.09805}}].

\bibitem{Moore:1988qv}
G.W.~Moore and N.~Seiberg, \emph{{Classical and Quantum Conformal Field Theory}}, \href{https://doi.org/10.1007/BF01238857}{\emph{Commun. Math. Phys.} {\bfseries 123} (1989) 177}.

\bibitem{Zamolodchikov:1995aa}
A.B.~Zamolodchikov and A.B.~Zamolodchikov, \emph{{Structure constants and conformal bootstrap in Liouville field theory}}, \href{https://doi.org/10.1016/0550-3213(96)00351-3}{\emph{Nucl. Phys. B} {\bfseries 477} (1996) 577} [\href{https://arxiv.org/abs/hep-th/9506136}{{\ttfamily hep-th/9506136}}].

\bibitem{connections}
H.~Desiraju, \emph{private communication}, .

\bibitem{Wang:1989}
Z.X.~Wang and D.R.~Guo, \emph{{Special Functions}}, World Scientific Publishing, Singapore (1989).

\bibitem{Gaiotto:2009ma}
D.~Gaiotto, \emph{{Asymptotically free $\mathcal{N} = 2$ theories and irregular conformal blocks}}, \href{https://doi.org/10.1088/1742-6596/462/1/012014}{\emph{J. Phys. Conf. Ser.} {\bfseries 462} (2013) 012014} [\href{https://arxiv.org/abs/0908.0307}{{\ttfamily 0908.0307}}].

\bibitem{Nekrasov:2003rj}
N.~Nekrasov and A.~Okounkov, \emph{{Seiberg-Witten theory and random partitions}}, \href{https://doi.org/10.1007/0-8176-4467-9_15}{\emph{Prog. Math.} {\bfseries 244} (2006) 525} [\href{https://arxiv.org/abs/hep-th/0306238}{{\ttfamily hep-th/0306238}}].

\bibitem{nakajima2005instanton}
H.~Nakajima and K.~Yoshioka, \emph{{Instanton counting on blowup. 1.}}, \href{https://doi.org/10.1007/s00222-005-0444-1}{\emph{Invent. Math.} {\bfseries 162} (2005) 313} [\href{https://arxiv.org/abs/math/0306198}{{\ttfamily math/0306198}}].

\bibitem{Jeong:2020uxz}
S.~Jeong and N.~Nekrasov, \emph{{Riemann-Hilbert correspondence and blown up surface defects}}, \href{https://doi.org/10.1007/JHEP12(2020)006}{\emph{JHEP} {\bfseries 12} (2020) 006} [\href{https://arxiv.org/abs/2007.03660}{{\ttfamily 2007.03660}}].

\bibitem{bershtein2013coupling}
M.~Bershtein, B.~Feigin and A.~Litvinov, \emph{{Coupling of two conformal field theories and Nakajima-Yoshioka blow-up equations}}, \href{https://doi.org/10.1007/s11005-015-0802-x}{\emph{Lett. Math. Phys.} {\bfseries 106} (2016) 29} [\href{https://arxiv.org/abs/1310.7281}{{\ttfamily 1310.7281}}].

\bibitem{Shchechkin:2020ryb}
A.~Shchechkin, \emph{{Blowup relations on $\mathbb{C}^2/\mathbb{Z}_2$ from Nakajima{\textendash}Yoshioka blowup relations}}, \href{https://doi.org/10.1134/S0040577921020070}{\emph{Teor. Mat. Fiz.} {\bfseries 206} (2021) 225} [\href{https://arxiv.org/abs/2006.08582}{{\ttfamily 2006.08582}}].

\bibitem{Bonelli:2019yjd}
G.~Bonelli, F.~Del~Monte, P.~Gavrylenko and A.~Tanzini, \emph{{Circular quiver gauge theories, isomonodromic deformations and $W_N$ fermions on the torus}}, \href{https://doi.org/10.1007/s11005-020-01343-4}{\emph{Lett. Math. Phys.} {\bfseries 111} (2021) 83} [\href{https://arxiv.org/abs/1909.07990}{{\ttfamily 1909.07990}}].

\bibitem{Oleg-Pasha-Misha}
M.~Bershtein, P.~Gavrylenko, O.~Lisovyy and Y.~Zhuravlov, ``{Logarithimic Painlevé functions and Mathieu stability chart}.'' to appear.

\bibitem{OlegTalk}
O.~Lisovyy, ``{Logarithmic Painlevé functions and Mathieu stability chart}.'' \url{https://indico.sissa.it/event/41/contributions/741/attachments/300/390/Lisovyi-Dubrovin-2021.pdf}, June, 2021.

\bibitem{OlegMovie}
O.~Lisovyy, ``{Logarithmic Painlevé functions and Mathieu stability chart}.'' \url{https://www.youtube.com/watch?v=_onGDVsnTvI}, July, 2023.

\bibitem{Seiberg:1994rs}
N.~Seiberg and E.~Witten, \emph{{Electric - magnetic duality, monopole condensation, and confinement in N=2 supersymmetric Yang-Mills theory}}, \href{https://doi.org/10.1016/0550-3213(94)90124-4}{\emph{Nucl. Phys. B} {\bfseries 426} (1994) 19} [\href{https://arxiv.org/abs/hep-th/9407087}{{\ttfamily hep-th/9407087}}].

\bibitem{Seiberg:1994aj}
N.~Seiberg and E.~Witten, \emph{{Monopoles, duality and chiral symmetry breaking in N=2 supersymmetric QCD}}, \href{https://doi.org/10.1016/0550-3213(94)90214-3}{\emph{Nucl. Phys. B} {\bfseries 431} (1994) 484} [\href{https://arxiv.org/abs/hep-th/9408099}{{\ttfamily hep-th/9408099}}].

\bibitem{Nekrasov:2017rqy}
N.~Nekrasov, \emph{{BPS/CFT correspondence IV: sigma models and defects in gauge theory}}, \href{https://doi.org/10.1007/s11005-018-1115-7}{\emph{Lett. Math. Phys.} {\bfseries 109} (2019) 579} [\href{https://arxiv.org/abs/1711.11011}{{\ttfamily 1711.11011}}].

\bibitem{Nekrasov:2017gzb}
N.~Nekrasov, \emph{{BPS/CFT correspondence V: BPZ and KZ equations from qq-characters}},  \href{https://arxiv.org/abs/1711.11582}{{\ttfamily 1711.11582}}.

\bibitem{Alday:2009fs}
L.F.~Alday, D.~Gaiotto, S.~Gukov, Y.~Tachikawa and H.~Verlinde, \emph{{Loop and surface operators in N=2 gauge theory and Liouville modular geometry}}, \href{https://doi.org/10.1007/JHEP01(2010)113}{\emph{JHEP} {\bfseries 01} (2010) 113} [\href{https://arxiv.org/abs/0909.0945}{{\ttfamily 0909.0945}}].

\bibitem{Kanno:2011fw}
H.~Kanno and Y.~Tachikawa, \emph{{Instanton counting with a surface operator and the chain-saw quiver}}, \href{https://doi.org/10.1007/JHEP06(2011)119}{\emph{JHEP} {\bfseries 06} (2011) 119} [\href{https://arxiv.org/abs/1105.0357}{{\ttfamily 1105.0357}}].

\bibitem{Bershtein:2018zcz}
M.~Bershtein and A.~Shchechkin, \emph{{Painlev\'e equations from Nakajima\textendash{}Yoshioka blowup relations}}, \href{https://doi.org/10.1007/s11005-019-01198-4}{\emph{Lett. Math. Phys.} {\bfseries 109} (2019) 2359} [\href{https://arxiv.org/abs/1811.04050}{{\ttfamily 1811.04050}}].

\bibitem{Bruzzo:2002xf}
U.~Bruzzo, F.~Fucito, J.F.~Morales and A.~Tanzini, \emph{{Multiinstanton calculus and equivariant cohomology}}, \href{https://doi.org/10.1088/1126-6708/2003/05/054}{\emph{JHEP} {\bfseries 05} (2003) 054} [\href{https://arxiv.org/abs/hep-th/0211108}{{\ttfamily hep-th/0211108}}].

\bibitem{Flume:2002az}
R.~Flume and R.~Poghossian, \emph{{An Algorithm for the microscopic evaluation of the coefficients of the Seiberg-Witten prepotential}}, \href{https://doi.org/10.1142/S0217751X03013685}{\emph{Int. J. Mod. Phys. A} {\bfseries 18} (2003) 2541} [\href{https://arxiv.org/abs/hep-th/0208176}{{\ttfamily hep-th/0208176}}].

\bibitem{Katz:1997eq}
S.~Katz, P.~Mayr and C.~Vafa, \emph{{Mirror symmetry and exact solution of 4-D N=2 gauge theories: 1.}}, \href{https://doi.org/10.4310/ATMP.1997.v1.n1.a2}{\emph{Adv. Theor. Math. Phys.} {\bfseries 1} (1998) 53} [\href{https://arxiv.org/abs/hep-th/9706110}{{\ttfamily hep-th/9706110}}].

\bibitem{Gaiotto:2009we}
D.~Gaiotto, \emph{{N=2 dualities}}, \href{https://doi.org/10.1007/JHEP08(2012)034}{\emph{JHEP} {\bfseries 08} (2012) 034} [\href{https://arxiv.org/abs/0904.2715}{{\ttfamily 0904.2715}}].

\bibitem{Nekrasov:2012xe}
N.~Nekrasov and V.~Pestun, \emph{{Seiberg-Witten Geometry of Four-Dimensional $\mathcal N=2$ Quiver Gauge Theories}}, \href{https://doi.org/10.3842/SIGMA.2023.047}{\emph{SIGMA} {\bfseries 19} (2023) 047} [\href{https://arxiv.org/abs/1211.2240}{{\ttfamily 1211.2240}}].

\bibitem{doi:10.1142/S0219530504000023}
H.~Volkmer, \emph{Four remarks on the eigenvalues of lam\'e's equation}, \href{https://doi.org/10.1142/S0219530504000023}{\emph{Analysis and Applications} {\bfseries 02} (2004) 161}.

\bibitem{practicalQM}
S.~Flügge, \emph{{Practical Quantum Mechanics}}, vol.~1 of \emph{Classics in Mathematics}, Springer Berlin, Heidelberg (1998), \href{https://doi.org/10.1007/978-3-642-61995-3}{10.1007/978-3-642-61995-3}.

\bibitem{Poghossian:2009mk}
R.~Poghossian, \emph{{Recursion relations in CFT and N=2 SYM theory}}, \href{https://doi.org/10.1088/1126-6708/2009/12/038}{\emph{JHEP} {\bfseries 12} (2009) 038} [\href{https://arxiv.org/abs/0909.3412}{{\ttfamily 0909.3412}}].

\end{thebibliography}\endgroup
%% or
%% [B] Manual formatting (see below)
%% (i) We suggest to always provide author, title and journal data or doi:
%% in short all the informations that clearly identify a document.
%% (ii) please avoid comments such as "For a review'', "For some examples",
%% "and references therein" or move them in the text. In general, please leave only references in the bibliography and move all
%% accessory text in footnotes.
%% (iii) Also, please have only one work for each \bibitem.

\end{document}